\documentclass[10pt,logo]{nvidiatechreport}

\usepackage{graphicx}
\usepackage[acronym]{glossaries}
\usepackage{siunitx}
\usepackage{algorithm}
\usepackage{algpseudocode}
\usepackage{xcolor}
\usepackage{subcaption}
\usepackage{nicefrac}
\usepackage{framed}
\usepackage{amssymb}
\usepackage{mathtools}
\renewcommand*{\glossaryentrynumbers}[1]{}
\makeglossaries

\DeclareSymbolFont{matha}{OML}{txmi}{m}{it}
\DeclareMathSymbol{\varv}{\mathord}{matha}{118}

\definecolor{shadecolor}{rgb}{0.95, 0.95, 0.95}

\newenvironment{highlightbox}[1]
  {
    \begin{shaded}
    \noindent\textbf{\underline{\large #1}}\par\vspace{0.5em}
  }
  {
    \end{shaded}
  }

\newacronym{CIR}{CIR}{channel impulse response}
\newacronym{RIS}{RIS}{reconfigurable intelligent surface}
\newacronym{JIT}{JIT}{just-in-time}
\newacronym{EM}{EM}{electromagnetic}
\newacronym{SBR}{SBR}{shooting and bouncing of rays}
\newacronym{GCS}{GCS}{global coordinate system}
\newacronym{BSDF}{BSDF}{bidirectional scattering distribution function}
\newacronym{LoS}{LoS}{line-of-sight}
\newacronym{CFR}{CFR}{channel frequency response}
\renewcommand{\vec}[1]{\mathbf{#1}}

\newcommand{\dv}{\vec{d}}
\newcommand{\ev}{\vec{e}}

\newcommand{\kv}{\vec{k}}

\newcommand{\nv}{\vec{n}}
\newcommand{\ov}{\vec{o}}
\newcommand{\pv}{\vec{p}}
\newcommand{\qv}{\vec{q}}
\newcommand{\rv}{\vec{r}}
\newcommand{\sv}{\vec{s}}
\newcommand{\tv}{\vec{t}}
\newcommand{\uv}{\vec{u}}
\newcommand{\vv}{\vec{v}}

\newcommand{\zerov}{\vec{0}}

\newcommand{\Cm}{\vec{C}}

\newcommand{\Em}{\vec{E}}

\newcommand{\Hm}{\vec{H}}
\newcommand{\Id}{\vec{I}}

\newcommand{\Rm}{\vec{R}}

\newcommand{\Tm}{\vec{T}}

\newcommand{\Wm}{\vec{W}}

\newcommand{\Ac}{{\cal A}}

\newcommand{\Dc}{{\cal D}}
\newcommand{\Ec}{{\cal E}}

\newcommand{\Ic}{{\cal I}}

\newcommand{\Lc}{{\cal L}}
\newcommand{\Mc}{{\cal M}}

\newcommand{\Pc}{{\cal P}}
\newcommand{\Qc}{{\cal Q}}
\newcommand{\Rc}{{\cal R}}
\newcommand{\Sc}{{\cal S}}
\newcommand{\Tc}{{\cal T}}
\newcommand{\Uc}{{\cal U}}

\newcommand{\Vc}{{\cal V}}

\newcommand{\CC}{\mathbb{C}}

\newcommand{\NN}{\mathbb{N}}
\newcommand{\RR}{\mathbb{R}}

\newcommand{\htp}{^{\mathsf{H}}}
\newcommand{\tp}{^{\mathsf{T}}}

\newcommand{\LB}{\left(}
\newcommand{\RB}{\right)}
\newcommand{\LP}{\left\{}
\newcommand{\RP}{\right\}}
\newcommand{\LSB}{\left[}
\newcommand{\RSB}{\right]}

 \newcommand{\SRT}{Sionna~RT}

\newcommand\norm[1]{\left\lVert#1\right\rVert_2}
\newcommand\abs[1]{\left| #1 \right|}

\renewcommand{\reportversion}{2.1}

\begin{document}


\title{Sionna RT: Technical Report}
\author{Fayçal Aït Aoudia, Jakob Hoydis, Merlin Nimier-David, Baptiste Nicolet, Sebastian Cammerer, and Alexander Keller}
\correspondingauthor{faitaoudia@nvidia.com}

\maketitle


\begin{abstract}
	Sionna\texttrademark~is an open-source, GPU-accelerated library that, as of version 0.14, incorporates a ray tracer, Sionna~RT, for simulating radio wave propagation.
	A unique feature of \SRT{} is differentiability, enabling the calculation of gradients for the \glspl{CIR}, radio maps, and other related metrics with respect to system and environmental parameters, such as material properties, antenna patterns, and array geometries.
	The release of Sionna~1.0 provided a complete overhaul of the ray tracer, significantly improving its speed, memory efficiency, and extensibility.
	This document details the algorithms employed by \SRT{} to simulate radio wave propagation efficiently, while also addressing their current limitations.
	Given that the computation of \glspl{CIR} and radio maps requires distinct algorithms, these are detailed in separate sections.
	For \glspl{CIR}, \SRT{} integrates \gls{SBR} with the image method and uses a hashing-based mechanism to efficiently eliminate duplicate paths. Radio maps are computed using a purely \gls{SBR}-based approach for the non-diffracted component, complemented by a stage for diffracted paths.
\end{abstract}

\glsresetall

\abscontent


\newpage

\tableofcontents

\newpage



\printglossary[type=\acronymtype, title={List of Acronyms}]

\newpage

\section{Introduction}
\label{sec:introduction}
Ray tracing for simulating radio wave propagation has received renewed attention in recent years, fueled by the growing interest in creating digital twins for wireless communication systems and research topics that require simulating spatially consistent \glspl{CIR}.
In 2023, we released \SRT{}~\cite{10465179}, the world's first fully differentiable ray tracer for radio propagation modeling, as part of version 0.14 of the Sionna open-source library for research in communication systems~\cite{sionna}.
Over the following years, we have expanded the capabilities of \SRT{} with additional features, including support for diffraction and diffuse reflection (version 0.15), mobility (version 0.17), and \glspl{RIS} (version 0.18).

\SRT{} is built on top of the \gls{JIT} compiler Dr.Jit~\cite{Jakob2022DrJit} and the differentiable physically-based renderer Mitsuba~\cite{Mitsuba3}.
\SRT{} implements ray-tracing algorithms to model the propagation of \gls{EM} radio waves, enabling the computation of \glspl{CIR} and radio maps (also known as \emph{coverage maps} or \emph{power maps}).
Additionally, it includes radio material models to simulate the interaction of radio waves with scatterers in a radio propagation environment.
The release of Sionna~1.0 marks a significant milestone as it features a complete revision of the ray-tracing module.\footnote{
        As of version 2.1, \SRT{} does not support \glspl{RIS}.
        It remains available as part of Sionna version 0.19.2.
}
The ray tracer is now interoperable with major deep learning frameworks, including PyTorch~\cite{pytorch} and TensorFlow~\cite{tensorflow2015-whitepaper}.
Moreover, the computation of \glspl{CIR} and radio maps has been substantially accelerated while maintaining full differentiability of the ray-tracing process by leveraging the automatic differentiation capabilities of Dr.Jit.
The ray tracer has also been made easier to extend to facilitate research on various aspects of radio propagation and digital twins.

\begin{figure}[ht!]
    \center
    \includegraphics[scale=0.7]{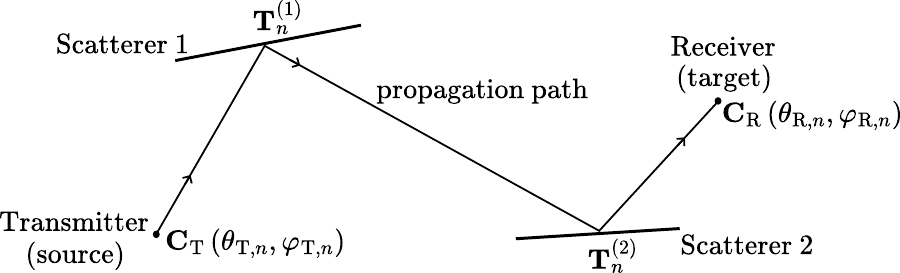}
    \caption{A propagation path, shown in 2D for clarity.\label{fig:intro-path}}
\end{figure}

The primary objective of \SRT{} is to identify the \emph{propagation paths} that link radio devices within a given propagation environment.
As depicted in Figure~\ref{fig:intro-path}, a propagation environment is defined by a collection of scatterers. A radio device functions as either a \emph{transmitter}, which emits radio waves, or a \emph{receiver}, which captures them, and is equipped with one or more antennas. The \emph{channel frequency response} $H(f)$ characterizes the propagation of radio waves from a transmitting antenna to a receiving antenna for the frequency $f$, and is expressed as the ratio of the received voltage to the voltage at the input of the transmitting antenna:
\begin{equation}
    H(f) = \frac{V_\text{R}}{V_\text{T}}.
\end{equation}
As explained in Appendix~\ref{sec:primer-em}, the channel frequency response is the Fourier transform of the \gls{CIR}, which is defined as:
\begin{equation}
    h(\tau) = \sum_{n=1}^N a_n \delta(\tau - \tau_n).
\end{equation}
Here, $N$ represents the number of \emph{propagation paths} connecting the transmit antenna to the receive antenna. The term $a_n$ is the \emph{complex-valued path coefficient} for the $n$-th path, and $\tau_n = \nicefrac{d_n}{c_0}$ is the \emph{propagation delay}, where $d_n$ is the distance the radio wave travels, and $c_0$ is the speed of light in vacuum. This leads to:
\begin{equation}
    H(f) = \int_{-\infty}^\infty h(\tau) e^{-j2\pi f\tau} \, d\tau = \sum_{n=1}^N a_n e^{-j2\pi f\tau_n}.
\end{equation}
Note that each $a_n$ is evaluated at a fixed carrier frequency denoted by $f_c$ and assumed constant over the simulated bandwidth.
The path coefficient $a_n$ is calculated by considering the sequence of matrices that model the transformations caused by interactions with scatterers along the $n$-th propagation path, as well as the characteristics of the transmit and receive antennas:
\begin{equation}
    \label{eq:path-coefficient}
    a_n = \frac{\lambda_0}{4\pi} \mathbf{C}_\text{R}(\theta_{\text{R},n}, \varphi_{\text{R},n})^{\mathsf{H}} A_n \mathbf{T}_n^{(\ell)} \cdots \mathbf{T}_n^{(1)} \mathbf{C}_\text{T}(\theta_{\text{T},n}, \varphi_{\text{T},n}).
\end{equation}
Here, $\lambda_0=c_0/f_c$ is the free-space wavelength, $\ell$ is the number of scatterers along the $n$-th path, $\mathbf{C}_\text{T}$ is the complex-valued vector of dimension 2 representing the \emph{transmit antenna pattern} evaluated for the angles of departure of the path $(\theta_{\text{T},n}, \varphi_{\text{T},n})$, $\mathbf{C}_\text{R}$ is the complex-valued vector of dimension 2 representing the \emph{receive antenna pattern} evaluated for the angles of arrival of the path $(\theta_{\text{R},n}, \varphi_{\text{R},n})$, and $\mathbf{T}_n^{(\ell')}$ is the $2 \times 2$ complex-valued matrix modeling the interaction with the $\ell'$-th scatterer. The rightmost matrix $\mathbf{T}_n^{(1)}$ acts first on the transmitted field, followed by the remaining matrices in increasing interaction order. For a path without interactions, the matrix product is the identity. The scalar $A_n$ is the spreading factor. Further details on the computation of the path coefficient are provided in Section~\ref{sec:path-solver}.
The corresponding \emph{non-coherent channel gain} is
\begin{equation}
    \label{eq:channel-gain}
    g = \sum_{n=1}^N \abs{a_n}^2.
\end{equation}
Calculating the path coefficients and delays requires first determining the intersection points of the paths with the scene geometry, referred to as the \emph{path vertices}. This is because computing the path coefficients~\eqref{eq:path-coefficient} requires the knowledge of the geometrical features of the paths. Specifically, the antenna patterns are evaluated at the angles of departure and arrival of the paths, and computing the matrices $\mathbf{T}_n^{(\ell')}$ requires the angles of incidence and scattering of waves at the scatterers as well as the traveled distances. Once the path vertices have been determined, the path coefficients and delays are computed, considering the characteristics of the antennas and scatterers' materials.

Paths are determined between two endpoints: a \emph{source} and a \emph{target}. Ideally, paths would be traced between every pair of transmitter and receiver antennas, designating the sources and targets as the set of transmit and receive antennas, respectively. However, when radio devices are equipped with a large number of antennas, computing paths for all antenna pairs becomes computationally demanding. To mitigate this issue, \SRT{} introduces a \emph{synthetic array} feature, which calculates paths only between radio devices instead of each antenna pair. Phase shifts due to the array geometry are then applied synthetically, as detailed in Section~\ref{sec:path-solver}. Note that the use of synthetic arrays can be disabled. Throughout this document, the endpoints of a path will be referred to as sources and targets, which can be either radio devices or antennas. Although the use of synthetic arrays does not change the pathfinding process, it does affect the computation of the path coefficients.

\begin{figure}[ht!]
    \center
    \includegraphics[scale=0.7]{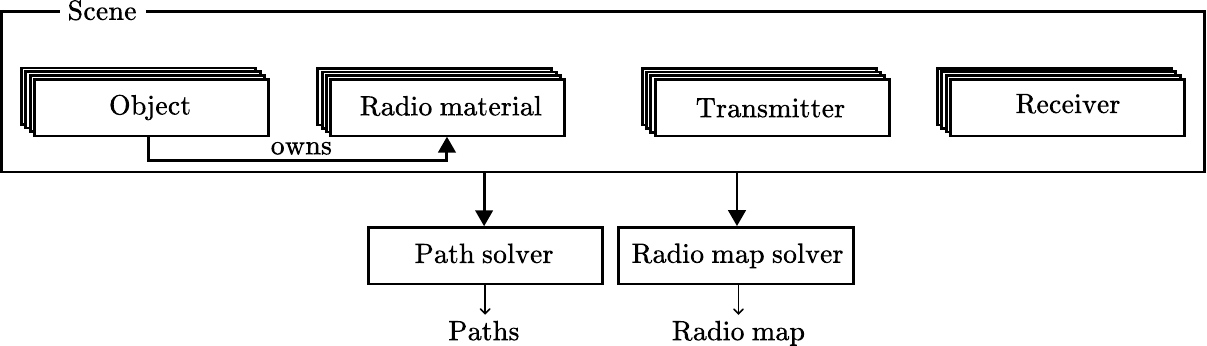}
    \caption{\SRT~architecture overview.\label{fig:basic-arch}}
\end{figure}

Figure~\ref{fig:basic-arch} illustrates the overall architecture of \SRT{}.
A \emph{scene} represents a radio propagation environment, comprising a set of \emph{objects}, \emph{radio materials}, and \emph{radio devices}.
Each object acting as a scatterer is associated with a radio material that determines its interaction with incident radio waves.
Multiple objects can share the same radio material. Radio materials are akin to \glspl{BSDF} in the field of computer graphics, see e.g.~\cite[Ch.~4.3.1]{PBRT4e}.
Note that an \emph{object} alone defines only the geometry of a scatterer and does not contain any information about the radio material.
Radio devices are either transmitters or receivers.
A \emph{solver} implements an algorithm for simulating radio wave propagation within a scene.
Given the distinct algorithmic requirements for \glspl{CIR} and radio maps, \SRT{} provides two types of solvers:
\begin{itemize}
    \item The \emph{path solver} computes a set of paths connecting a source to a target, which can then be used to compute a \gls{CIR}.
    \item The \emph{radio map solver} computes a radio map, values over a partitioned measurement surface that approximate the channel gain~\eqref{eq:channel-gain} that a receiver located on the grid would observe.
\end{itemize}
While \SRT{} provides built-in solvers and radio material models, it is designed to facilitate the implementation of custom alternatives.

\paragraph{Contributions of the Technical Report}

Beyond serving as a technical reference, this technical report provides theoretical contributions by detailing the key methods implemented in \SRT{} for simulating radio wave propagation by ray tracing.
For computing paths, a method that combines \gls{SBR} with the image method is presented. Although this approach existed in earlier versions of \SRT{}, it has been significantly refined in Sionna~1, notably by incorporating a hashing-based mechanism to efficiently remove duplicate paths.
The technical report also introduces a formal definition of radio maps, which forms the foundation for the radio map solver implemented in \SRT{}, as it enables efficient computation through \gls{SBR} alone for the non-diffracted component. It is worth noting that \gls{SBR} is already extensively used in the graphics community (see e.g.~\cite{PBRT4e}).
Moreover, we include discussions on the limitations of the current algorithms and offer perspectives for future improvements.

\paragraph{Structure of the Technical Report}

Section~\ref{sec:essential-concepts} introduces the essential concepts and terminology.
Sections~\ref{sec:path-solver} and~\ref{sec:radio-map-solver} provide an in-depth look at the built-in path solver and radio map solver of \SRT{}, respectively.
These sections primarily focus on path tracing algorithms, specifically the methods used to determine the intersection points of paths with scene geometry.
Appendix~\ref{sec:primer-em} provides a primer on \gls{EM} theory and explains the computation of the electric field along a path. The remaining appendices contain further details and derivations.

\newpage
\section{Essential Concepts and Terminology}
\label{sec:essential-concepts}
\subsection{Scene Objects and Meshes}
\label{sec:essential-concepts-scene-objects-and-meshes}

\begin{figure}[ht!]
    \centering
    \includegraphics[width=0.6\textwidth]{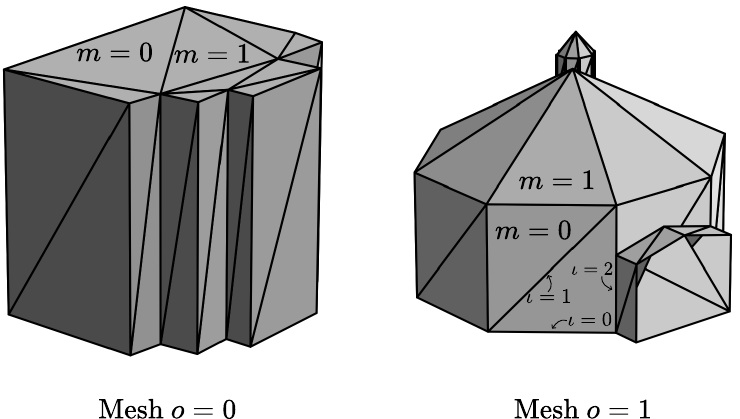}
    \caption{A mesh is composed of triangular primitives connected by their edges and vertices. For clarity, only the indices of two primitives are displayed for each mesh. These meshes were downloaded from OpenStreetMap~\cite{OpenStreetMap} using Blosm~\cite{blosm}.\label{fig:essential-concepts-mesh}}
\end{figure}

A scene object is represented as a \emph{mesh}, which is composed of a set of triangles, also known as \emph{primitives}, connected by their edges and vertices, as shown in Figure~\ref{fig:essential-concepts-mesh}.
Each object is identified by an index $o \in \NN$, and each primitive within an object is identified by an index $m \in \NN$.
Each primitive possesses three edges, identified locally by an index $\iota \in \{0, 1, 2\}$.
Primitives from different objects may share indices, therefore a primitive is uniquely identified by the pair $(o, m) \in \NN^2$.
Similarly, an edge is uniquely identified by the triplet $(o, m, \iota) \in \NN^3$.

\subsection{Rays and Paths}

\begin{figure}[ht!]
    \centering
    \includegraphics[width=0.50\textwidth]{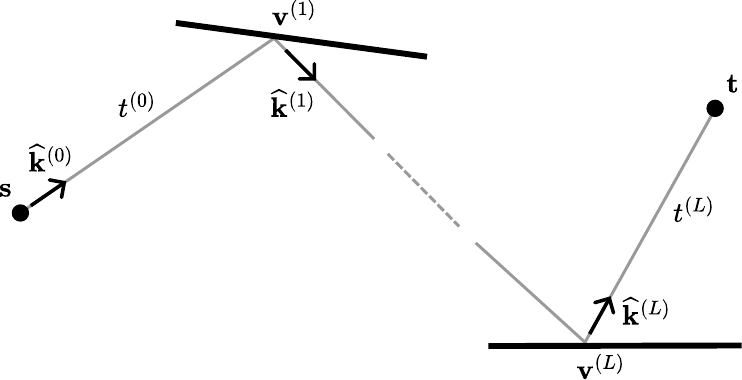}
    \caption{A path consists of a sequence of ray segments connecting a source point $\sv$ to a target point $\tv$.\label{fig:essential-concepts-path}}
\end{figure}

A \emph{ray} $\LB \vv, \widehat{\kv}, t \RB$ is characterized by an origin $\vv \in \RR^3$, a direction $\widehat{\kv} \in \RR^3$, and a length $t \geq 0$. The direction $\widehat{\kv}$ is a unit vector, meaning $\norm{\widehat{\kv}} = 1$. If the length is unspecified, the ray is considered infinite, i.e. $t = \infty$. A \emph{path} $p^{(L)}$ with \emph{depth} $L \in \NN$ is a sequence of $L+1$ rays that connect a \emph{source} point $\sv \in \RR^3$ to a \emph{target} point $\tv \in \RR^3$, as illustrated in Figure~\ref{fig:essential-concepts-path},
\begin{equation}
    p^{(L)} = \LB \underbrace{\LB \vv^{(0)}, \widehat{\kv}^{(0)}, t^{(0)} \RB}_{\text{ray 0}}, \cdots, \underbrace{\LB \vv^{(L)}, \widehat{\kv}^{(L)}, t^{(L)} \RB}_{\text{ray $L$}} \RB
\end{equation}
where the superscript $(L)$ indicates the depth of the path, $\vv^{(0)} = \sv$, $\vv^{(i+1)} = \vv^{(i)} + t^{(i)} \widehat{\kv}^{(i)}$, and $\tv = \vv^{(L)} + t^{(L)} \widehat{\kv}^{(L)}$. The ray origins together with the target are the path vertices. A path can therefore also be defined by a sequence of $L+2$ vertices:
\begin{equation}
    p^{(L)} = \left( \vv^{(0)} = \sv, \vv^{(1)}, \cdots, \vv^{(L)}, \vv^{(L+1)} = \tv \right)
\end{equation}
In this case, the rays have directions $\widehat{\kv}^{(i)} = \frac{\vv^{(i+1)} - \vv^{(i)}}{\norm{\vv^{(i+1)} - \vv^{(i)}}}$ and lengths $t^{(i)} = \norm{\vv^{(i+1)} - \vv^{(i)}}$ for $0 \leq i \leq L$.

The \emph{suffix} of a path, denoted by $p^{(\ell:)}$ where $0 \leq \ell \leq L$, is defined as the sub-path consisting of all the rays from the $\ell$-th one onward, i.e.
\begin{equation}
    p^{(\ell:)} = \left( \left(\vv^{(\ell)}, \widehat{\kv}^{(\ell)}, t^{(\ell)}\right), \cdots, \left(\vv^{(L)}, \widehat{\kv}^{(L)}, t^{(L)}\right) \right).
\end{equation}
Similarly, the corresponding \emph{prefix} is:
\begin{equation}
    p^{(:\ell)} = \left( \left(\vv^{(0)}, \widehat{\kv}^{(0)}, t^{(0)}\right), \cdots, \left(\vv^{(\ell-1)}, \widehat{\kv}^{(\ell-1)}, t^{(\ell-1)}\right) \right)
\end{equation}
where $p^{(:0)}$ corresponds to the empty path.

\subsection{Interactions with Scene Objects}

\begin{figure}[ht!]
    \centering
    \begin{subfigure}[b]{0.4\textwidth}
        \centering
        \includegraphics[width=\textwidth]{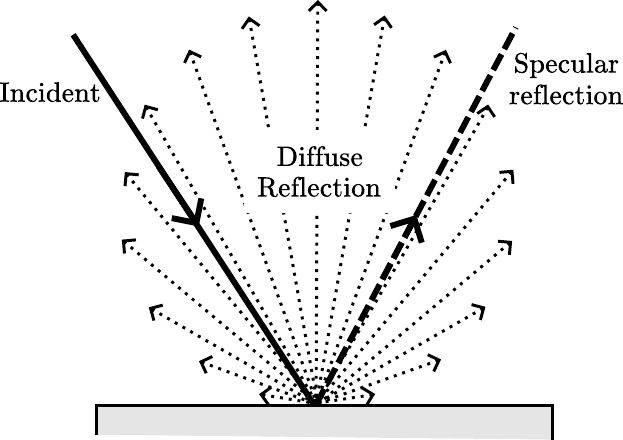}
        \caption{Specular (dashed line) and diffuse reflection (dotted lines). Inspired by a figure from~\cite{Wikipedia-Diffuse-Reflection}.\label{fig:path-solver-ref}}
    \end{subfigure}
    \hfill
    \begin{subfigure}[b]{0.4\textwidth}
        \centering
        \includegraphics[width=0.65\textwidth]{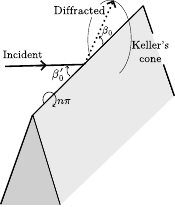}
        \caption{Diffraction of an incident wave by a wedge-shaped object.\label{fig:path-solver-diffraction-model}}
    \end{subfigure}
    \vspace{1em}

    \begin{subfigure}[b]{\textwidth}
        \centering
        \includegraphics[width=0.6\textwidth]{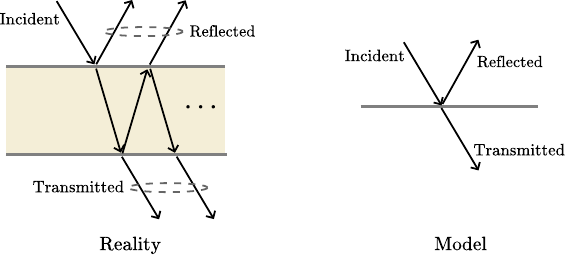}
        \caption{Transmission model.\label{fig:path-solver-trans-model}}
    \end{subfigure}
    \caption{\SRT{} currently supports specular reflection, diffuse reflection, refraction, and diffraction.}
\end{figure}

\SRT{} currently supports four types of interactions with scene objects:
\begin{description}
    \item[Specular reflection ($\Rc$):] The wave is reflected with an angle of reflection equal to the angle of incidence, as illustrated in Figure~\ref{fig:path-solver-ref}.
    \item[Diffuse reflection ($\Sc$):] The wave is reflected in multiple directions, as illustrated in Figure~\ref{fig:path-solver-ref}.
    \item[Refraction ($\Tc$):] The wave propagates into the scattering medium. The solvers assume that object surfaces are thin enough that their effect on transmitted rays (i.e. rays that traverse the surfaces through double refraction) can be modeled by a single transmitted ray. The transmitted rays are traced without angular deflection. Surfaces like walls should be modeled as single flat surfaces, as illustrated in Figure~\ref{fig:path-solver-trans-model}.
    However, when computing the transmitted and reflected fields, the thickness of the traversed object is considered.
    \item[Diffraction ($\Dc$):] The wave encounters a wedge-shaped object and is diffracted, as shown in Figure~\ref{fig:path-solver-diffraction-model}. Diffraction occurs in all directions along the \emph{Keller cone}~\cite{Keller:62}, where the angle between the incident ray and the wedge edge, denoted by $\beta_0'$, equals the angle between the diffracted ray and the wedge edge, denoted by $\beta_0$. The wedge exterior angle is denoted by $n\pi$, such that $n \in [1, 2]$. If the wedge's exterior angle $n\pi$ equals $2\pi$, i.e., $n = 2$, the wedge is termed an \emph{edge}.
\end{description}
The set of possible interaction types is denoted by $\Ic = \{\Rc, \Sc, \Tc, \Dc\}$.

\subsection{Ray Tubes}
\label{sec:essential-concepts-ray-tube}

\begin{figure}[ht!]
    \centering
    \includegraphics[width=0.4\textwidth]{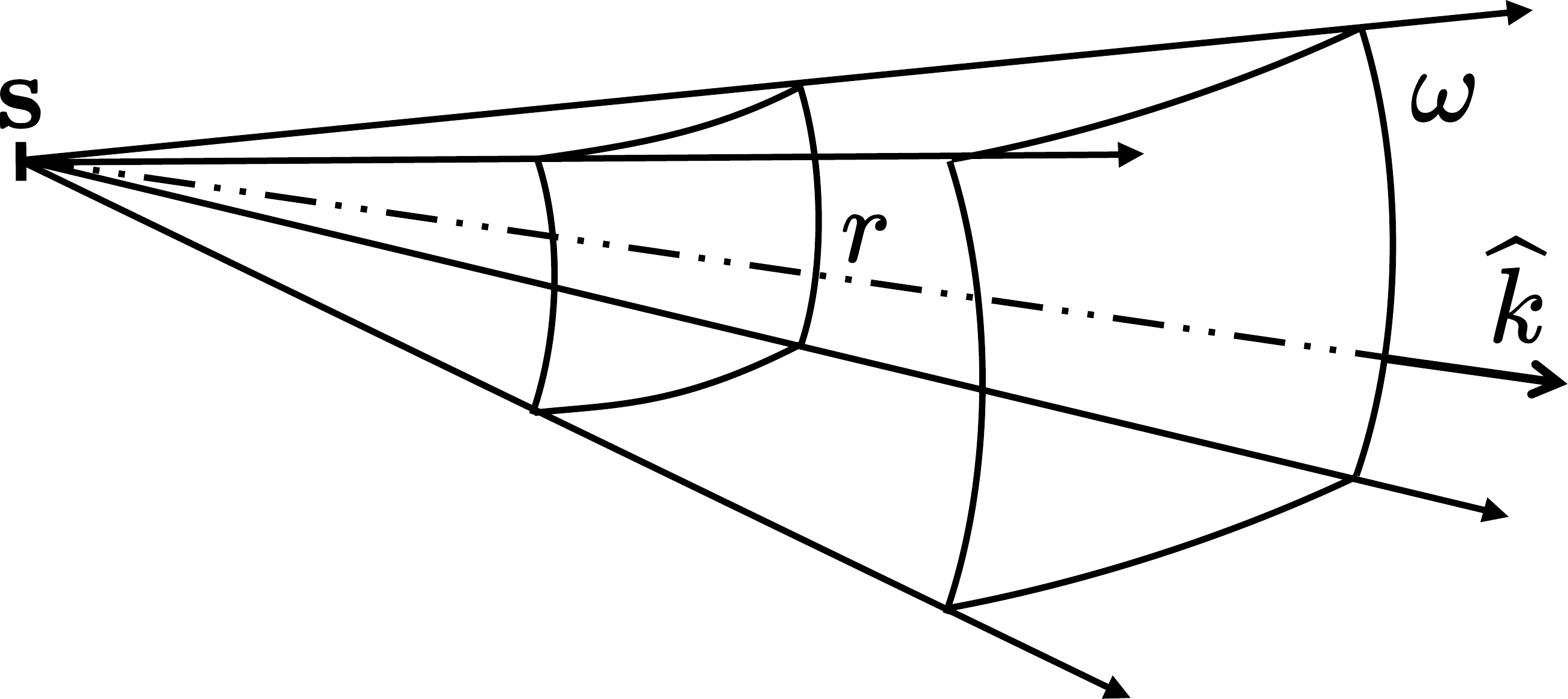}
    \caption{A ray tube is a bundle of rays adjacent to an axial ray.\label{fig:essential-ray-tube}}
\end{figure}

While a more detailed background on the propagation of \gls{EM} waves is provided in Appendix~\ref{sec:primer-em}, we introduce here some essential \gls{EM} concepts that are required for understanding the process of ray tracing. The source $\sv$ is modeled as a \emph{point source}, an infinitesimally small emitter of \gls{EM} waves that radiates in all directions according to a user-defined \emph{transmitter antenna pattern}.
Note that modeling transmit antennas as point sources is a valid approximation when the distances to scatterers and targets are significantly greater than the wavelength of the propagating waves.
Each ray traced from the source $\sv$ serves as the axial ray of a \emph{ray tube}~\cite[Chapter~2]{McNamara90}, which is a bundle of rays adjacent to the axial one. These ray tubes originate at the source and are characterized by a length $r$ and a solid angle $\omega$ (in steradians), as illustrated in Figure~\ref{fig:essential-ray-tube}. Importantly, specular reflection and refraction through planar surfaces only alter the ray direction. Diffuse reflection and diffraction alter the shape of the wavefront. As detailed in Sections~\ref{sec:path-solver} and~\ref{sec:radio-map-solver}, this process requires specific computations to determine the path vertices and corresponding fields. For a given path $p^{(L)}$ with depth $L$, we denote by $\boldsymbol{\chi}^{(L)} = \left(\chi^{(1)}, \cdots, \chi^{(L)}\right) \in \Ic^L$ the sequence of interaction types along the path. A key quantity for the remainder of this document is the depth of the last diffuse reflection, denoted by $\ell_d \in \NN$ and defined as:
\begin{equation}
    \label{eq:essential-concepts-ell_d}
    \ell_d =
    \begin{cases}
        \max_{\substack{1 \leq \ell \leq L}} \left\{ \ell~:~\chi^{(\ell)} = \Sc \right\} & \text{ if } \exists \ell : \chi^{(\ell)} = \Sc \\
        0 & \text{ otherwise.}
    \end{cases}
\end{equation}
The dependency of $\ell_d$ on $\boldsymbol{\chi}^{(L)}$ is omitted for brevity. The case where $\ell_d = 0$ corresponds to paths that do not contain any diffuse reflection, known as \emph{specular chains}. If $\ell_d > 0$, then the suffix $p^{(\ell_d:)}$ is referred to as the \emph{specular suffix} of $p$. The value $\ell_d$ represents the depth index from which the path is composed solely of specular reflections, refractions, and diffraction, and is termed the \emph{specular suffix index} of $p$.

\newpage
\section{Path Solver}
\label{sec:path-solver}

\begin{figure}[ht!]
    \centering
    \includegraphics[width=0.7\linewidth]{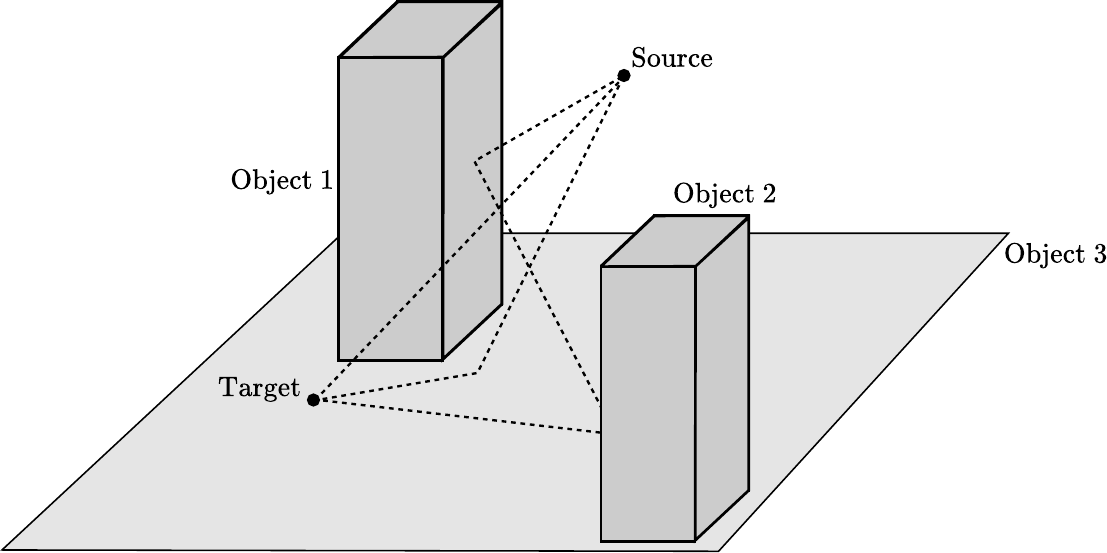}
    \caption{The path solver constructs propagation paths that connect a source and a target in a scene.\label{fig:paths}}
\end{figure}

A path solver aims to determine a set of paths that connect two endpoints in a scene, as illustrated in Figure~\ref{fig:paths}.
From this set of paths, a \gls{CIR} or \gls{CFR} can be computed.
The path solver leverages \gls{SBR} to efficiently compute paths in a scene.
\Gls{SBR} operates as follows:
First, rays\footnote{The processing of the rays is parallelized for accelerated computation.} are traced from the source into directions specified by a lattice on the unit sphere (see Section~\ref{sec:path-solver-fibonacci}).
Then, an \gls{SBR} loop iteratively tests the intersection of the rays with the scene geometry.
At every intersection, an interaction type is selected from $\Ic$, and a new ray is spawned according to the selected interaction type.
The method for choosing interaction types is detailed in Section~\ref{sec:path-solver-interaction-sampling}.
The loop terminates when no more rays remain \emph{active}, meaning they have either bounced out of the scene, reached a predefined maximum number of bounces, or satisfied another termination criterion.
Notably, at each interaction of a ray with a surface in the scene, only one scattered ray is spawned, as spawning multiple rays would result in an exponential increase in the number of rays, quickly overwhelming the available computational resources.

\begin{figure}[ht!]
    \centering
    \begin{subfigure}[b]{0.45\textwidth}
        \centering
        \includegraphics[width=\textwidth]{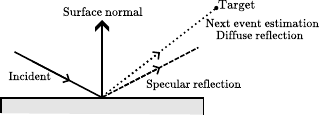}
        \subcaption{\label{fig:sbr-specular-diffuse}}
    \end{subfigure}
    \hspace{1em}
    \begin{subfigure}[b]{0.45\textwidth}
        \centering
        \includegraphics[width=\textwidth]{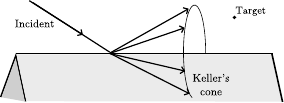}
        \subcaption{\label{fig:sbr-diffraction}}
    \end{subfigure}
    \caption{(a) When a diffuse reflection occurs, next event estimation tests whether the \gls{LoS} between the intersection point and the target is occluded (represented by the dotted line). This process is not applicable for specular reflection (represented by the dashed line) and refraction, as only one direction for the scattered ray is valid. (b) Similarly, for diffraction, the probability that the target is on the surface of the Keller cone is zero, and a single ray is spawned in a random direction along the Keller cone to continue the sample.}
\end{figure}

When a diffuse reflection occurs during the \gls{SBR} loop, the solver checks whether there is a \gls{LoS} to the target, meaning that the ray segment from the interaction point to the target point is not occluded. If such a connection exists, the path is deemed valid and is returned by the solver. This process is known as \emph{next-event estimation}. Subsequently, a new ray is spawned in a random direction within the hemisphere defined by the surface normal of the primitive involved. This new ray may lead to the discovery of additional paths.
Next-event estimation is only possible for diffuse reflections that scatter energy into all directions.
For specular reflection and refraction, only a single ray is spawned in the only possible direction.
Consequently, the probability of finding paths connecting the source to the target that end with a specular reflection or refraction by \gls{SBR} is effectively zero, as depicted in Figure~\ref{fig:sbr-specular-diffuse}.
This is because hitting a specific point (the target) with a specular reflection or refraction requires perfect alignment of the incident ray, which has zero probability in continuous space.
Likewise, for diffraction, the probability that the target is on the surface of the Keller cone is zero, as shown in Figure~\ref{fig:sbr-diffraction}.
When diffraction occurs, a single ray is spawned in a random direction along the Keller cone to continue the sample.
Paths ending with a specular suffix can therefore not be determined solely by \gls{SBR}.
In particular, specular chains cannot be computed by \gls{SBR}.
Since specular chains typically carry a significant amount of the transported energy, \gls{SBR} alone is insufficient for computing \glspl{CIR}.

\begin{figure}[ht!]
    \centering
    \includegraphics[width=0.65\textwidth]{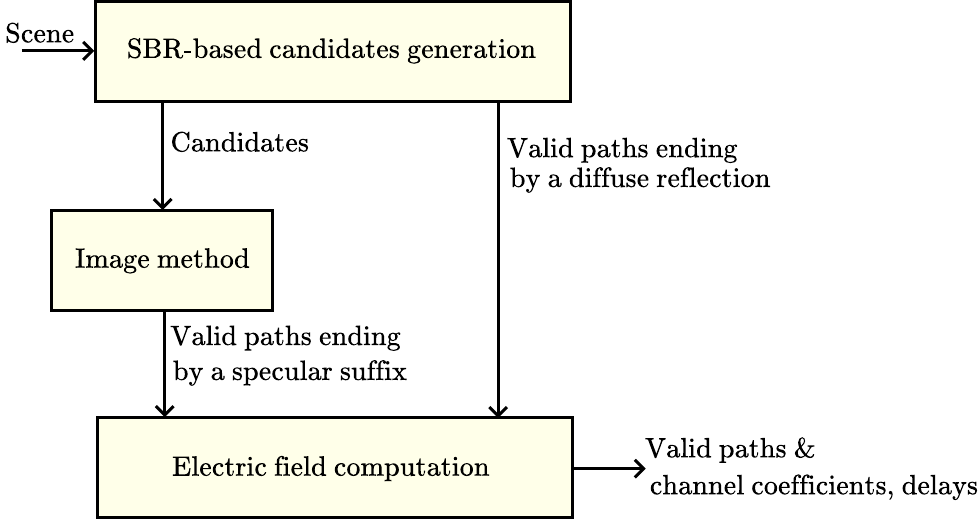}
    \caption{Main steps of the \SRT{} path solver.\label{fig:path-solver-archi}}
\end{figure}

To compute paths ending with a specular suffix, including specular chains, the path solver uses an \emph{image method}-based algorithm.
However, a brute-force implementation of the image method that would test all possible combinations of primitives and wedges that constitute the scene is prohibitive even for moderately large scenes.
Therefore, the path solver uses \gls{SBR} as a heuristic to find \emph{path candidates}.
A path candidate consists of a sequence of primitives or wedges and interaction types ending with a specular reflection, refraction, or diffraction. As the path vertices determined during the \gls{SBR} do not form a path that connects to the target, the image method is used to adjust the vertex locations to establish the connection.
Intuitively, \gls{SBR} pre-selects path candidates in the vicinity of the transmitter and receiver for the image method to compute corresponding valid paths to be returned by the solver.
These steps of the \SRT{} path solver are depicted in Figure~\ref{fig:path-solver-archi}.
Notably, only path candidates ending with a specular suffix (including specular chains) require further processing by the image method algorithm.
Finally, for each valid path identified, the path solver computes the corresponding channel coefficients and propagation delays that characterize the \gls{EM} wave propagation along that path.

Alternative methods for identifying paths that include specular chains or specular suffixes may involve detection spheres centered at the targets to capture paths ending with a specular suffix, and/or extending the direction of specular reflection (or refraction) to a lobe around the specular, refracted, or diffracted direction. Such methods only approximate the path vertex locations and the directions of the scattered rays, yielding inaccurate path coefficients and resulting in incorrect \glspl{CIR}.

The remainder of this section follows the architecture of the path solver shown in Figure~\ref{fig:path-solver-archi}.
First, Section~\ref{sec:path-solver-sbr} presents the \gls{SBR}-based candidate generator.
Section~\ref{sec:path-solver-im} then describes the image method-based algorithm that processes candidates to find valid paths with specular suffixes.
Section~\ref{sec:path-solver-em} details how the solver computes the complex-valued channel coefficients and propagation delays, which characterize the radio channel.
Finally, Section~\ref{sec:path-solver-limitations} discusses current limitations and potential improvements of the path solver.

\subsection{Generating Candidates by Shooting and Bouncing of Rays (SBR)}
\label{sec:path-solver-sbr}

\begin{figure}[ht!]
    \centering
    \includegraphics[width=0.95\textwidth]{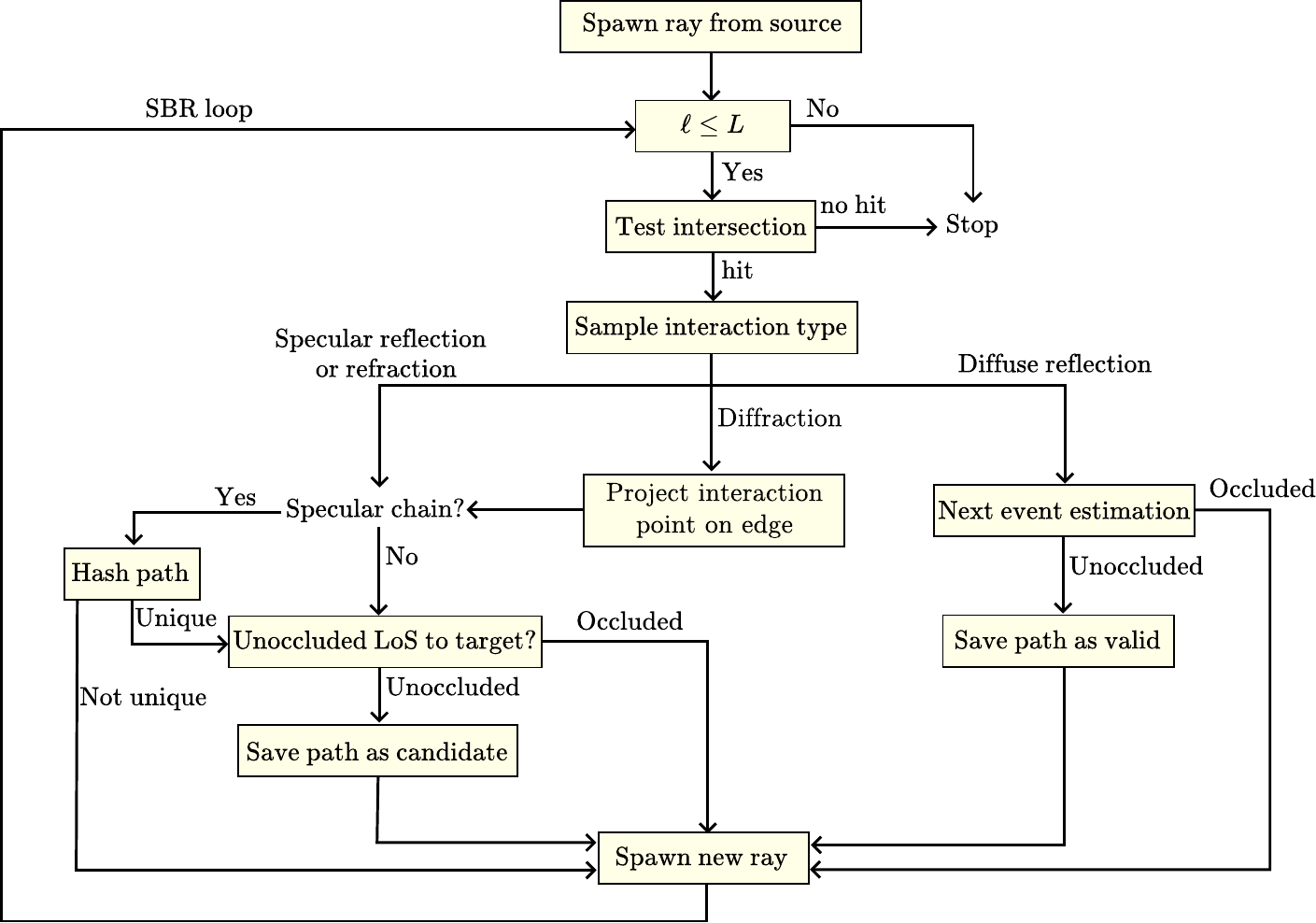}
    \caption{Algorithm for one sample as implemented in the \gls{SBR}-based candidate generator. In practice, many samples are processed in parallel.}
    \label{fig:path-solver-sbr}
\end{figure}

\begin{figure}[ht!]
    \centering
    \includegraphics[width=0.75\textwidth]{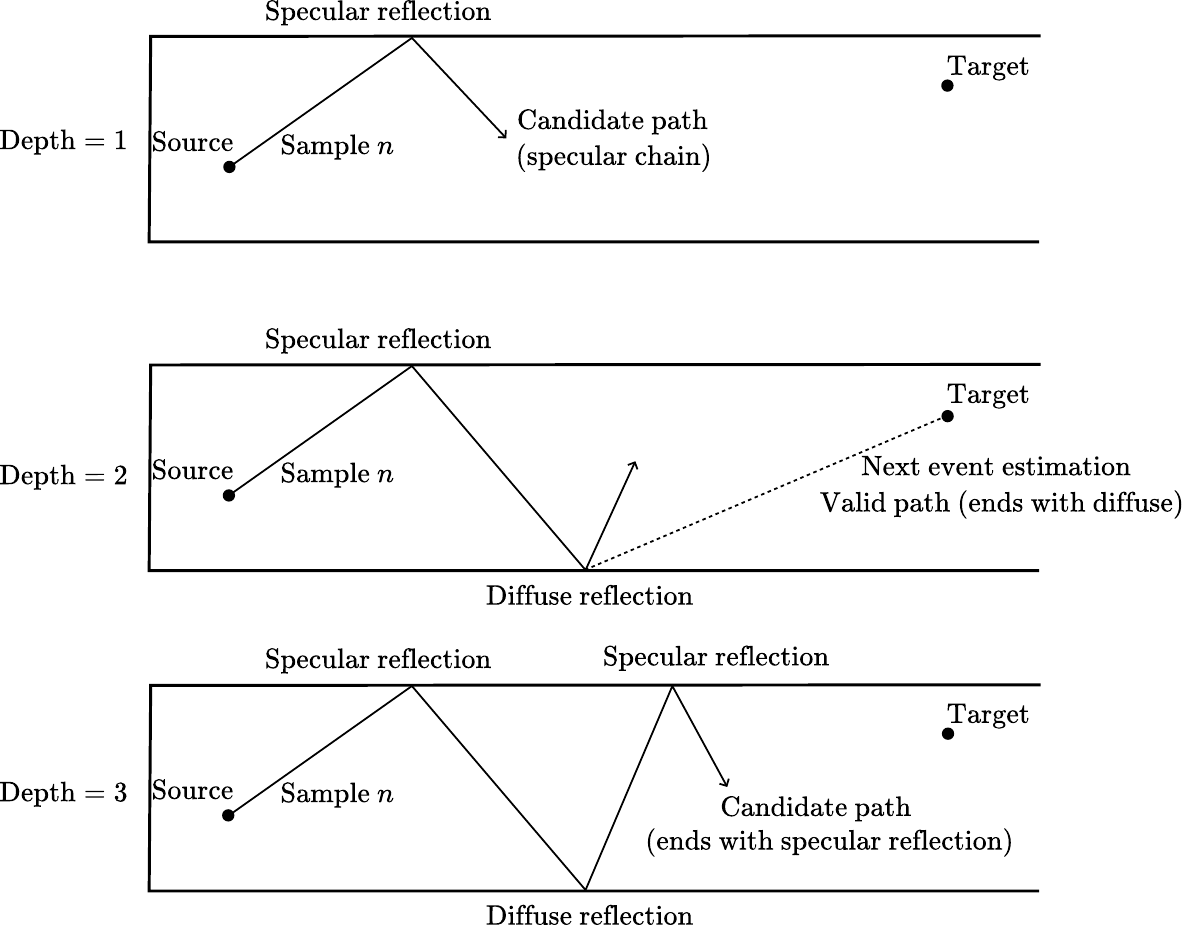}
    \caption{\gls{SBR} procedure for a single sample $n$ at depths 1, 2, and 3. At each interaction (i.e. iteration of the \gls{SBR} loop), the sample may result in a candidate or valid path, and a new ray is spawned from the current interaction point to prolong the sample path. Shown in 2D for clarity.\label{fig:path-solver-sbr-samples}}
\end{figure}

The candidate generator utilizes the \gls{SBR}-based method as illustrated in Figure~\ref{fig:path-solver-sbr}.
We first provide an overview of the \gls{SBR} procedure, followed by a detailed explanation.
We consider a scene with a single source located at $\sv \in \RR^3$ and $N_T$ targets at $\tv_k \in \RR^3$, where $k = 1, \dots, N_T$.
Although the path solver built into \SRT{} can handle multiple sources, we focus on a single source for the sake of clarity. The process is identical for all sources.
The paths traced by the candidate generator during the \gls{SBR} procedure are referred to as \emph{samples} in this section to avoid confusion with the valid paths returned to the user and the candidate paths forwarded to the image method-based algorithm.
A sample therefore consists of a sequence of rays connected end-to-end, traced by the candidate generator.
As we will see, a single sample can result in multiple valid or candidate paths, as illustrated in Figure~\ref{fig:path-solver-sbr-samples}.

The \gls{SBR} procedure begins by spawning a user-specified number of rays, denoted by $N_S$,\footnote{In the interface of \SRT{}, $N_S$ corresponds to the $\texttt{samples\_per\_src}$ parameter.} from the source $\sv$ in directions $\widehat{\kv}_n^{(0)}$ on a Fibonacci lattice, where $n = 1, \dots, N_S$.
Each ray starts a sample that is traced by the candidate generator.
Details on the implemented Fibonacci lattice can be found in Section~\ref{sec:path-solver-fibonacci}.
The \gls{SBR} loop then iterates until either all samples bounce out of the scene or reach a user-specified maximum number of bounces, denoted as the \emph{maximum depth} $L$.
We denote the depth of the samples by $\ell = 0, \dots, L$.
The first step of the \gls{SBR} loop involves testing the intersection of the samples with the scene geometry.
If a sample does not intersect any object, it has bounced out of the scene, and is deactivated.
When a sample $n$ with depth $\ell$ intersects the scene, we denote by $\vv_n^{(\ell)} \in \RR^3$ the intersection point, by $\widehat{\nv}_n^{(\ell)} \in \RR^3~\LB \norm{\widehat{\nv}_n^{(\ell)}} = 1 \RB$ the surface normal at the intersection point oriented towards the half-space containing the incident field, by $o_n^{(\ell)} \in \NN$ the index of the intersected object, and by $m_n^{(\ell)} \in \NN$ the index of the specific intersected primitive within that object.
The interaction type, denoted by $\chi_n^{(\ell)}$, is then sampled from $\Ic$, determining how the sample will continue to propagate.
Details on the sampling of the interaction type are provided in Section~\ref{sec:path-solver-interaction-sampling}.
The rest of the loop depends on the sampled interaction type.

\paragraph{Diffuse Reflection}
As previously indicated, diffuse reflection is assumed to scatter energy in all directions of the half-space containing the incident field, which enables next event estimation.
Therefore, if a diffuse reflection is sampled (i.e. $\chi_n^{(\ell)} = \Sc$), the \gls{LoS} between the intersection point and each target $\tv_1, \cdots, \tv_{N_T}$ is tested.
For each target $\tv_k$ with an unobstructed \gls{LoS} to the intersection point, a path connecting the source to the target
\begin{equation}
    \label{eq:path-solver-sample-path}
    p_n^{(\ell)} = \left(\sv, \vv_n^{(1)}, \cdots, \vv_n^{(\ell)}, \tv_k\right)
\end{equation}
is recorded as a valid path to be returned to the user.
Finally, a new ray is spawned to continue the sample path, with its direction selected uniformly at random from the hemisphere defined by $\widehat{\nv}_n^{(\ell)}$, i.e. $\widehat{\kv}_{n}^{(\ell)} \sim \Uc_{2\pi}(\widehat{\nv}_n^{(\ell)})$, where $\Uc_{2\pi}(\widehat{\nv}_n^{(\ell)})$ represents the uniform distribution on the hemisphere.

\begin{figure}[ht!]
    \centering
    \includegraphics[width=0.55\textwidth]{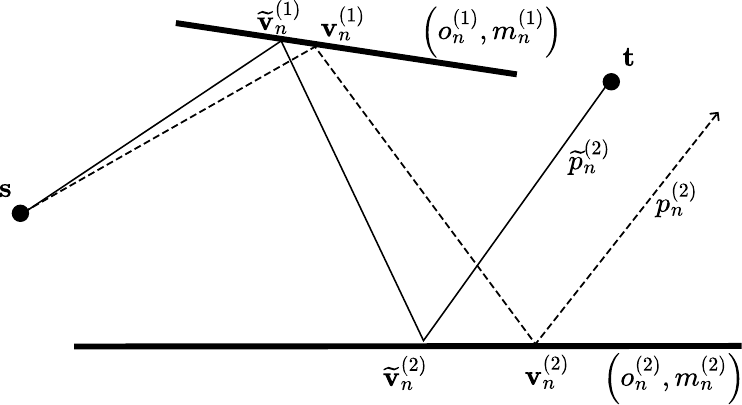}
    \caption{Processing a candidate specular chain of depth $\ell = 2$ ($p_n^{(2)}$, dashed line) to produce a valid path ($\widetilde{p}_n^{(2)}$, solid line). The same process applies to specular suffixes. Shown in 2D for clarity.\label{fig:path-solver-sbr-cand-im}}
\end{figure}

\paragraph{Specular Reflection and Refraction}
When a specular reflection or refraction is sampled (i.e. $\chi_n^{(\ell)} = \Rc$ or $\chi_n^{(\ell)} = \Tc$), the path~\eqref{eq:path-solver-sample-path} is a valid path for target $\tv_k$, where $k \in \left\{1, \dots, N_T\right\}$, only if
\begin{equation}
    \label{eq:path-solver-spec-cond}
    \frac{\tv_k - \vv_n^{(\ell)}}{\norm{\tv_k - \vv_n^{(\ell)}}} = \widehat{\kv}_n^{(\ell)},
\end{equation}
where
\begin{equation}
    \label{eq:path-solver-scatter-dir}
    \widehat{\kv}_n^{(\ell)} =
    \begin{cases}
        \frac{\widehat{\kv}_n^{(\ell-1)} - 2 \LB\LB\widehat{\kv}_n^{(\ell-1)}\RB\tp \widehat{\nv}_n^{(\ell)}\RB \widehat{\nv}_n^{(\ell)}}{\norm{\widehat{\kv}_n^{(\ell-1)} - 2 \LB\LB\widehat{\kv}_n^{(\ell-1)}\RB\tp \widehat{\nv}_n^{(\ell)}\RB \widehat{\nv}_n^{(\ell)}}} & \text{if } \chi_n^{(\ell)} = \Rc  \\
        \widehat{\kv}_n^{(\ell-1)} & \text{if } \chi_n^{(\ell)} = \Tc
    \end{cases}
\end{equation}
and where $\uv\tp$ denotes the transpose of $\uv$.
Since targets are modeled as points, condition~\eqref{eq:path-solver-spec-cond} effectively occurs with zero probability, and, consequently, the path~\eqref{eq:path-solver-sample-path} is not a valid path.
However, the specular suffix $p_n^{(\ell_d:)}$, where $\ell_d$ is defined as in~\eqref{eq:essential-concepts-ell_d}, can be further processed to determine vertices $\left(\widetilde{\vv}_n^{(\ell_d+1)}, \cdots, \widetilde{\vv}_n^{(\ell)}\right)$ such that
\begin{equation}
    \label{eq:path-solver-complete-cand}
    \widetilde{p}_n^{(\ell)} = \LB\sv, \vv_n^{(1)}, \cdots,
    \underbrace{\widetilde{\vv}_n^{(\ell_d+1)}, \cdots, \widetilde{\vv}_n^{(\ell)}, \tv_k}_{\text{Processed specular suffix}}\RB
\end{equation}
is a valid path, as illustrated in Figure~\ref{fig:path-solver-sbr-cand-im}.
This additional processing is performed using the image method in the algorithm described in Section~\ref{sec:path-solver-im}.

Importantly, for a sequence of primitives and interaction types
\begin{equation}
    \label{eq:path-solver-cand}
    \LB \vv_n^{(\ell_d)}, \LB o_n^{(\ell_d+1)}, m_n^{(\ell_d+1)}, \chi_n^{(\ell_d+1)} \RB, \cdots, \LB o_n^{(\ell)}, m_n^{(\ell)}, \chi_n^{(\ell)} \RB, \tv_k\RB
\end{equation}
where $\vv_n^{(0)} = \sv$, there exists \emph{at most a single} set of vertices $\widetilde{\vv}_n^{(\ell_d+1)},\cdots,\widetilde{\vv}_n^{(\ell)}$ such that $\widetilde{p}_n^{(\ell)}$ is a valid path.
Consequently, if $\ell_d = 0$, i.e. $p_n^{(\ell)}$ is a specular chain, then further processing by the image method of multiple identical candidates~\eqref{eq:path-solver-cand} would result in multiple identical valid paths, which would lead to erroneous \glspl{CIR}.
Therefore, if the sample $p_n^{(\ell)}$ is a specular chain, a hashing-based de-duplication mechanism, detailed in Section~\ref{sec:path-solver-hashing},  ensures that specular chain candidates are only evaluated once.
Notably, the path hashing mechanism discards redundant candidates ``on-the-fly'', significantly reducing memory and compute requirements.
However, when $\ell_d > 0$, then $\vv_n^{(\ell_d)}$ corresponds to a diffuse reflection point that is the result of a randomly sampled ray direction at the last previous diffuse reflection point.
Since the probability of two different paths having identical randomly sampled diffuse reflection points is effectively zero, path uniqueness is guaranteed in these cases without requiring the de-duplication mechanism.
As shown in Figure~\ref{fig:path-solver-sbr}, a \gls{LoS} occlusion test to the target is used as a heuristic to discard path candidates that have little chance of resulting in a valid path after processing by the image method-based algorithm.
Finally, a new ray is traced into the direction~\eqref{eq:path-solver-scatter-dir} to continue the sample path.

\paragraph{Diffraction}
\begin{figure}[ht!]
    \centering
    \includegraphics[width=0.75\textwidth]{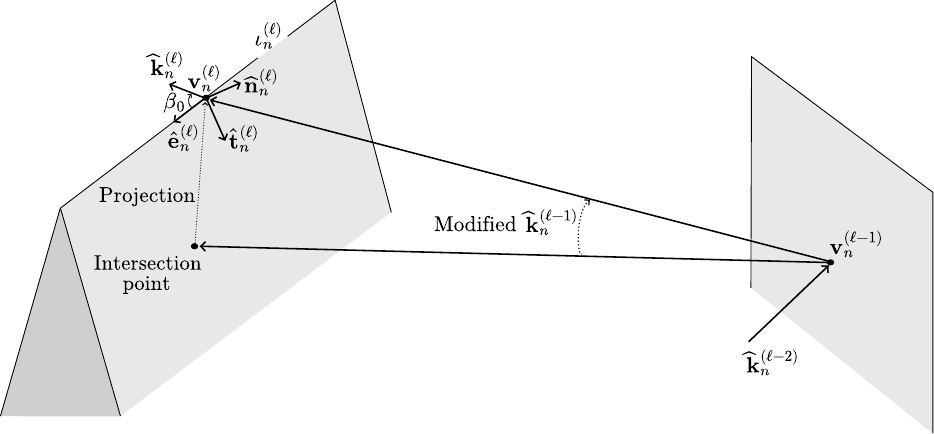}
    \caption{Diffracting edges are identified by projecting the intersection point onto the edges of the intersected primitive.\label{fig:path-solver-diffr-proj}}
\end{figure}
When tracing rays, the probability that a ray exactly intersects the edge of a wedge is zero.
Therefore, when a ray intersects a primitive and diffraction is sampled (i.e., $\chi_n^{(\ell)} = \Dc$), the intersection point between the ray $\left( \vv_n^{(\ell-1)}, \widehat{\kv}_n^{(\ell-1)}\right)$ and the primitive $\left(o_n^{(\ell)}, m_n^{(\ell)}\right)$ is projected onto one of the primitive's edges, yielding the vertex $\vv_n^{(\ell)}$, as illustrated in Figure~\ref{fig:path-solver-diffr-proj}.
The index of the edge onto which the intersection point is projected is denoted by $\iota^{(\ell)}_n$.

The path~\eqref{eq:path-solver-sample-path} is a valid path to a target $\tv_k$ (with $k \in \{1, \dots, N_T\}$) only if there exists an angle $\phi \in [0, n\pi]$, where $n\pi$ is the exterior angle of the wedge, such that
\begin{equation}
    \label{eq:path-solver-diffr-cond}
    \frac{\tv_k - \vv_n^{(\ell)}}{\norm{\tv_k - \vv_n^{(\ell)}}} = \widehat{\kv}_n^{(\ell)},
\end{equation}
with the diffraction direction given by
\begin{equation}
    \label{eq:path-solver-k_diffr}
    \widehat{\kv}_n^{(\ell)} = \sin{\left(\beta_0\right)}\cos{\phi} \widehat{\mathbf{t}}_n^{(\ell)}
                               + \sin{\left(\beta_0\right)}\sin{\phi} \widehat{\nv}_n^{(\ell)}
                               + \cos{\left(\beta_0\right)} \widehat{\mathbf{e}}_n^{(\ell)},
\end{equation}
where $\beta_0$ is the angle between the incident ray direction $\widehat{\kv}_n^{(\ell-1)}$ and the wedge vector $\widehat{\mathbf{e}}_n^{(\ell)}$, and $\left(\widehat{\mathbf{t}}_n^{(\ell)}, \widehat{\nv}_n^{(\ell)}, \widehat{\mathbf{e}}_n^{(\ell)}\right)$ is an orthonormal basis such that $\widehat{\mathbf{t}}_n^{(\ell)} = \widehat{\nv}_n^{(\ell)} \times \widehat{\mathbf{e}}_n^{(\ell)}$ (see Figure~\ref{fig:path-solver-diffr-proj}).
Note that $\widehat{\tv}_n^{(\ell)}$ is tangent to the wedge face.
In~\eqref{eq:path-solver-k_diffr}, ensuring that the angle between the diffracted ray and the edge, denoted by $\beta_0$, matches the angle between the incident ray and the edge, places the ray on the Keller cone.
As with specular reflection and refraction, since targets are modeled as points, condition~\eqref{eq:path-solver-diffr-cond} is satisfied with zero probability for all $\phi \in [0, n\pi]$, rendering the sampled path invalid.
Therefore, candidate paths ending with a diffraction are further refined using the image method-based algorithm described in Section~\ref{sec:path-solver-im}, just as for paths ending with specular reflection or refraction.
As shown in Figure~\ref{fig:path-solver-sbr}, the processing of diffraction thus follows the same steps as for specular reflection and refraction: a hashing-based de-duplication mechanism ensures uniqueness of specular chains, and an occlusion test to the target is used as a heuristic to eliminate candidates unlikely to produce valid paths after refinement with the image method.
Finally, a new ray is traced in the direction~\eqref{eq:path-solver-k_diffr} to continue the sample path, with $\phi$ sampled uniformly from $[0, n\pi]$.

As illustrated in Figure~\ref{fig:path-solver-diffr-proj}, projecting the intersection point onto an edge alters the scattered ray direction $\widehat{\kv}_n^{(\ell-1)}$ departing from $\vv_n^{(\ell-1)}$.
This modification invalidates any preceding specular reflections or refractions, meaning that path segments that are not suffixes would require refinement if a diffraction is followed by a diffuse reflection.
To avoid significantly increased computational complexity from refining non-suffix path segments, the path solver prohibits sampling paths that contain both diffraction and diffuse reflection.
Such paths typically contribute negligibly to the total transported energy and therefore have little impact on the accuracy of the \glspl{CIR}.
Moreover, \SRT{} currently supports only first-order diffraction, meaning that paths may contain at most one diffraction event.

The candidate generator depicted in Figure~\ref{fig:path-solver-sbr} thus yields a set of valid paths connecting the source to all targets that do not require further processing (except for the computation of the corresponding path coefficients and delays) and end with a diffuse reflection.
Additionally, it produces a set of candidate paths ending with specular suffixes, which require further processing to be either discarded or refined into valid paths.

\begin{figure}[htbp]
    \centering
    \begin{subfigure}[b]{0.3\textwidth}
        \centering
        \includegraphics[width=\textwidth]{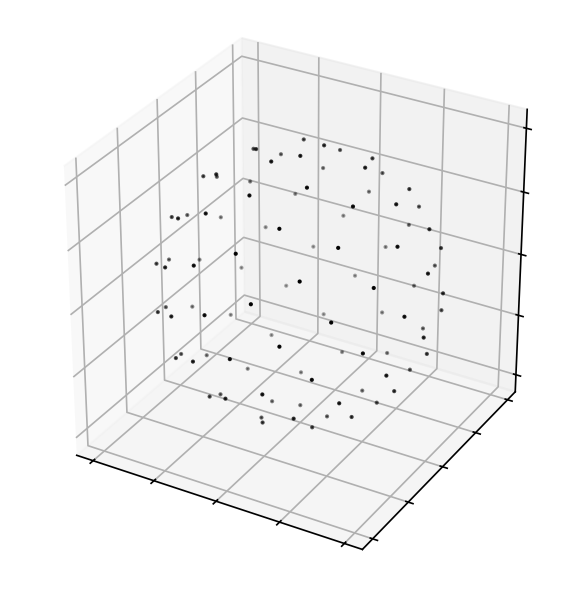}
        \caption{100 points.}
    \end{subfigure}
    \hfill
    \begin{subfigure}[b]{0.3\textwidth}
        \centering
        \includegraphics[width=\textwidth]{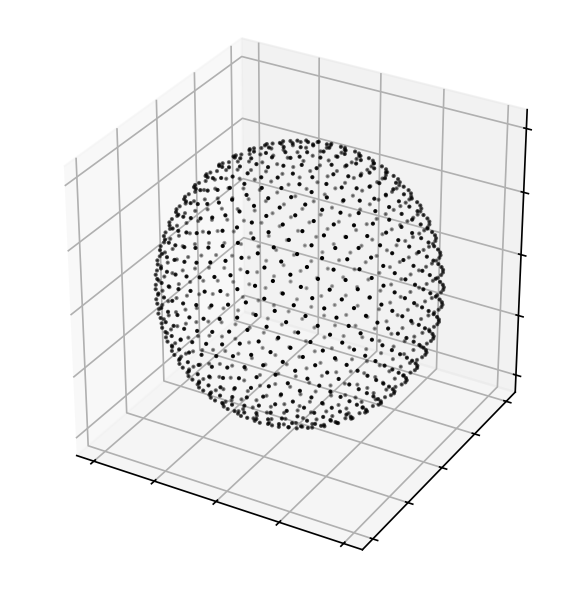}
        \caption{1000 points.}
    \end{subfigure}
    \hfill
    \begin{subfigure}[b]{0.3\textwidth}
        \centering
        \includegraphics[width=\textwidth]{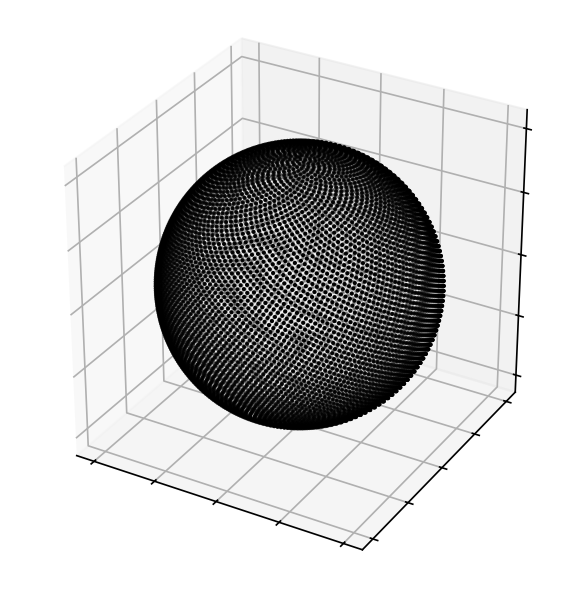}
        \caption{10000 points.}
        \end{subfigure}
    \caption{Spherical Fibonacci lattice.\label{fig:path-solver-fibonacci}}
\end{figure}

\subsubsection{Sampling Initial Ray Directions}
\label{sec:path-solver-fibonacci}

The \gls{SBR} procedure starts by spawning rays from the source with directions obtained from a spherical Fibonacci lattice with $N_S$ points, which is defined as follows:
\begin{equation}
    \widehat{\kv}_n^{(0)} =
    \begin{bmatrix}
        \sin{\theta_n}\cos{\varphi_n}\\
        \sin{\theta_n}\sin{\varphi_n}\\
        \cos{\theta_n}
    \end{bmatrix}
\end{equation}
where
\begin{eqnarray*}
    \theta_n &=& \arccos\left(1-\frac{2n}{N_S-1}\right)\\
    \varphi_n &=& 2\pi\left(\frac{n}{\varphi_g}
        -\left\lfloor\frac{n}{\varphi_g}\right\rfloor\right)
\end{eqnarray*}
for $n=0,\dots,N_S-1$ and $N_S\geq 2$, where $\varphi_g = \frac{1 + \sqrt{5}}{2}$ is the golden ratio and $\left\lfloor \cdot \right\rfloor$ denotes the floor function.
For $N_S=1$, $\theta_0=\frac{\pi}{2}$ and $\varphi_0=\pi$.
Sampling rays from the Fibonacci lattice results in ray tubes sharing approximately the same solid angle $\frac{4\pi}{N_S}$.

The Fibonacci lattice was selected for its simple construction and its benefits over other lattices for numerical integration~\cite{gonzalez2010measurement,hannay2004fibonacci}.
Figure~\ref{fig:path-solver-fibonacci} shows the spherical Fibonacci lattice with 100, 1000, and 10000 points.
The \SRT{} path solver uses $10^6$ points as the default value for $N_S$, and higher values generally increase the probability of finding additional valid paths.

\subsubsection{Sampling Interaction Types}
\label{sec:path-solver-interaction-sampling}

Within the \gls{SBR} loop, when a ray intersects the scene geometry, the algorithm generates at most one scattered ray. This constraint is necessary because generating multiple rays at each intersection, such as both reflected and refracted rays, would lead to an exponential increase in computational complexity and memory usage with the depth of the path. Instead, a single interaction type is randomly chosen based on a distribution:
\begin{equation}
    \label{eq:path-solver-int-dist}
\Qc  = \LP q\LB\Rc\RB, q\LB\Sc\RB, q\LB\Tc\RB, q\LB\Dc\RB \RP,\qquad \text{where } q\LB\Rc\RB + q\LB\Sc\RB + q\LB\Tc\RB + q\LB\Dc\RB = 1 \text{ and } q\LB\cdot\RB \geq 0.
\end{equation}
Here, $q \LB \chi \RB$ represents the probability of selecting the interaction type $\chi \in \Ic$. Importantly, the choice of distribution $\Qc$ does not affect the calculation of the paths' electric fields, which are solely dependent on the propagation environment. However, the choice of distribution $\Qc$ significantly impacts the \emph{sample efficiency} of the ray tracer, which affects the number of samples $N_S$ needed to compute paths that connect a source to a target and capture the majority of the transported energy.
Sampling only one interaction type per intersection may appear limiting, but this is offset by the large number of samples generated from each source. In most scenarios, a high sample count allows identifying all significant paths.

For each sample $n$ at depth $\ell$, the interaction type is sampled independently for every interaction along every path.
For specular reflection, diffuse reflection, and refraction, the probabilities are set using a heuristic based on the squared amplitudes of the corresponding Fresnel field coefficients.
Diffraction, on the other hand, is sampled using a fixed probability $q\LB\Dc\RB \in [0, 1]$.
This choice is motivated by the fact that diffraction typically carries much less energy compared to reflection and refraction, which would result in very low probabilities for sampling diffraction if its probability were also based on the diffraction coefficients.
The resulting distribution for the interaction types is:
\begin{equation}
    \label{eq:path-solver-used-dist-event}
    \begin{aligned}
        q\LB\Rc\RB &= \left(1 - q\LB\Dc\RB\right)R^2\frac{\abs{r_{\perp}}^2 + \abs{r_{\parallel}}^2}{\abs{r_{\perp}}^2 + \abs{r_{\parallel}}^2 + \abs{t_{\perp}}^2 + \abs{t_{\parallel}}^2}\\
        q\LB\Sc\RB &= \left(1 - q\LB\Dc\RB\right)S^2\frac{\abs{r_{\perp}}^2 + \abs{r_{\parallel}}^2}{\abs{r_{\perp}}^2 + \abs{r_{\parallel}}^2 + \abs{t_{\perp}}^2 + \abs{t_{\parallel}}^2}\\
        q\LB\Tc\RB &= \left(1 - q\LB\Dc\RB\right)\frac{\abs{t_{\perp}}^2 + \abs{t_{\parallel}}^2}{\abs{r_{\perp}}^2 + \abs{r_{\parallel}}^2 + \abs{t_{\perp}}^2 + \abs{t_{\parallel}}^2}\\
        q\LB\Dc\RB &\in [0, 1]
    \end{aligned}
\end{equation}
In \SRT{}, $q\LB \Dc \RB$ is set to $0.2$.
Here, $r_{\perp}$, $r_{\parallel}$, $t_{\perp}$, and $t_{\parallel}$ are the Fresnel coefficients~\eqref{eq:fresnel_slab}, $S$ is the scattering coefficient (see Section~\ref{sec:primer-em-diffuse}), and $R$ is the reflection reduction factor~\eqref{eq:R}. This approach biases sampling toward interaction types with larger squared field coefficients, a technique known as \emph{importance sampling}~\cite{Kloek1978, Wikipedia_Importance_Sampling}.
This is similar to importance sampling techniques for light transport paths commonly used in computer graphics.
A brief overview of importance sampling in the context of ray tracing for the simulation of radio wave propagation is provided in Appendix~\ref{sec:importance-sampling}.

\begin{figure}[t]
    \centering
    \begin{subfigure}[b]{0.48\textwidth}
        \centering
        \includegraphics[width=\textwidth]{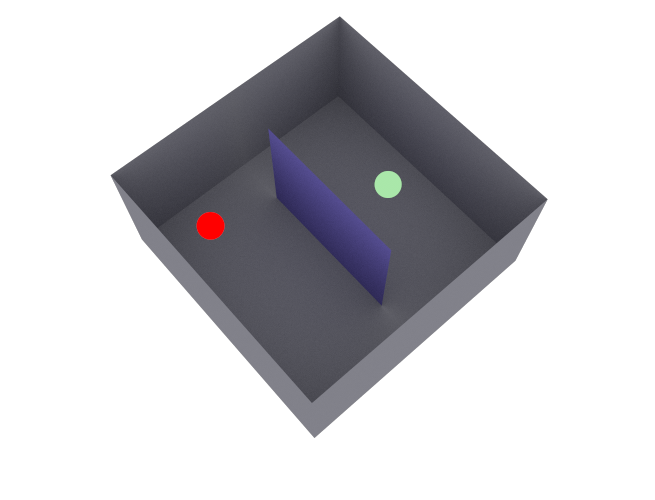}
        \caption{Scene used for the comparison.}
        \label{fig:path-solver-sampling-scene}
    \end{subfigure}
    \hfill
    \begin{subfigure}[b]{0.48\textwidth}
        \centering
        \includegraphics[width=\textwidth]{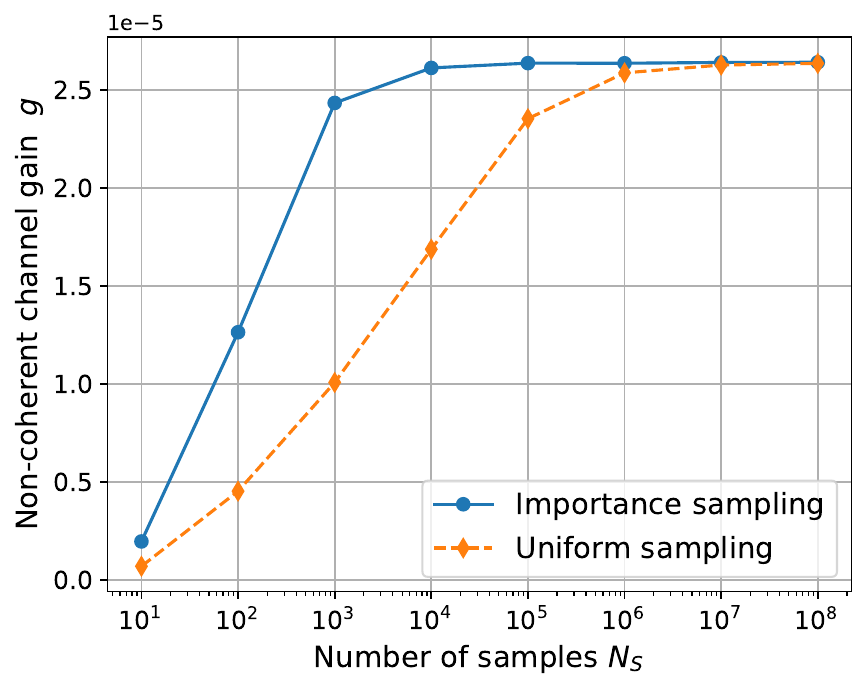}
        \caption{Channel gain at the target.}
        \label{fig:path-solver-sampling-importance}
    \end{subfigure}
    \caption{Comparison of sampling strategies for interaction types. The scene consists of a closed metallic box containing a glass screen, with a source (red ball) and target (green ball) on opposite sides. The box is shown open to visualize the interior. Importance sampling achieves faster convergence of the channel gain compared to uniform sampling.}
    \label{fig:path-solver-sampling}
\end{figure}

To illustrate the importance of optimized sampling strategies, we compare uniform sampling, where specular reflection, diffuse reflection, and refraction are selected with equal probability, to importance sampling using the scene shown in Figure~\ref{fig:path-solver-sampling-scene}. The scene consists of a metallic box containing a glass screen, with a source and target positioned on opposite sides.
Diffuse reflection, specular reflection, refraction, and diffraction are enabled, and the maximum depth $L$ is set to $5$.
As shown in Figure~\ref{fig:path-solver-sampling-importance}, using importance sampling leads to better sample efficiency when compared to uniform sampling, as convergence requires approximately $100\times$ fewer samples to capture most of the received energy. This demonstrates the benefit of selecting interactions according to the material properties of the scatterers.

\subsubsection{Specular Chain Candidates Deduplication}
\label{sec:path-solver-hashing}

As mentioned in Section~\ref{sec:path-solver-sbr} and Figure~\ref{fig:path-solver-sbr}, to avoid storing multiple identical specular chain candidates, a path hashing method is used to discard duplicates of previously found candidates ``on-the-fly''.
A specular chain candidate is defined by a sequence of interactions with the scene, each characterized by the intersected object $o_n^{(\ell)} \in \NN$, the corresponding intersected primitive $m_n^{(\ell)} \in \NN$, the diffracted edge $\iota_n^{(\ell)} \in \{0, 1, 2\}$ (if $\chi_n^{(\ell)} = \Dc$), and the interaction type $\chi_n^{(\ell)} \in \LP \Rc, \Tc, \Dc \RP$ with $\ell_d = 0$, i.e.
\begin{equation}
    c^{(\ell)}_n = \LB \sv, \LB o_n^{(1)}, m_n^{(1)}, \iota_n^{(1)}, \chi_n^{(1)} \RB, \cdots, \LB o_n^{(\ell)}, m_n^{(\ell)}, \iota_n^{(\ell)}, \chi_n^{(\ell)} \RB \RB.
\end{equation}
For specular chain candidates, we do not need to store the path vertices. If the candidate leads to a valid path, these vertices will be computed. Invalid candidates are discarded.

\begin{figure}[ht!]
    \centering
    \includegraphics[width=0.51\textwidth]{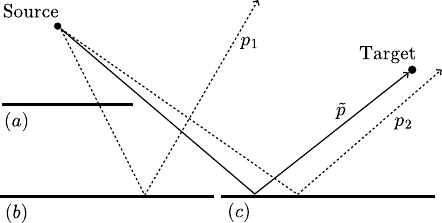}
    \caption{Two candidate paths, $p_1$ and $p_2$, which differ by being reflected by different coplanar primitives and by encountering a refraction (in $p_1$ only), yield the same path $\widetilde{p}$ once refined by the image method. Shown in 2D for clarity.\label{fig:path-solver-hashing-coplanar}}
\end{figure}

The hashing process is designed to assign the same identifier to specular chain candidates that correspond to the same path geometry after refinement with the image method (see Section~\ref{sec:path-solver-im}).
The hash value is computed from the sequence of primitives constituting the path.
Primitives that are coplanar (i.e., lie in the same plane) are treated as equivalent during hashing.
This is because the image method assumes all reflection surfaces extend infinitely, making the exact in-plane position of a primitive irrelevant to the path geometry.
Refraction can be omitted from the hashing procedure since it does not change the ray direction and therefore does not alter the geometry of the path, as refraction is modeled without angular deflection.
Figure~\ref{fig:path-solver-hashing-coplanar} illustrates this concept: two candidate paths, $p_1$ and $p_2$, yield the same refined path $\widetilde{p}$, and thus they should share the same hash value.
This holds even though only one candidate is refracted by primitive $(a)$, and both are reflected by different but coplanar primitives $(b)$ and $(c)$, respectively.

\begin{algorithm}
    \begin{center}
    \begin{algorithmic}[1]
        \State $\texttt{hash}_n^{(0)} \gets 0$
        \State $\ell \gets 1$
        \While{\text{<Stop condition>}} \Comment{\gls{SBR} loop}
            \State $\cdots$
            \If{$\chi_n^{(\ell)} \in \{\Rc,\Dc\}$}
                \State $\texttt{inter\_hash}_{n}^{(\ell)} \gets \Call{hash\_intersection}{o_n^{(\ell)}, m_n^{(\ell)}, \iota_n^{(\ell)}, \chi_n^{(\ell)}}$
                \State $\texttt{hash}_n^{(\ell)} \gets \Call{hash\_update}{\texttt{hash}_n^{(\ell-1)}, \texttt{inter\_hash}_{n}^{(\ell)}}$
            \Else
                \State $\texttt{hash}_n^{(\ell)}
                    \gets \texttt{hash}_n^{(\ell-1)}$
                    \Comment{Omit refraction from the hash}
            \EndIf
            \State $\cdots$
            \State $\ell \gets \ell + 1$
        \EndWhile
        \State
        \Procedure{hash\_intersection}{$o, m, \iota, \chi$}
            \If{$\chi = \Rc$}
                \Return $\Call{hash\_plane}{o,m}$
            \ElsIf{$\chi = \Dc$}
                \Return $\Call{hash\_edge}{o,m, \iota}$
            \EndIf
        \EndProcedure
        \State
        \Procedure{hash\_plane}{$o, m$}
            \State $\nv \gets \Call{get\_normal}{o, m}$ \Comment{Read the primitive normal}
            \State $\nv \gets \Call{canonicalize\_sign}{\nv}$ \Comment{Enforce a deterministic orientation}
            \State $\pv \gets \Call{get\_vertex}{o, m}$ \Comment{Read a primitive vertex}
            \State $d \gets \nv\tp \pv$ \Comment{Offset of the plane to the origin}
            \State $h \gets \Call{FNV1a}{\texttt{Q}(\nv_x), \texttt{H}}$
            \State $h \gets \Call{FNV1a}{\texttt{Q}(\nv_y), h}$
            \State $h \gets \Call{FNV1a}{\texttt{Q}(\nv_z), h}$
            \State $h \gets \Call{FNV1a}{\texttt{Q}(d), h}$
            \State \Return $h$
        \EndProcedure
        \State
        \Procedure{hash\_edge}{$o, m, \iota$}
            \State $\pv_1, \pv_2 \gets \Call{get\_endpoints}{o, m, \iota}$ \Comment{Read the edge endpoints}
            \State $\mathbf q_1 \gets \LB\texttt{Q}(\pv_{1,x}),\texttt{Q}(\pv_{1,y}),\texttt{Q}(\pv_{1,z})\RB$
            \State $\mathbf q_2 \gets \LB\texttt{Q}(\pv_{2,x}),\texttt{Q}(\pv_{2,y}),\texttt{Q}(\pv_{2,z})\RB$
            \If{$(q_{2,z},q_{2,y},q_{2,x}) <_{\mathrm{lex}} (q_{1,z},q_{1,y},q_{1,x})$}
                \State $\mathbf q_1,\mathbf q_2 \gets \mathbf q_2,\mathbf q_1$ \Comment{Enforce a deterministic endpoint order}
            \EndIf
            \State $h \gets \Call{FNV1a}{q_{1,x}, \texttt{H}}$
            \State $h \gets \Call{FNV1a}{q_{1,y}, h}$
            \State $h \gets \Call{FNV1a}{q_{1,z}, h}$
            \State $h \gets \Call{FNV1a}{q_{2,x}, h}$
            \State $h \gets \Call{FNV1a}{q_{2,y}, h}$
            \State $h \gets \Call{FNV1a}{q_{2,z}, h}$
            \State \Return $h$
        \EndProcedure
        \State
        \Procedure{hash\_update}{$\texttt{curr\_hash}, \texttt{inter\_hash}$}
            \State \Return $(1373 \cdot \texttt{curr\_hash} + \texttt{inter\_hash})$ \Comment{Update rolling hash}
        \EndProcedure
    \end{algorithmic}
    \caption{Hashing of specular chains\label{alg:path-solver-hashing}}
    \end{center}
\end{algorithm}

Algorithm~\ref{alg:path-solver-hashing} depicts how path hashing works during ray tracing to avoid computing duplicate specular chain candidates.
It starts by initializing a hash value $\texttt{hash}_n^{(0)}$ to $0$ for each sample $n = 1,\dots,N_S$ being traced.
Then, as the sample interacts with objects in the scene, the algorithm computes a hash value $\texttt{inter\_hash}_n^{(\ell)}$ for each intersection, and updates the path hash $\texttt{hash}_n^{(\ell)}$ by combining it with the hash of the intersection, creating an identifier for the entire sample up to that point.
The computation of the interaction hash depends on the interaction type.
For specular reflection, hashing is performed based on the primitive normal vector and an offset positioning the plane with respect to the origin.
An integer hash value is then computed by sequentially combining the normal components and the offset using the FNV-1a hash function~\cite{fnv1a}.
As numerical imprecision can lead to slight variations in the normal vectors of coplanar primitives, the real-valued normal components are quantized to a fixed precision and then converted to integers using a function $\texttt{Q}(\cdot)$ prior to hashing.
Discussion on the choice of the quantization function $\texttt{Q}(\cdot)$ is provided below.
For diffraction, hashing is performed on the two edge endpoints by applying the FNV-1a hash function sequentially to the components of the endpoints.
In both $\Call{hash\_plane}$ and $\Call{hash\_edge}$, the first call to the FNV-1a hash function uses a large constant $\texttt{H}$ as seed.
Prior to hashing, plane normals are canonicalized by giving them a deterministic orientation.
Similarly, edge endpoints are quantized and lexicographically ordered.

Once the interaction hash is computed, the \textsc{hash\_update} function updates the sample hash value $\texttt{hash}_n^{(\ell)}$ through a polynomial rolling hash function~\cite{Wikipedia_Rolling_hash} with a prime base of $1373$.
For each new intersection, it updates the running hash by multiplying the current hash value by the prime number and adding the new intersection hash.
When a specular chain candidate is considered for a target $k \in \left\{1, \dots, N_T\right\}$, the hash value of the candidate is paired with the target index $k$ to form a new hash value
\begin{equation}
    \texttt{hash}_{n,k}^{(\ell)} = \Call{FNV1a}{\texttt{hash}_n^{(\ell)}, k}.
\end{equation}
This target-specific hash value is used to determine whether the candidate should be discarded.

\paragraph{Practical Consideration: Hash Array}
To track information about previously encountered specular chains, an array of integers of size $N_H$ is allocated in memory and initialized to zero, referred to as the \emph{hash array}.
When considering a specular chain candidate for target $k$ with corresponding hash value $\texttt{hash}_{n,k}^{(\ell)}$, the corresponding index $i_{n,k}^{(\ell)}$ is computed as
\begin{equation}
    \label{eq:path-solver-hash-mod}
    i_{n,k}^{(\ell)} = \texttt{hash}_{n,k}^{(\ell)} \mod N_H.
\end{equation}
The hash array entry at index $i_{n,k}^{(\ell)}$ is atomically incremented and its previous value returned. If the previous value is greater than zero, it indicates that this path has already been encountered and should be discarded.
Otherwise, the candidate is retained for further processing by the image method.
The index $i_{n,k}^{(\ell)}$ is not guaranteed to be unique for different specular chains. When two different specular chains share the same index value despite being different, the path solver will discard one of them. Such an event is referred to as a \emph{collision}, and can result in the discarding of valid paths.
However, when two candidates are identical, their index values will also be identical.

\begin{figure}[ht!]
    \centering
    \includegraphics[width=0.8\textwidth]{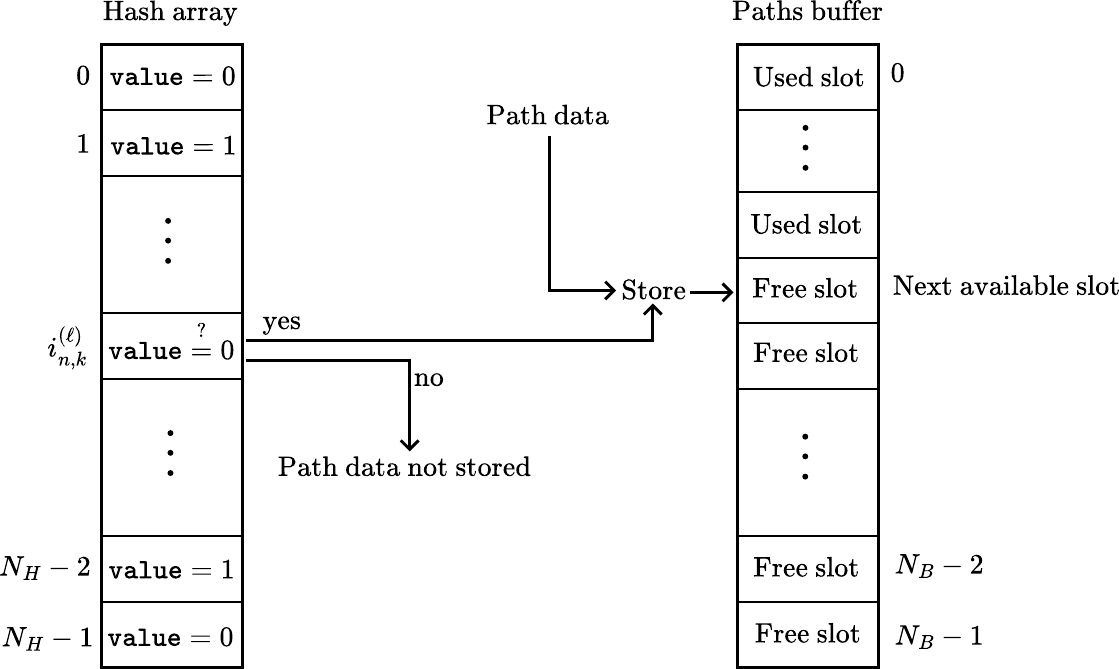}
    \caption{The hash array is used to track encountered path candidates. New candidates are stored in the buffer if they haven't been encountered before. Non-specular chains are added to the buffer without a uniqueness check (not depicted in the figure).\label{fig:path-solver-hash-buffer}}
\end{figure}

Valid paths and candidates are stored in a buffer of size $N_B$. This buffer holds, for each path, the sequence of interaction types, vertices, intersected objects, primitives, and more. When a valid path is found or a candidate not previously encountered, indicated by the atomic read-and-increment operation for entry $i_{n,k}^{(\ell)}$ returning zero, the path data are stored in the buffer.
The index $i_{n,k}^{(\ell)}$ is solely used for indexing the hash array and not for storing path data in the buffer. Instead, path data are stored contiguously in the next available slot of the path buffer, as shown in Figure~\ref{fig:path-solver-hash-buffer}. If the buffer becomes full, any newly found paths or candidates are discarded.
Discarding paths rather than overwriting existing ones is motivated by the fact that paths with lower depth, which are added to the buffer first as the \gls{SBR} loop advances from $\ell = 0$ to $\ell = L$, typically carry more energy and should not be replaced by higher depth paths when the buffer becomes full.
Consequently, the buffer size $N_B$ determines the maximum number of paths and candidates that can be stored.

Due to the substantial amount of data stored for each path, setting $N_B$ to excessively high values can lead to memory issues. This requires separating the hash array size $N_H$ from the buffer size $N_B$. This is because a small hash array size can increase the probability of collisions. To reduce the risk of collisions, $N_H$ is set to
\begin{equation}
    N_H = \max\LB N_B, 10^6 \RB
\end{equation}
ensuring a minimum array size of $10^6$ elements. In the interface of the \SRT{} built-in path solver, $N_B$ corresponds to the $\texttt{max\_num\_paths\_per\_src}$ parameter.

\paragraph{Practical Consideration: Quantization Function}
An important practical consideration is the design of the quantization function $\texttt{Q}(\cdot)$.
When quantizing floating-point values to integers, values that are very close but lie on opposite sides of a quantization boundary can be mapped to different integers and therefore yield different hash values. For example, with rounding as the quantization function, $\texttt{Q}(1.4999)=1$ and $\texttt{Q}(1.5001)=2$.
This can cause identical candidates, which differ only due to numerical imprecision, to be treated as distinct, resulting in duplicate paths.
To mitigate this issue, multiple hash values are computed for each sample, each using a different quantization function. A path is only considered unique if all of its corresponding hash values are unique.
In the current implementation of \SRT{}, two quantization functions are used: rounding and floor, leading to two hash values being computed per candidate.
This approach can be interpreted as a form of locality-sensitive hashing~\cite{Wikipedia_Locality_Sensitive_Hashing}, as it aims to group together candidates that differ only due to numerical imprecision by using multiple hash functions to reduce the risk of misclassification.

\subsection{Image Method-based Candidate Processing}
\label{sec:path-solver-im}

The image method processes path candidates to determine valid paths that are specular chains, or end with specular suffixes. Consider a candidate path
\begin{equation}
    c_n^{(\ell)} = \left(\sv, \left(o_n^{(1)}, m_n^{(1)}, \iota_n^{(1)}, \chi_n^{(1)}\right), \cdots,  \underbrace{\left(o_n^{(\ell_d+1)}, m_n^{(\ell_d+1)}, \iota_n^{(\ell_d+1)}, \chi_n^{(\ell_d+1)}\right), \cdots, \left(o_n^{(\ell)}, m_n^{(\ell)}, \iota_n^{(\ell)}, \chi_n^{(\ell)}\right), \tv}_{\text{Specular suffix}}\right)
\end{equation}
where $\ell$ denotes the path depth, $\ell_d$ is the specular suffix index defined as in~\eqref{eq:essential-concepts-ell_d}, and $\tv$ is the target of the path.
Recall that the sequence of interactions $\boldsymbol{\chi}^{(\ell)} = \LB \chi_n^{(1)}, \cdots, \chi_n^{(\ell)} \RB$ was determined by the \gls{SBR}-based candidate generator.
For path candidates that end with a specular suffix but are not specular chains (i.e. $\ell_d > 0$), the image method processes only the specular suffix. In these cases, the vertex of the last diffuse reflection $\vv_n^{(\ell_d)}$ serves as the source point for the image method calculations.
Since the processing algorithm is identical for both specular chains and specular suffixes, we treat them equivalently in the following discussion. For simplicity of notation, we will refer to both types as specular chains and assume $\ell_d = 0$ throughout the remainder of this section.

\begin{figure}[htbp]
    \centering
    \begin{subfigure}[b]{0.45\textwidth}
        \centering
        \includegraphics[width=0.8\textwidth]{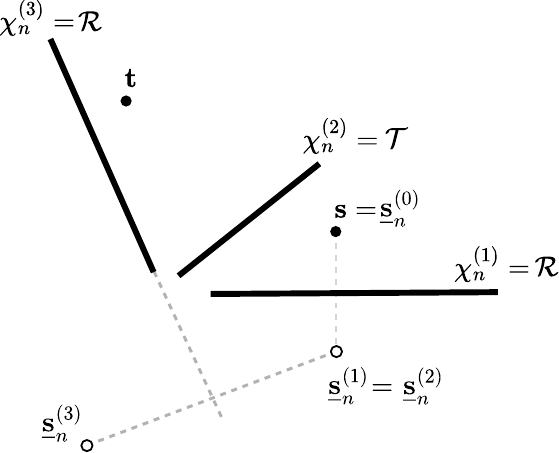}
        \caption{\label{fig:path-solver-im-step-1}}
    \end{subfigure}
    \hfill
    \begin{subfigure}[b]{0.45\textwidth}
        \centering
        \includegraphics[width=0.8\textwidth]{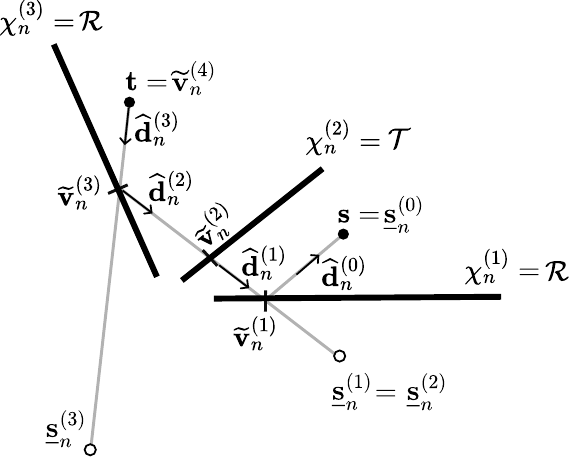}
        \caption{\label{fig:path-solver-im-step-2}}
    \end{subfigure}
    \caption{The two steps of the image method for paths without diffraction: (a) compute the images of the source and (b) backtrack the path. Shown in 2D for clarity.\label{fig:path-solver-im}}
\end{figure}

\paragraph{Paths without Diffraction}
We begin by describing the image method process for candidates that contain no diffraction. As illustrated in Figure~\ref{fig:path-solver-im}, this process consists of two main steps. First, the method iteratively computes images of the source by reflecting it across each plane at which a specular reflection occurs.
Since refracted paths are modeled without angular deflection, incorporating refraction into the image method is straightforward: refractions may simply be skipped when computing the source images, as they do not alter the ray's direction.
Then, the algorithm backtracks from the target to determine the actual path vertices by intersecting lines between successive image sources.

Formally, the image method iteratively computes images of the source $\underline{\sv}_n^{(i)}$ for $i = 0, \dots, \ell$, starting from the original source $\sv$:
\begin{equation}
    \underline{\sv}_n^{(i)} = \begin{cases}
        \sv & \text{if } i = 0\\
        \underline{\sv}_n^{(i-1)} - 2\left( \left(\underline{\sv}_n^{(i-1)} - \pv_n^{(i)}\right)\tp \widehat{\nv}_n^{(i)}\right) \widehat{\nv}_n^{(i)}& \text{if } \chi_n^{(i)} = \Rc\\
        \underline{\sv}_n^{(i-1)} & \text{if } \chi_n^{(i)} = \Tc
    \end{cases}
\end{equation}
Here, $\pv_n^{(i)}$ represents any point on the plane containing the primitive $\LB o_n^{(i)}, m_n^{(i)} \RB$, and $\widehat{\nv}_n^{(i)}$ denotes the surface's normal vector. For specular reflections ($\chi_n^{(i)} = \Rc$), the image is computed by reflecting the previous image across the surface plane, while for refractions ($\chi_n^{(i)} = \Tc$), the image position remains unchanged.

During backtracking, the algorithm traces rays iteratively from the target $\tv$ back to the source $\sv$ by intersecting lines between successive image points.
The intersection points $\widetilde{\vv}_n^{(i)}$ are the refined vertices of the path.
Specifically, a sequence of $\ell$ rays $\LB \widetilde{\vv}_n^{(i+1)}, \widehat{\dv}_n^{(i)}\RB$, $i = \ell, \dots, 1$, is traced, where $\widetilde{\vv}_n^{(\ell+1)} = \tv$ and
\begin{equation}
    \label{eq:path-solver-im-backtracking}
    \begin{aligned}
        \widehat{\dv}_n^{(i)} &= \frac{\underline{\sv}_n^{(i)} - \widetilde{\vv}_n^{(i+1)}}{\norm{\underline{\sv}_n^{(i)} - \widetilde{\vv}_n^{(i+1)}}}\\
        \widetilde{\vv}_n^{(i)} &= \texttt{intersection}\left(\LB \widetilde{\vv}_n^{(i+1)}, \widehat{\dv}_n^{(i)} \RB, \LB o_n^{(i)}, m_n^{(i)} \RB\right), \text{only if } i > 0
    \end{aligned}
\end{equation}
where $\widetilde{\vv}_n^{(i)}$ is the intersection point between the ray $\LB \widetilde{\vv}_n^{(i+1)}, \widehat{\dv}_n^{(i)} \RB$ and the primitive $\LB o_n^{(i)}, m_n^{(i)} \RB$, as shown in Figure~\ref{fig:path-solver-im-step-2}.
The path is considered valid only if two conditions are met: (i) each ray $\LB \widetilde{\vv}_n^{(i+1)}, \widehat{\dv}_n^{(i)} \RB$ successfully intersects a primitive that is coplanar with the primitive $\LB o_n^{(i)}, m_n^{(i)} \RB$, and (ii) the line segment between $\widetilde{\vv}_n^{(i+1)}$ and $\widetilde{\vv}_n^{(i)}$ is unobstructed by any other scene geometry. If either condition fails, then the path is discarded.
The case $i = 0$ corresponds to the final segment of the path, which connects to the source $\sv$.
If the path is valid, then the returned path will be~\eqref{eq:path-solver-complete-cand}.

\paragraph{Paths with Diffraction}
To handle paths involving diffraction, we adopt an approach inspired by~\cite{6815738}. Since \SRT{} currently only supports first-order diffraction, each sample $n$ contains at most a single diffracting edge, whose position within the path is denoted by $i_{\Dc} \in \LP 1, \dots, \ell \RP$. For notational simplicity, we omit the explicit dependence of $i_{\Dc}$ on $n$.

The core idea is to use image theory to compute the diffraction point in image space and subsequently map it back to the original space.
The motivation for this approach is that, in the image space, all specular reflections and refractions are effectively ``removed'', leaving only the diffraction event.
This reduces the problem to a simpler form: finding the appropriate diffraction point for a path of the shape source $\rightarrow$ diffraction $\rightarrow$ target.
More precisely, for every sample $n$, the diffracting edge at depth $i_{\Dc}$, which is characterized by an endpoint $\ov_n$ and an edge vector $\widehat{\ev}_n$, has its images computed by reflecting its endpoint and edge vector across each plane at which a specular reflection occurs over the following interactions $i = i_{\Dc}+1, \dots, \ell$:
\begin{equation}
    \begin{aligned}
        \underline{\ov}_n^{(i)} &=
        \begin{cases}
            \ov_n & \text{if } i = i_{\Dc} \\
            \underline{\ov}_n^{(i-1)} & \text{if } \chi_n^{(i)} = \Tc \\
            \underline{\ov}_n^{(i-1)} - 2\left( \left(\underline{\ov}_n^{(i-1)} - \pv_n^{(i)}\right)\tp \widehat{\nv}_n^{(i)}\right) \widehat{\nv}_n^{(i)} & \text{if } \chi_n^{(i)} = \Rc
        \end{cases}\\
        \widehat{\underline{\ev}}_n^{(i)} &=
        \begin{cases}
            \widehat{\ev}_n & \text{if } i = i_{\Dc} \\
            \widehat{\underline{\ev}}_n^{(i-1)} & \text{if } \chi_n^{(i)} = \Tc \\
            \widehat{\underline{\ev}}_n^{(i-1)} - 2\left(\left(\widehat{\underline{\ev}}_n^{(i-1)}\right)\tp \cdot \widehat{\nv}_n^{(i)}\right) \widehat{\nv}_n^{(i)} & \text{if } \chi_n^{(i)} = \Rc
        \end{cases}
    \end{aligned}
\end{equation}
where $\pv_n^{(i)}$ represents any point on the plane containing the primitive $\LB o_n^{(i)}, m_n^{(i)} \RB$, and $\widehat{\nv}_n^{(i)}$ denotes the surface's normal vector.
Applying image theory, specular reflections and refractions can be ignored when searching for the diffraction point in image space.
The problem thus reduces to finding the diffraction point in image space, $\underline{\widetilde{\vv}}_n^{(i_{\Dc}, \ell)}$, of the path $\underline{\sv}_n^{(\ell)} \rightarrow \underline{\widetilde{\vv}}_n^{(i_{\Dc}, \ell)} \rightarrow \tv$.
According to Fermat's principle, the diffraction point minimizes the total path length, denoted by $\Lc(x_n)$.
Specifically,
\begin{align}
    \Lc(x_n) &=
        \norm{\tv - \underline{\widetilde{\vv}}_n^{(i_{\Dc}, \ell)}}
        +
        \norm{\underline{\sv}_n^{(\ell)} - \underline{\widetilde{\vv}}_n^{(i_{\Dc}, \ell)}} \label{eq:path-solver-diffraction-image-problem}
        \\
    \underline{\widetilde{\vv}}_n^{(i_{\Dc}, \ell)} &= \underline{\ov}_n^{(\ell)} + x_n \widehat{\underline{\ev}}_n^{(\ell)}
\end{align}
where $x_n \in \RR$.
For first-order diffraction, this minimization has a closed-form solution (see Appendix~\ref{sec:diffraction-first-order}).
The candidate is discarded if the diffraction point lies outside the finite diffracting edge $\iota_n^{(i_{\Dc})}$.

\begin{figure}[htbp]
    \centering
    \begin{subfigure}[b]{0.40\textwidth}
        \centering
        \includegraphics[width=\textwidth]{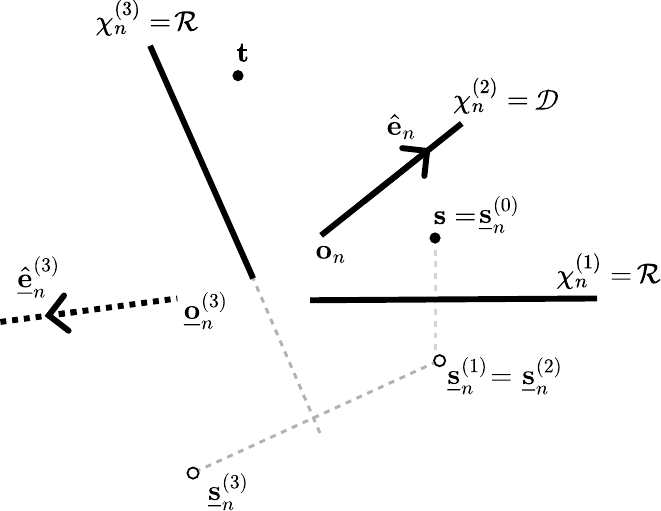}
        \caption{\label{fig:path-solver-im-diffraction-step-1}}
    \end{subfigure}
    \hfill
    \begin{subfigure}[b]{0.18\textwidth}
        \centering
        \includegraphics[width=0.75\textwidth]{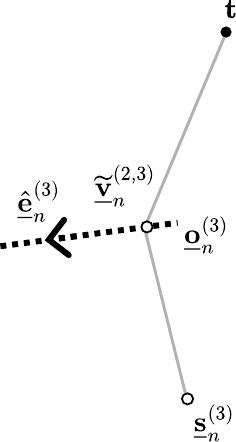}
        \caption{\label{fig:path-solver-im-diffraction-step-2}}
    \end{subfigure}
    \hfill
    \begin{subfigure}[b]{0.40\textwidth}
        \centering
        \includegraphics[width=\textwidth]{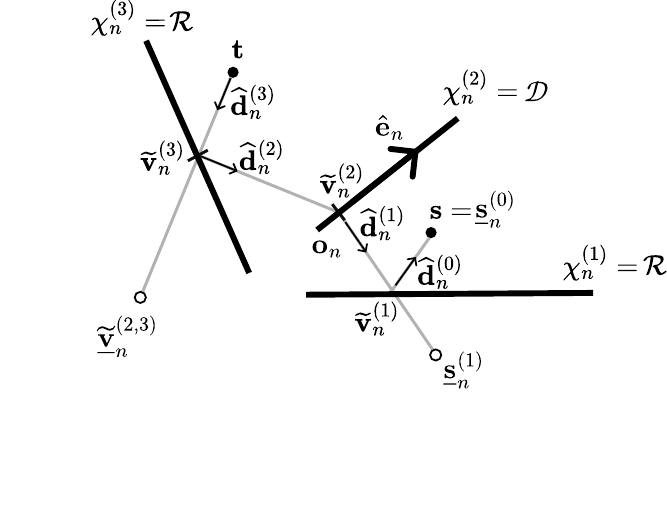}
        \caption{\label{fig:path-solver-im-diffraction-step-3}}
    \end{subfigure}
    \caption{The three steps of the image method for paths with diffraction: (a) compute the images of the source and the edge, (b) compute the diffraction point in image space, and (c) backtrack the path. Shown in 2D for clarity.\label{fig:path-solver-im-step-diffraction}}
\end{figure}

Once the diffraction point is found in the image space, the refined path vertices $\widetilde{\vv}_n^{(i)}$, $i = 1, \dots, \ell$, are determined by backtracking as in the case without diffraction. However, for depths $i = i_{\Dc} + 1, \dots, \ell$, the algorithm uses the intermediate images of the diffraction point:
\begin{equation}
    \widetilde{\underline{\vv}}_n^{(i_{\Dc}, i)} = \underline{\ov}_n^{(i)} + x_n \widehat{\underline{\ev}}_n^{(i)}
\end{equation}
in place of $\underline{\sv}_n^{(i)}$ in~\eqref{eq:path-solver-im-backtracking}.
For $i = 1, \dots, i_{\Dc} - 1$, $\underline{\sv}_n^{(i)}$ is used as in the non-diffraction case.
The diffraction point is simply $\widetilde{\vv}_n^{(i_{\Dc})} = \ov_n + x_n \widehat{\ev}_n$.
This process is depicted in Figure~\ref{fig:path-solver-im-step-diffraction}.

\begin{figure}[ht!]
    \centering
    \begin{subfigure}[b]{0.45\textwidth}
        \centering
        \includegraphics[width=\textwidth]{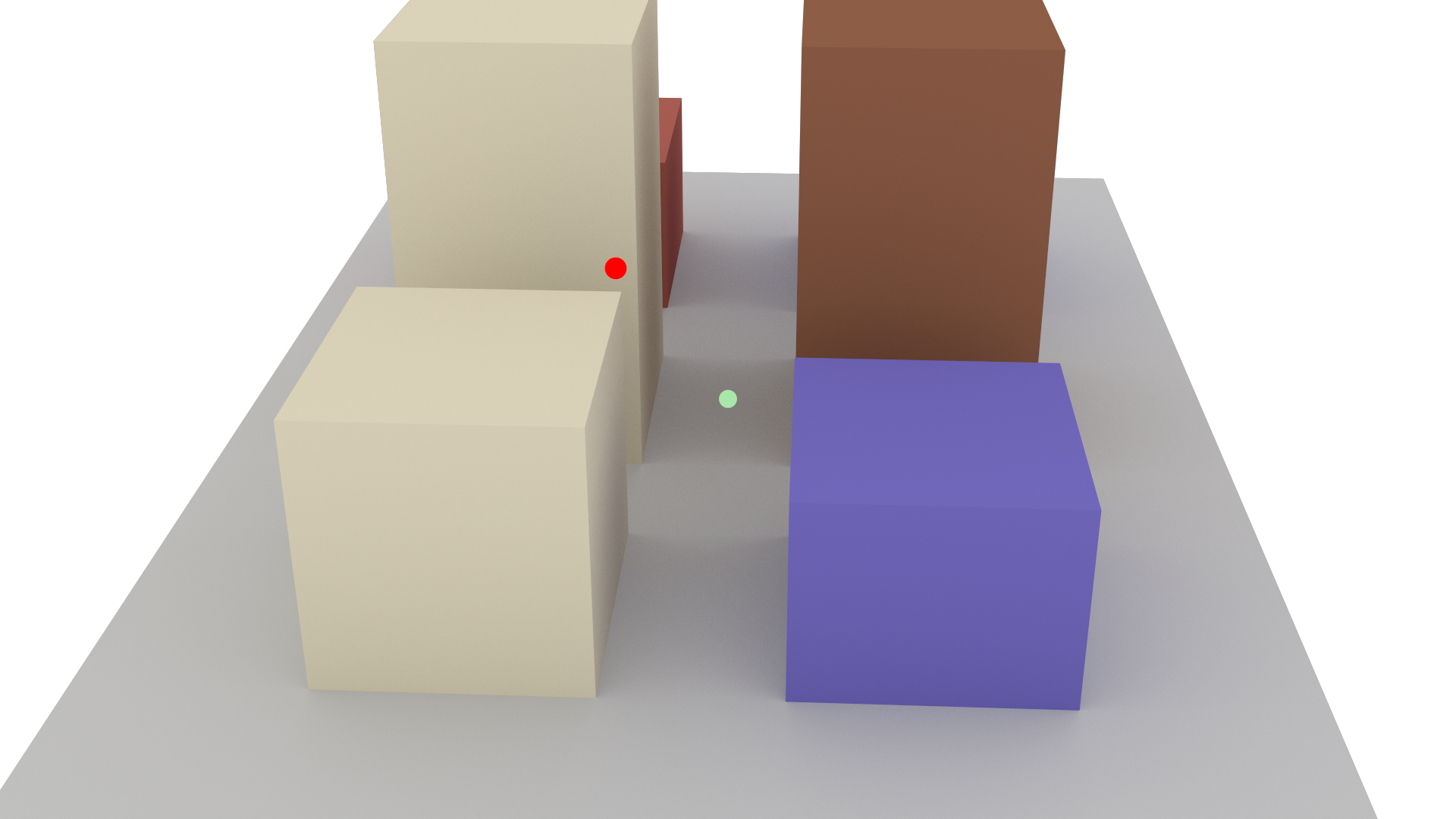}
        \caption{Scene for this experiment. Red ball: source, green ball: target.\label{fig:path-solver-im-scene}}
    \end{subfigure}
    \hfill
    \begin{subfigure}[b]{0.45\textwidth}
        \centering
        \includegraphics[width=\textwidth]{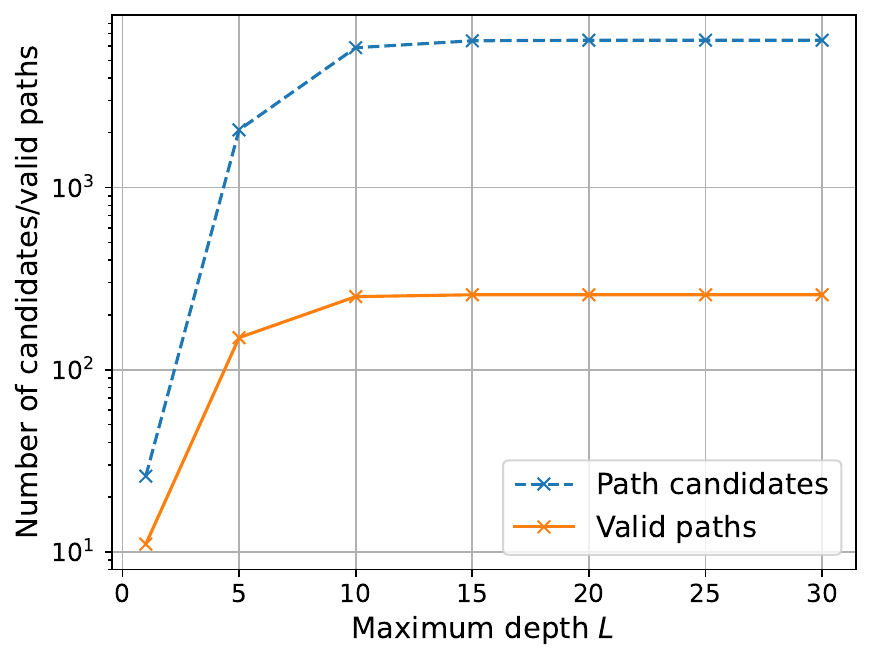}
        \caption{Number of candidates and valid paths.\label{fig:path-solver-gen-valid}}
    \end{subfigure}
    \caption{The image method typically discards most candidates.\label{fig:path-solver-sbr-im}}
\end{figure}

Typically, most of the candidates are discarded by the image method. This is illustrated in Figure~\ref{fig:path-solver-sbr-im}, which shows the number of candidates discarded as a function of the path depth for a single source and target. The considered scene is shown in Figure~\ref{fig:path-solver-im-scene}. The number of samples $N_S$ is $10^6$, diffuse reflection is disabled, and diffraction is enabled.
As seen in Figure~\ref{fig:path-solver-gen-valid}, most of the candidates are discarded by the image method because only specular chains are generated by the candidate generator, and most of them do not result in valid paths.

\subsection{Channel Coefficients, Delays, and Doppler Shifts Computation}
\label{sec:path-solver-em}

The final step of the path solver computes complex-valued channel coefficients and real-valued delays for each valid path (see Figure~\ref{fig:path-solver-archi}).
For clarity, we will focus our discussion on a single valid path with index $n$ and depth $\ell$, though the same procedure applies to all valid paths computed by both the \gls{SBR}-based candidate generator and image method.
We denote the path vertices as $\vv_n^{(i)},~i = 1, \dots, \ell$, which may correspond to vertices previously computed by the image method (denoted earlier as $\widetilde{\vv}_n^{(i)}$).
The directions of propagation between vertices are represented by unit vectors $\widehat{\kv}_n^{(i)},~i = 0, \dots, \ell$, defined as
\begin{equation}
    \widehat{\kv}_n^{(i)} = \frac{\vv_n^{(i+1)} - \vv_n^{(i)}}{\norm{\vv_n^{(i+1)} - \vv_n^{(i)}}},\qquad i = 0, \dots, \ell
\end{equation}
where $\vv_n^{(0)} = \sv$ is the source and $\vv_n^{(\ell+1)} = \tv$ is the target. Note that these propagation directions may differ from those used during the initial \gls{SBR} loop described in Section~\ref{sec:path-solver-sbr} due to the refinement by the image method.

\begin{figure}[ht!]
    \centering
    \includegraphics[width=0.6\textwidth]{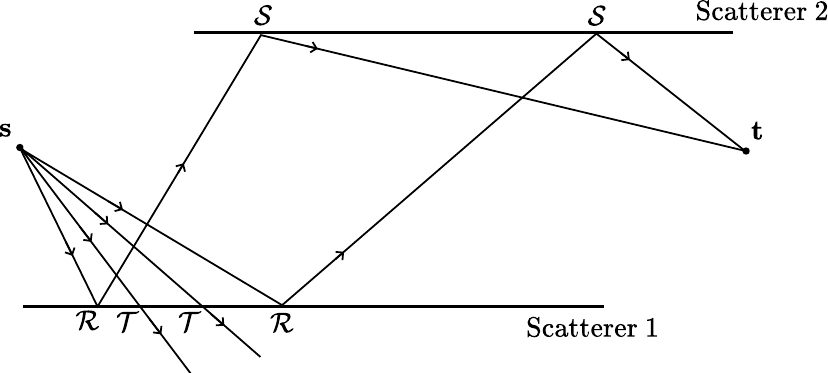}
    \caption{Due to the random sampling of interaction types, only some of the ray tubes that should reach the diffusely reflective surface (Scatterer 2) are traced. In this example, two out of four initial rays are transmitted when interacting with Scatterer 1, while the other two are reflected and reach the diffusely reflective surface.\label{fig:path-solver-energy}}
\end{figure}

\paragraph{Ensuring Energy Conservation}
As explained in Section~\ref{sec:path-solver-interaction-sampling}, at each interaction point, an interaction type is randomly chosen based on the probability distribution $\Qc$~\eqref{eq:path-solver-int-dist}. Initially, the energy emitted by the source is distributed across $N_S$ samples according to the transmit antenna pattern (see Section~\ref{sec:primer-em-tx-far-field}). Since only one interaction type is selected at each interaction point, ray tubes that do not match the chosen interaction type are discarded, as illustrated in Figure~\ref{fig:path-solver-energy}. The remaining ray tubes must therefore compensate for the discarded ones to maintain energy conservation. Given that there is a finite number of specular chains with a specific depth, the channel gain~\eqref{eq:channel-gain} observed at the target due to specular chains is accurate if all specular chains are identified by the path solver, which is assumed to be the case. This assumption will be reasonable if the number of samples $N_S$ is sufficiently large.

In the context of diffuse reflections, an incident ray tube scatters an amount of energy that is determined by its footprint on the diffusely reflective surface (see Section~\ref{sec:primer-em-diffuse}). Due to the random sampling of interaction types, only a portion of the ray tubes that should reach a diffusely reflective surface are actually traced (see Figure~\ref{fig:path-solver-energy}), which requires compensating for the untraced ones. For a given path $p^{(\ell)}$ that includes a diffuse reflection and has a specular suffix index $\ell_d$~\eqref{eq:essential-concepts-ell_d}, we denote by $\Pr \LB \boldsymbol{\chi}_n^{(\ell_d)} \RB = \prod_{\ell'=1}^{\ell_d} q \LB \chi_n^{(\ell')}\RB$ the probability of sampling the prefix leading to the last diffuse reflection point, where $q\LB \cdot \RB$ is defined in~\eqref{eq:path-solver-int-dist}. Consequently, on average, only a fraction $\Pr \LB \boldsymbol{\chi}_n^{(\ell_d)} \RB$ of the ray tubes that should reach the diffusely reflective surface are actually traced. To account for this, the electric field is scaled by $\nicefrac{1}{\sqrt{\Pr \LB \boldsymbol{\chi}_n^{(\ell_d)} \RB}}$ at the point of diffuse reflection.

\begin{algorithm}[ht!]
\caption{Electric field computation\label{alg:path-solver-electric-field}}
\begin{algorithmic}[1]
\State $\Em_n \gets \Call{tx\_pattern}{\widehat{\kv}^{(0)}_n}$ \Comment{Electric field vector}
\State $\tau_n \gets 0$ \Comment{Cumulative delay}
\State $r_n \gets 0$ \Comment{Length of the ray tube}
\State $\gamma_n \gets 1$ \Comment{Cumulative path probability}
\State $s_n \gets 0$ \Comment{Length of the incident ray tube on the diffracting edge}
\For{$\ell' = 1, \dots, \ell$}
    \State $\Tm_n^{(\ell')} \gets \Call{Evaluate\_Material}{o_n^{(\ell')}, \chi_n^{(\ell')}, \dots}$
    \State $\Em_n \gets \Tm_n^{(\ell')} \Em_n$
    \State $\gamma_n \gets \gamma_n \cdot q\LB \chi_n^{(\ell')}\RB$
    \State $r_n \gets r_n + \norm{\vv_n^{(\ell')} - \vv_n^{(\ell'-1)}}$
    \If{$\chi_n^{(\ell')} = \Sc$}
        \State $\Em_n \gets \frac{\Em_n}{\sqrt{\gamma_n}}$
        \State $\gamma_n \gets 1$
        \State $r_n \gets 0$
    \ElsIf{$\chi_n^{(\ell')} = \Dc$}
        \State $s_n \gets r_n$
        \State $r_n \gets 0$
    \EndIf
    \State $\tau_n \gets \tau_n + \frac{\norm{\vv_n^{(\ell')} - \vv_n^{(\ell'-1)}}}{c_0}$
\EndFor
\State $r_n \gets r_n + \norm{\tv - \vv_n^{(\ell)}}$
\State $\tau_n \gets \tau_n + \frac{\norm{\tv - \vv_n^{(\ell)}}}{c_0}$
\State $\Em_n^{\text{rx}} \gets \Call{rx\_pattern}{-\widehat{\kv}_n^{(\ell)}}$
\State $a_n \gets \frac{\lambda_0}{4\pi} \LB \Em_n^{\text{rx}} \RB\htp \Em_n$
\If{Diffracted path} \label{line:sf-start}
    \State $a_n \gets  \frac{a_n}{\sqrt{s_nr_n(s_n + r_n)}}$
\Else
    \State $a_n \gets \frac{a_n}{r_n}$
\EndIf \label{line:sf-end}
\State \Return $a_n,~\tau_n$
\end{algorithmic}
\end{algorithm}

\paragraph{Path Coefficient and Delay Computation}
Algorithm~\ref{alg:path-solver-electric-field} outlines the computation of the electric field along a path, based on the \gls{EM} background in Appendix~\ref{sec:primer-em}.
The algorithm calculates the electric field vector for each path by iterating through its sequence of interactions.
Computation begins by initializing the electric field $\Em_n$ with the transmit antenna pattern, evaluated in the initial departure direction $\widehat{\kv}_n^{(0)}$.
For dual polarization, two independent electric field vectors are tracked per path.
The variable $\tau_n$ accumulates the total propagation delay, and $r_n$ tracks the ray tube length. Both are initialized to zero.
The cumulative path probability $\gamma_n$, used for energy normalization, is initialized to one (see previous paragraph) and updated at each interaction.
The variable $s_n$ is only used for diffracted paths. For these paths, it accumulates the length of the incident ray tube on the diffracting edge, which is required to compute the diffraction spreading factor, i.e., the field-amplitude spreading factor due to free space propagation (see Section~\ref{sec:primer-em-diffraction}).

At each interaction, the electric field is updated by the transfer matrix $\Tm_n^{(\ell')}$, which depends on the intersected object $o_n^{(\ell')}$, the interaction type $\chi_n^{(\ell')}$, and other parameters (omitted for brevity). The path probability $\gamma_n$ is multiplied by the current interaction probability $q\LB \chi_n^{(\ell')}\RB$.
For diffuse reflections, the electric field is divided by $\sqrt{\gamma_n}$ to ensure energy conservation and $\gamma_n$ and $r_n$ are reset to one and zero, respectively, as diffuse reflection spawns a new ray tube.
For diffractions, the length of the incident ray tube on the diffracting edge $r_n$ is stored in $s_n$ and $r_n$ is reset to zero.
Each segment’s travel time is added to $\tau_n$, with $c_0$ the speed of light in vacuum.

After processing all interactions, the algorithm adds the final segment to the target to both $\tau_n$ and $r_n$.
The receive antenna pattern is then evaluated for the incoming direction, and the channel coefficient is computed following Section~\ref{sec:primer-em-freq-ir}.
Finally, the spreading factor (lines~\ref{line:sf-start}--\ref{line:sf-end}) is computed and applied to the channel coefficient depending on whether the path involves diffraction or not.

\paragraph{Remark}
The transfer matrices $\Tm_n^{(\ell')}$ and the electric field vectors $\Em_n$ are complex-valued, but are implemented using their real-valued representations defined as
\begin{equation}
    \begin{bmatrix}
        \Re \LB \Tm_n^{(\ell')} \RB & -\Im \LB \Tm_n^{(\ell')} \RB\\
        \Im \LB \Tm_n^{(\ell')} \RB & \Re \LB \Tm_n^{(\ell')} \RB
    \end{bmatrix}
\end{equation}
for matrices and
\begin{equation}
    \begin{bmatrix}
        \Re\LB \Em_n \RB\\
        \Im\LB \Em_n \RB
    \end{bmatrix}
\end{equation}
for vectors, where $\Re{\LB \cdot \RB}$ and $\Im{\LB \cdot \RB}$ denote the real and imaginary parts, respectively.

\subsubsection{Handling Multiple Antennas at the Transmitter and Receiver}
\label{sec:path-solver-antenna-arrays}

As explained in the introduction (Section~\ref{sec:introduction}), \SRT{} supports antenna arrays of any size at both the transmitter and receiver, utilizing either synthetic or non-synthetic arrays. In the case of non-synthetic arrays, each transmit antenna is treated as a source and each receive antenna as a target, meaning paths are traced from every transmit antenna, and the solver calculates paths for each transmit-receive antenna pair. For synthetic arrays, paths originate from a single ``virtual'' source at the center of the transmitter array, and a single ``virtual'' target is considered at the center of every receiver array. Channel coefficients are then computed by applying phase shifts synthetically. This method allows for efficient scaling with the number of antennas, but it requires plane wave approximation on each array, which is valid only when $D^2 \ll R \lambda_0$, where $D$ is the array total aperture and $R$ is the distance from the array center to the first interaction at transmitter or the last interaction at receiver.
Formally, for synthetic arrays, the channel matrix for the $n$-th path is expressed as:
\begin{equation}
    \Hm_{\text{array}, n} = a_n \uv_A^{\text{rx}} \LB \uv_A^{\text{tx}} \RB\tp
\end{equation}
where $a_n$ is the channel coefficient for the $n$-th path, calculated using Algorithm~\ref{alg:path-solver-electric-field}. The array response vectors for the transmit ($\uv_A^{\text{tx}}$) and receive ($\uv_A^{\text{rx}}$) antenna arrays are defined as:
\begin{equation}
    \label{eq:path-solver-array-response-vector}
    \begin{aligned}
        \uv_A^{\text{tx}} &= \LSB e^{j \frac{2\pi}{\lambda_0}\LB \widehat{\kv}_n^{(0)} \RB \tp\dv_1^{\text{tx}}}, \cdots, e^{j \frac{2\pi}{\lambda_0}\LB \widehat{\kv}_n^{(0)} \RB \tp\dv_{N_A^{\text{tx}}}^\text{tx}} \RSB\tp,\\
        \uv_A^{\text{rx}} &= \LSB e^{j \frac{2\pi}{\lambda_0}\LB -\widehat{\kv}_n^{(\ell)} \RB \tp\dv_1^{\text{rx}}}, \cdots, e^{j \frac{2\pi}{\lambda_0}\LB -\widehat{\kv}_n^{(\ell)} \RB \tp\dv_{N_A^{\text{rx}}}^\text{rx}} \RSB\tp.
    \end{aligned}
\end{equation}
Here, $\widehat{\kv}_n^{(0)}$ ($\widehat{\kv}_n^{(\ell)}$) represents the departure (arrival) direction of the $n$-th path, $N_A^{\text{tx}}$ ($N_A^{\text{rx}}$) denotes the number of antennas in the transmit (receive) array, and $\dv_i^{\text{tx}}$ ($\dv_i^{\text{rx}}$) indicates the relative position of the $i$-th antenna element to the virtual source (target) in the transmit (receive) array.

\subsubsection{Doppler Shifts}

The impact of moving scene objects on the paths can be simulated in two ways.
The first approach involves moving objects in small incremental steps along their trajectories and recomputing the propagation paths at each step.
While this method provides the highest accuracy, it is computationally expensive.
However, when trajectory lengths are small (not exceeding a few wavelengths), a more efficient approach may be used.
In this alternative method, a velocity vector $\boldsymbol{\varv}\LB o \RB \in \RR^3$ is assigned to each object $o$ and the time evolution is computed by accumulating Doppler shifts along the propagation paths.

\begin{figure}[t]
    \centering
    \includegraphics[width=0.6\textwidth]{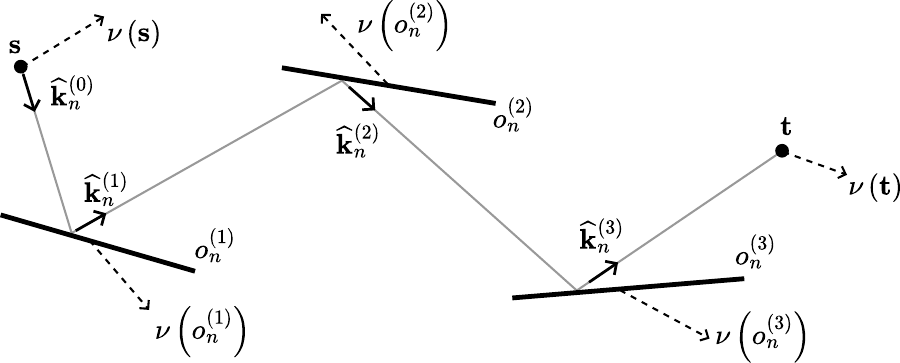}
    \caption{Each object as well as the source and target has an associated velocity vector that contributes to the total Doppler shift. Shown in 2D for clarity.\label{fig:path-solver-doppler}}
\end{figure}

Figure~\ref{fig:path-solver-doppler} illustrates this approach.
When replaying the valid paths using Algorithm~\ref{alg:path-solver-electric-field}, the Doppler shift for each path $n$ is computed by accumulating the contributions from individual objects:
\begin{equation}
    \nu_n^{(\ell')} =
    \begin{cases}
        0 & \text{if } \ell' = 0\\
        \nu_n^{(\ell'-1)} + \frac{\boldsymbol{\varv}\LB o_n^{(\ell')} \RB\tp \LB \widehat{\kv}_n^{(\ell')} - \widehat{\kv}_n^{(\ell'-1)} \RB}{\lambda_0} & \text{if } 1 \leq \ell' \leq \ell\\
    \end{cases}
\end{equation}
where $\lambda_0$ is the free-space wavelength. This results in the total Doppler shift $\nu_n^{(\ell)}$ caused by object movements along the path. The final Doppler shift observed at the target combines the accumulated shift from moving objects with the shifts caused by source and target motion:
\begin{equation}
    \nu_n = \nu_n^{(\ell)} + \frac{\boldsymbol{\varv}\LB \sv \RB\tp \widehat{\kv}_n^{(0)}}{\lambda_0} - \frac{\boldsymbol{\varv}\LB \tv \RB\tp \widehat{\kv}_n^{(\ell)}}{\lambda_0}
\end{equation}
where $\boldsymbol{\varv}\LB \sv \RB \in \RR^3$ and $\boldsymbol{\varv}\LB \tv \RB \in \RR^3$ denote the velocity vectors of the source and target, respectively.

\subsection{Current Limitations and Potential Improvements}
\label{sec:path-solver-limitations}

\begin{figure}[h!]
    \centering
    \includegraphics[width=0.55\textwidth]{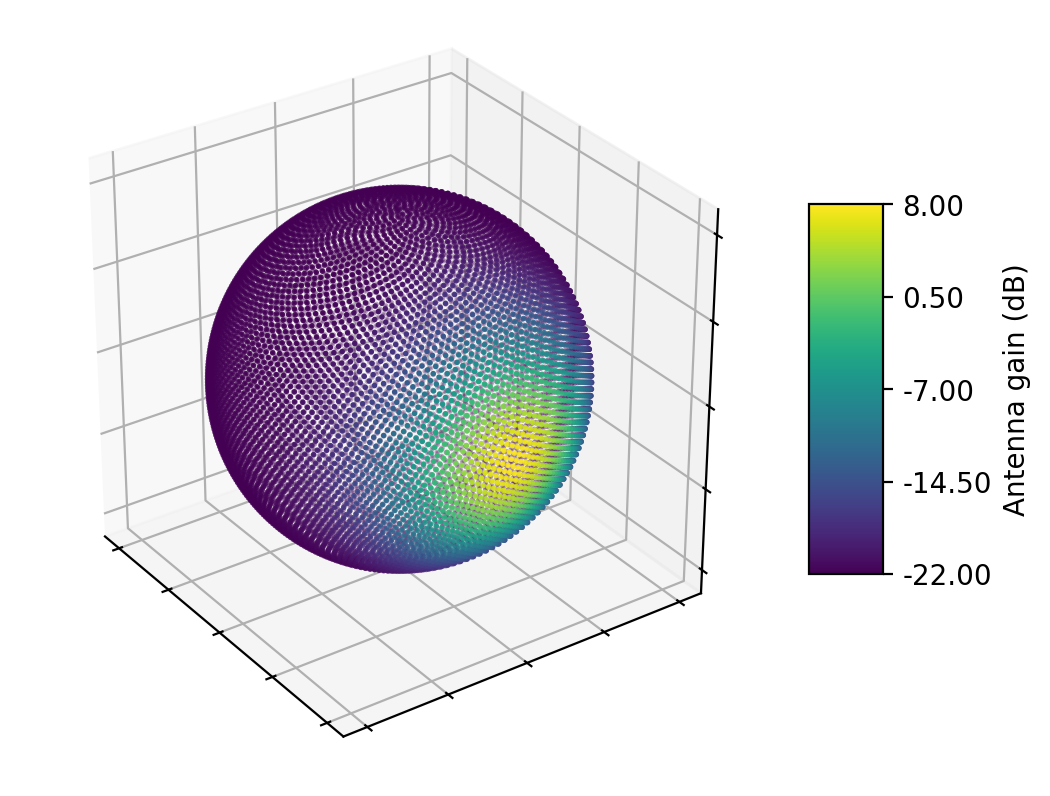}
    \caption{Gain of the 3GPP TR 38.901~\cite[Table~7.3-1]{TR38901} antenna pattern for each sample direction on a Fibonacci lattice with $10^4$ points. The majority of sample directions exhibit low gain, indicating that importance sampling of ray directions may be beneficial.\label{fig:path-solver-is-ant-pattern}}
\end{figure}

We discuss in this section some current limitations and potential improvements of the path solver.

\paragraph{Sampling of Ray Directions}

Improved sample efficiency could be achieved by sampling initial ray directions and scattered ray directions based on the transmit antenna pattern and the scattering pattern, respectively. This is motivated by the substantial sample efficiency gains shown in Section~\ref{sec:path-solver-interaction-sampling} through the use of importance sampling. By applying optimized sampling strategies for ray directions, further improvements may be obtained, as illustrated in Figure~\ref{fig:path-solver-is-ant-pattern}.

\paragraph{Handling Large Numbers of Targets and Sources}

\begin{figure}[h!]
    \centering
    \begin{subfigure}[b]{0.50\textwidth}
        \centering
        \includegraphics[width=\textwidth]{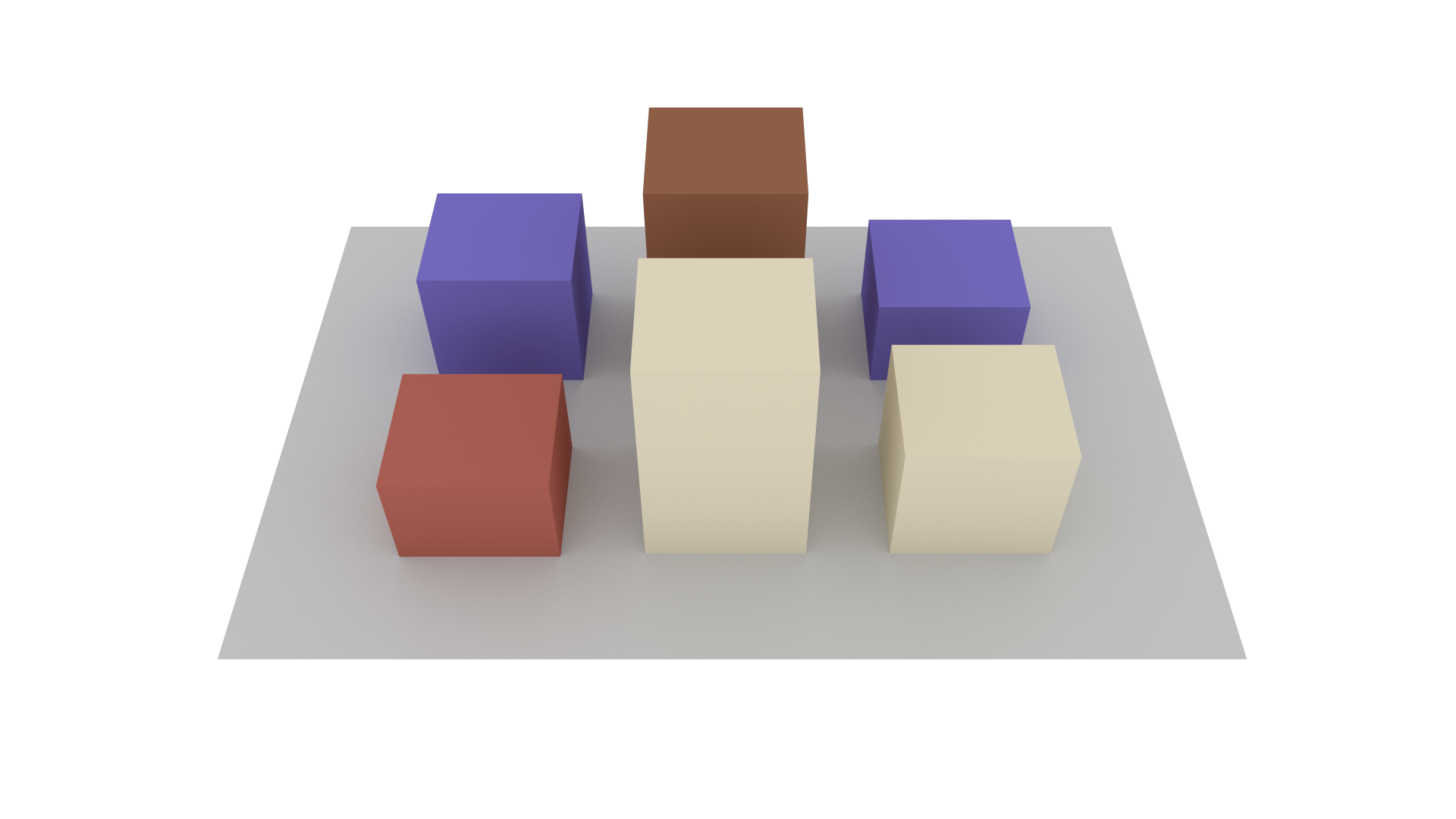}
        \caption{Considered scene.\label{fig:simple-street-canyon}}
    \end{subfigure}
    \hfill
    \begin{subfigure}[b]{0.49\textwidth}
        \centering
        \includegraphics[width=\textwidth]{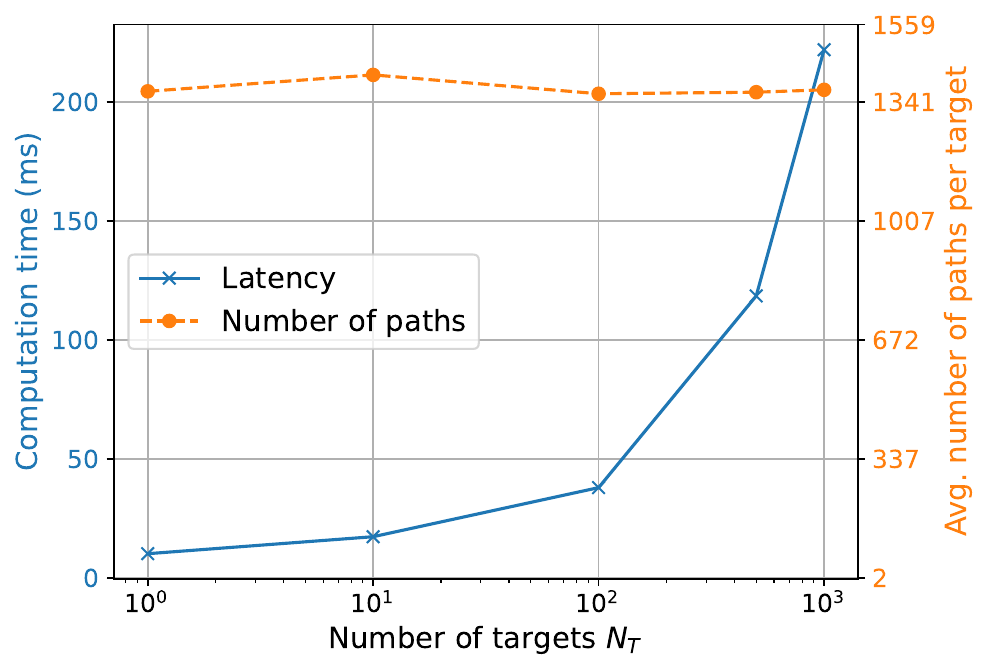}
        \caption{Computation time and number of candidates found vs. the number of targets.\label{fig:path-solver-sbr-tgt}}
    \end{subfigure}

    \begin{subfigure}[b]{0.49\textwidth}
        \centering
        \includegraphics[width=\textwidth]{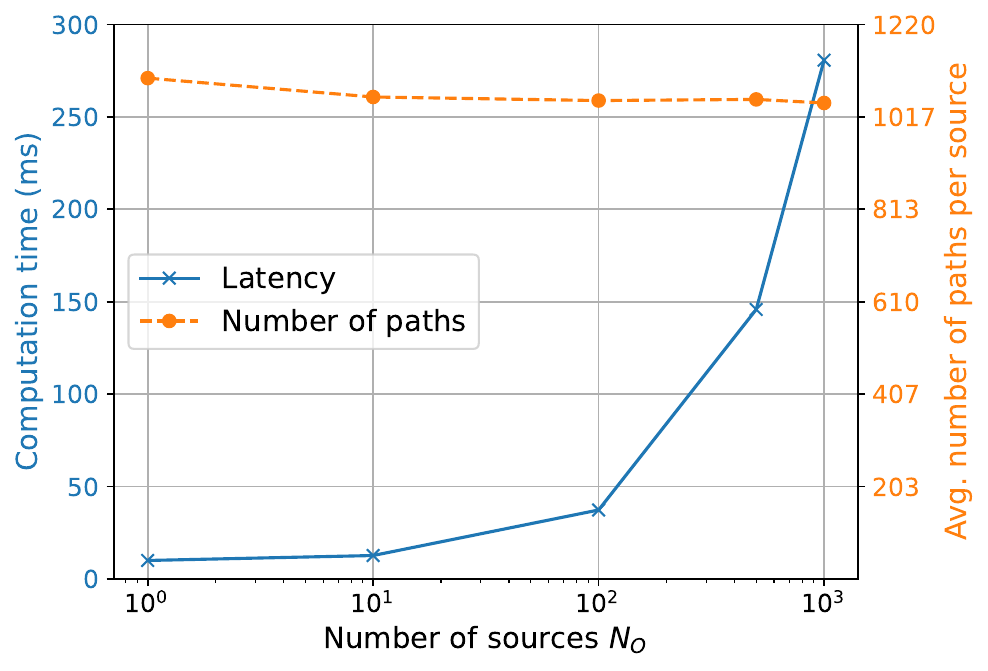}
        \caption{Computation time and number of candidates found vs. the number of sources and with a constant number of samples per source.\label{fig:path-solver-sbr-src-cst-samples-per-src}}
    \end{subfigure}
    \hfill
    \begin{subfigure}[b]{0.49\textwidth}
        \centering
        \includegraphics[width=\textwidth]{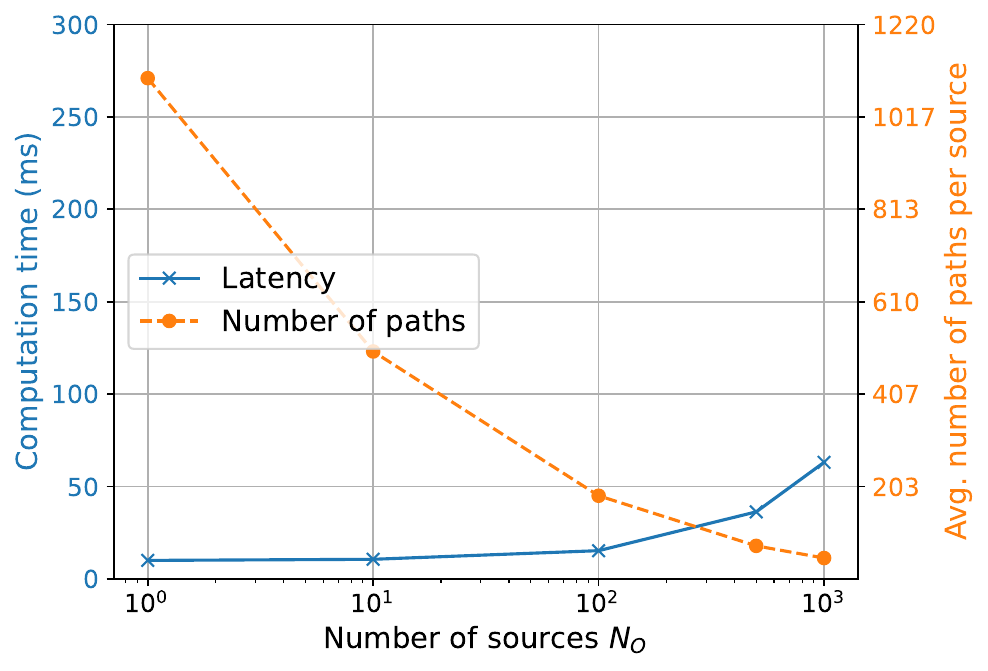}
        \caption{Computation time and number of candidates found vs. the total number of samples with a constant number of samples in total.\label{fig:path-solver-sbr-src-cst-total-samples}}
    \end{subfigure}
    \caption{Computation time and number of candidates found as a function of the number of sources and targets. The experiments are conducted on an NVIDIA RTX 4090 GPU.\label{fig:path-solver-sbr-src-tgt}}
\end{figure}

The current method for handling multiple targets involves iterating over all targets during each \gls{SBR} loop iteration. Figure~\ref{fig:path-solver-sbr-tgt} illustrates this, showing the computation time of the candidate generator and the average number of candidates per target found as the number of targets varies. In this experiment, the built-in \SRT{} scene ``simple street canyon'' depicted in Figure~\ref{fig:simple-street-canyon} was used. Targets were sampled uniformly at random on a plane parallel to the ground at an elevation of 1.5 meters. The maximum depth $L$ was set to 5, diffuse reflections were disabled, and the number of samples $N_S$ was set to $10^6$. As observed, the computation time increases with the number of targets due to the iteration over all targets in each \gls{SBR} loop iteration. However, the number of candidates per target remains stable, although with variations due to hash collisions. Note that, as discussed in Section~\ref{sec:path-solver-im}, most candidates do not result in valid paths.
The current approach may be optimized by selecting only targets near intersection points or the source, which will enhance scalability when dealing with a large number of targets.

Scaling with the number of sources presents its own set of challenges. When $N_S$ samples are generated per source, the total number of samples produced by the candidate generator becomes $N_S \cdot N_{\text{src}}$, where $N_{\text{src}}$ denotes the number of sources. This results in increased computation time as the number of sources grows, as illustrated in Figure~\ref{fig:path-solver-sbr-src-cst-samples-per-src}. However, the number of candidates per source remains stable. In this experiment, a single target is positioned at the center of the scene at an elevation of 1.5 meters, while sources are randomly sampled on a plane parallel to the ground at an elevation of 70 meters, above all buildings. Recall that variations in the number of candidates per source may occur due to collisions. For a comprehensive view, Figure~\ref{fig:path-solver-sbr-src-cst-total-samples} presents results where the total number of samples is kept constant at $10^6$, resulting in $N_S = \left\lfloor \frac{10^6}{N_{\text{src}}} \right\rfloor$ samples per source. In this scenario, the computation time increases at a much slower rate with the number of sources, but the number of candidates found per source decreases rapidly as the number of sources increases.

\paragraph{Higher-Order Diffraction and Diffuse Reflection}

The path solver supports only first-order diffraction and does not handle paths that contain both diffraction and diffuse reflection. Enabling support for such combined paths would require the refinement of non-suffix path segments, which is computationally expensive and is therefore not currently implemented. However, such paths typically transport a negligible amount of energy and therefore have little impact on the accuracy of the computed \glspl{CIR}.

Regarding high-order diffraction, algorithms exist for determining the diffraction vertices~\cite{4200889}, and multiple-edge approximations are available for the associated fields.
However, higher-order diffracted paths are not modeled by the current implementation. Selecting a robust and scalable method suitable for general scenes remains future work.

\paragraph{Collisions of Specular Chain Candidates}

The path solver does not handle collisions of specular chain candidates.
More precisely, when two different specular chains share the same index value in the hash array despite having a different hash due to the modulo operation~\eqref{eq:path-solver-hash-mod}, the path solver discards one of them. This could be avoided by using a collision resolution mechanism, and by dynamically adjusting the hash array size to reduce the probability of collisions.
Currently, it is assumed that the hash array size is set to a sufficiently large value to make the collision rate acceptable.

\clearpage

\begin{highlightbox}{Path Solver: Supported Path Families and Array Models}

    Let $L$ denote the user-configured maximum path depth, i.e.,
    the maximum number of interactions along a path. The path solver
    supports the following path families:

    \renewcommand{\arraystretch}{1.25}
    \begin{center}
    \begin{tabular}{@{}p{0.28\linewidth}p{0.66\linewidth}@{}}
    \hline
    \textbf{Path family} & \textbf{Supported interaction sequences} \\
    \hline

    \Gls{LoS}
    &
    A path of depth zero, containing no interaction.
    \\

    Specular chains
    &
    Any sequence of specular reflections, refractions, and at most one diffraction, with total depth at most $L$.
    \\

    Paths with diffuse reflection
    &
    Any sequence of specular reflections, refractions, and at least one diffuse reflection,
    with total depth at most $L$.
    \\

    \hline
    \end{tabular}
    \end{center}
    \renewcommand{\arraystretch}{1.0}
    Paths containing both diffuse reflection and diffraction are not
    supported. Higher-order diffraction, i.e., paths containing more
    than one diffraction event, is also not supported.

    \medskip
    \noindent
    \textbf{Array models.}
    The path solver supports both non-synthetic and synthetic antenna arrays:
    \begin{itemize}
        \item With a \emph{non-synthetic array}, every transmit antenna is
        treated as a source and every receive antenna as a target. Paths are
        therefore computed separately for every transmit--receive antenna pair,
        and no synthetic-array approximation is applied.

        \item With a \emph{synthetic array}, paths are computed between the
        centers of the transmit and receive arrays. Element-specific channel
        coefficients are then obtained by applying phase shifts determined by
        the array geometries and the path departure and arrival directions.
        The approximation requires the wavefront to be approximately planar across the array. A commonly used sufficient far-field criterion is $D^2 \ll R \lambda_0$, where $D$ is the array total aperture and $R$ is the distance from the array center to the first interaction at transmitter or the last interaction at receiver.
    \end{itemize}
    \end{highlightbox}

\newpage
\section{Radio Map Solver}
\label{sec:radio-map-solver}
In the previous sections, we have detailed the path solver, which computes propagation paths and the related electric fields from a source to a target. Now, we turn our attention to the radio map solver, which computes a \emph{radio map} (also known as \emph{coverage map} or \emph{power map}) for a given scene and source. While we describe the algorithm for a single source, \SRT{} can compute radio maps for multiple sources in parallel.

\begin{figure}[ht!]
    \centering
    \begin{subfigure}[b]{0.6\textwidth}
        \centering
        \includegraphics[width=\textwidth]{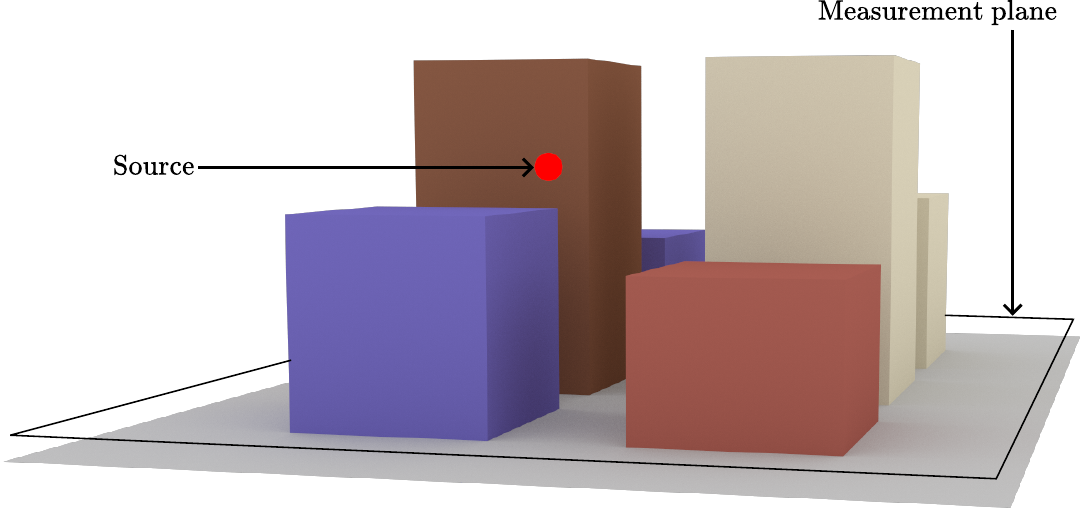}
        \caption{Scene with a measurement plane.}
    \end{subfigure}

    \begin{subfigure}[b]{0.6\textwidth}
        \centering
        \includegraphics[width=\textwidth]{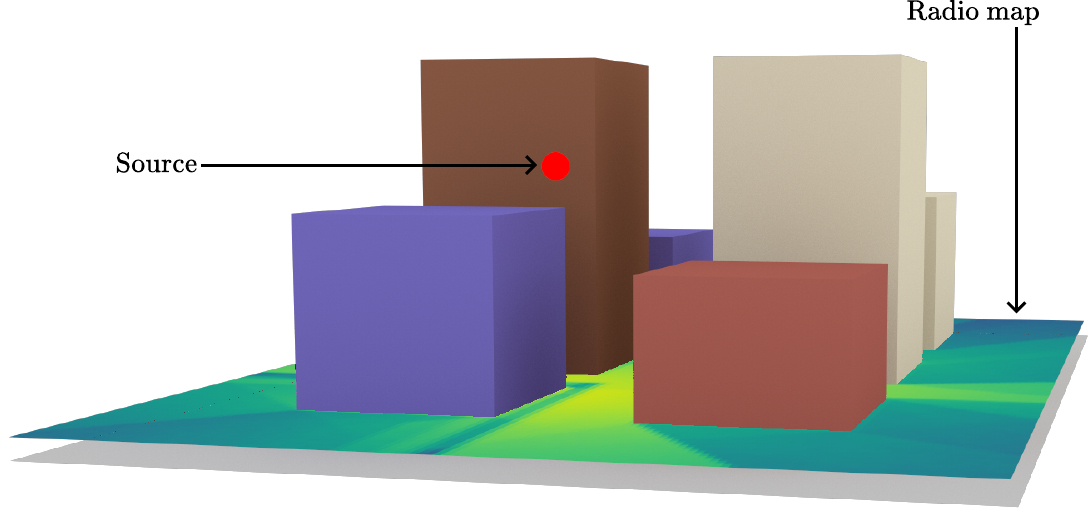}
        \caption{Radio map for the measurement plane in (a).}
    \end{subfigure}
    \caption{(a) A measurement surface, which is here a plane parallel to the ground, is placed in the scene to capture paths without affecting their propagation, (b) the resulting radio map showing the spatial distribution of channel gains.\label{fig:radio-map-solver-eg}}
\end{figure}

A radio map estimates the channel gain~\eqref{eq:channel-gain} from a source to each point on a \emph{measurement surface}, as illustrated in Figure~\ref{fig:radio-map-solver-eg}. The measurement surface is partitioned into \emph{measurement cells}, which are assumed to be planar. For each cell, the solver estimates the average channel gain that would be observed by a target placed within that cell.
Note that the measurement surface does not interact with the electromagnetic waves, but only serves to capture the paths that intersect it.
Moreover, a ray can intersect the measurement surface at multiple points.

In Section~\ref{sec:radio-map-solver-definition}, we present a definition of radio maps that allows for their efficient computation using \gls{SBR}. This approach makes the computation of radio maps both fast and scalable.

\subsection{Definition of Radio Map}
\label{sec:radio-map-solver-definition}

We define radio maps as path integrals, similarly to how measurements are defined in computer graphics~\cite[Section~3.7]{10.1145/15886.15902,Veach:1997:RMC}.
Consider a point source $\sv$ and a measurement surface $\Mc$, divided into $N_M$ cells $M_i$, for $i=1,\dots,N_M$, each of area $\abs{M_i}$.
For clarity, we ignore diffraction in this subsection. Diffraction is considered in Section~\ref{sec:radio-map-solver-diffraction}.

\begin{figure}[ht!]
    \centering
    \begin{subfigure}[b]{0.20\textwidth}
        \centering
        \includegraphics[width=\textwidth]{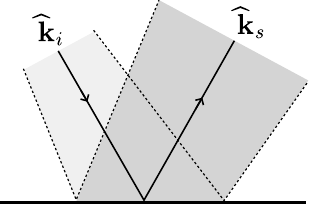}
        \caption{Specular reflection}
        \label{fig:radio-map-solver-sbr-1}
    \end{subfigure}%
    \hspace{7em}
    \begin{subfigure}[b]{0.20\textwidth}
        \centering
        \includegraphics[width=\textwidth]{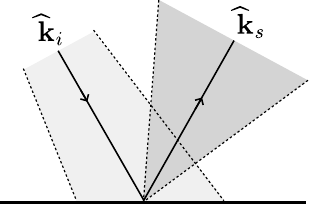}
        \caption{Diffuse reflection}
        \label{fig:radio-map-solver-sbr-2}
    \end{subfigure}
    \caption{
        A specular reflection (left) deviates the incident ray tube, whereas a diffuse reflection (right) spawns a new ray tube. Shown in 2D for clarity.
        \label{fig:radio-map-solver-sbr}
    }
\end{figure}

In the propagation model implemented by \SRT{}, diffuse reflections spawn new ray tubes, whereas specular reflections and refractions merely deviate incident ray tubes, as illustrated in Figure~\ref{fig:radio-map-solver-sbr}.
Ignoring diffuse reflections for now, $\sv$ is the only source of ray tubes, and every path therefore consists of a single ray tube.
The radio map value for the $i$-th cell due to specular reflections and refractions is
\begin{equation}
    \label{eq:radio-map-solver-ci-ell}
    C_i^{(\Rc,\Tc)}
    =
    \frac{1}{\abs{M_i}}
    \int_{M_i}
    e^{(\Rc,\Tc)}(\tv)\,dA(\tv),
\end{equation}
where $dA(\tv)$ is an infinitesimal surface element of $M_i$, and
\begin{equation}
    \label{eq:radio-map-solver-point-gain}
    e^{(\Rc,\Tc)}(\tv)
    =
    \sum_{p\in\Pc^{(\Rc,\Tc)}(\sv\to\tv)}
    e(p)\Vc\LB p\RB
\end{equation}
is the channel gain at $\tv$ due to specular reflections and refractions.
Here, $\Pc^{(\Rc,\Tc)}(\sv\to\tv)$ denotes the countable set of paths $p = \LB \sv, \vv^{(1)}, \dots, \vv^{(\ell)}, \tv \RB$ connecting $\sv$ to $\tv$ through specular reflections and refractions.
The visibility function $\Vc\LB p\RB$ equals $1$ if $p$ is not occluded and $0$ otherwise.
The channel gain associated with $p$ is
\begin{equation}
    \label{eq:radio-map-solver-path-gain}
    e(p)
    =
    \norm{
        \frac{\lambda_0}{4\pi}
        \Tm
        \Cm_{\mathrm T}
        (\theta_{\mathrm T},\varphi_{\mathrm T})
    }^2,
\end{equation}
where $\lambda_0$ is the free-space wavelength, $\Cm_{\mathrm T}$ is the two-dimensional complex transmit antenna pattern evaluated at the departure angles $(\theta_{\mathrm T},\varphi_{\mathrm T})$ of $p$, and $\Tm$ is a $2\times2$ complex matrix modeling the effects of the interactions and free-space propagation along $p$.
Unlike in~\eqref{eq:path-coefficient}, the receive antenna pattern is not applied.
Instead, the squared norm of the electric-field amplitude is used, corresponding to an isotropic receive antenna whose polarization is matched to that of the incident wave.
Because the squared amplitudes of the individual paths are summed, the radio map is non-coherent and does not capture interference between paths.

We now generalize the definition to diffuse reflections.
Let $\Ac$ denote the union of all diffuse reflecting surfaces in the scene.
These surfaces can be regarded as a continuum of point sources.
As detailed in Appendix~\ref{sec:primer-em-diffuse}, the squared amplitude of the diffuse scattered field is proportional to the area of the impinging ray-tube footprint on the scattering surface.
For a path $p$ containing a diffuse reflection at $\vv^{(\Sc)}\in\Ac$, its channel gain can therefore be written as
\begin{equation}
    \label{eq:radio-map-solver-diffuse-reflection}
    e^{(\Sc)}(p)
    =
    \widetilde e^{(\Sc)}(p)\,dA(\vv^{(\Sc)})
\end{equation}
where $\widetilde e^{(\Sc)}(p)$ is the channel gain normalized by the area of the impinging ray-tube footprint.
A path containing one diffuse reflection consists of two ray tubes: one originating at $\sv$ and one originating at $\vv^{(\Sc)}$.
The path may additionally contain specular reflections and refractions before or after $\vv^{(\Sc)}$.
The contribution of paths containing exactly one diffuse reflection is
\begin{equation}
    \label{eq:radio-map-solver-ci-diffuse-one}
    C_i^{(\Sc,1)}
    =
    \frac{1}{\abs{M_i}}
    \int_{M_i}
    \int_{\Ac}
    \widetilde e^{(\Sc,1)}
    \LB \vv^{(\Sc)},\tv\RB
    \,dA(\vv^{(\Sc)})\,dA(\tv)
\end{equation}
where
\begin{equation}
    \label{eq:radio-map-solver-diffuse-one-density}
    \widetilde e^{(\Sc,1)}
    \LB \vv^{(\Sc)},\tv\RB
    =
    \sum_{
        p\in
        \Pc^{(\Sc,1)}
        (\sv\to\vv^{(\Sc)}\to\tv)
    }
    \widetilde e^{(\Sc)}(p)
    \Vc\LB p\RB.
\end{equation}
The diffuse-reflection point $\vv^{(\Sc)}$ is held fixed when defining
$\Pc^{(\Sc,1)}(\sv\to\vv^{(\Sc)}\to\tv)$.
Consequently, this path set is countable, although the subsequent integration over $\vv^{(\Sc)}\in\Ac$ is continuous.

For an arbitrary number $n_{\Sc}$ of diffuse reflections, define
\begin{equation}
    \Ac^{n_{\Sc}}
    =
    \underbrace{
        \Ac\times\cdots\times\Ac
    }_{n_{\Sc}\text{ times}}
\end{equation}
as the set of ordered sequences $\vv_{\Sc}^{(n_{\Sc})} = \LB \vv^{(\Sc,1)}, \dots, \vv^{(\Sc,n_{\Sc})} \RB$ of diffuse-reflection points, and define the product surface element
\begin{equation}
    dA^{n_{\Sc}}
    \LB \vv_{\Sc}^{(n_{\Sc})}\RB
    =
    \prod_{i=1}^{n_{\Sc}}
    dA\LB\vv^{(\Sc,i)}\RB.
\end{equation}
Repeated application of~\eqref{eq:radio-map-solver-diffuse-reflection} gives
\begin{equation}
    e^{(\Sc,n_{\Sc})}(p)
    =
    \widetilde e^{(\Sc,n_{\Sc})}(p)
    \,dA^{n_{\Sc}}
    \LB \vv_{\Sc}^{(n_{\Sc})}\RB,
\end{equation}
where for fixed diffuse-reflection points $\vv_{\Sc}^{(n_{\Sc})}$, we define
\begin{equation}
        \widetilde e^{(\Sc,n_{\Sc})}
        \LB \vv_{\Sc}^{(n_{\Sc})},\tv\RB
         =
        \sum_{
            p\in
            \Pc^{(\Sc,n_{\Sc})}
            \LB
                \sv
                \to
                \vv^{(\Sc,1)}
                \to\cdots\to
                \vv^{(\Sc,n_{\Sc})}
                \to
                \tv
            \RB
        }
        \widetilde e^{(\Sc,n_{\Sc})}(p)
        \Vc\LB p\RB.
    \label{eq:radio-map-solver-diffuse-n-density}
\end{equation}
As in the single-reflection case, the path set in this expression is countable because all diffuse-reflection points are held fixed.
The radio map value is obtained by summing over all possible numbers of diffuse reflections:
\begin{equation}
    \label{eq:radio-map-solver-ci-all}
    C_i
    =
    \frac{1}{\abs{M_i}}
    \int_{M_i}
    \sum_{n_{\Sc}=0}^{\infty}
    \int_{\Ac^{n_{\Sc}}}
    \widetilde e^{(\Sc,n_{\Sc})}
    \LB \vv_{\Sc}^{(n_{\Sc})},\tv\RB
    \,dA^{n_{\Sc}}
    \LB \vv_{\Sc}^{(n_{\Sc})}\RB
    \,dA(\tv).
\end{equation}
For $n_{\Sc}=0$, the inner surface integral is omitted and $\widetilde e^{(\Sc,0)}(\tv) = e^{(\Rc,\Tc)}(\tv)$, so that the corresponding term is precisely~\eqref{eq:radio-map-solver-ci-ell}.

\paragraph{Remark}
Equation~\eqref{eq:radio-map-solver-ci-all} accounts for arbitrarily many diffuse reflections. In practice, the path depth is limited to a finite maximum, so the corresponding sum is always finite.

\begin{highlightbox}{Relation to computer graphics}
    Defining radio maps as path integrals closely parallels the measurement equation used in computer graphics (see, e.g.,~\cite[Eq.~3.18]{Veach:1997:RMC}).
    Following~\cite[Chapter~8]{Veach:1997:RMC}, the radio map definition~\eqref{eq:radio-map-solver-ci-all} can be written as
    \begin{equation}
        \label{eq:radio-map-solver-ci-all-integral}
        C_i
        =
        \frac{1}{\abs{M_i}}
        \int_{M_i}
        \int_{\Ac^{\infty}}
        \widetilde e^{(\Sc)}(q,\tv)
        \,d\mu^{(\Sc)}(q)
        \,dA(\tv),
    \end{equation}
    where
    \begin{align}
        \Ac^{\infty}
        &=
        \bigcup_{n_{\Sc}=0}^{\infty}
        \Ac^{n_{\Sc}},
        \label{eq:radio-map-solver-Ac-infty}
        \\
        \mu^{(\Sc)}(B)
        &=
        \sum_{n_{\Sc}=0}^{\infty}
        A^{n_{\Sc}}
        \LB
            B\cap\Ac^{n_{\Sc}}
        \RB.
        \label{eq:radio-map-solver-mu-P}
    \end{align}
    Here, $B\subseteq\Ac^{\infty}$ is a measurable set, and
    $A^{n_{\Sc}}$ denotes the product surface-area measure on
    $\Ac^{n_{\Sc}}$.
    An element $q\in\Ac^{n_{\Sc}}$ is an ordered sequence of
    $n_{\Sc}$ diffuse-reflection points, and
    \begin{equation}
        \widetilde e^{(\Sc)}(q,\tv)
        =
        \widetilde e^{(\Sc,n_{\Sc})}(q,\tv),
        \qquad
        q\in\Ac^{n_{\Sc}},
    \end{equation}
    is the corresponding channel-gain density defined in
    Section~\ref{sec:radio-map-solver-definition}.
    With these definitions, the sum over the number of diffuse reflections
    in~\eqref{eq:radio-map-solver-ci-all} is absorbed into the measure
    $\mu^{(\Sc)}$, resulting in the single integral
    in~\eqref{eq:radio-map-solver-ci-all-integral}.
    This reduced path-space formulation is analogous
    to~\cite[Eq.~8.5]{Veach:1997:RMC} and enables the use of computer
    graphics techniques for radio map computation, including methods for
    improving sampling efficiency.
\end{highlightbox}

While the definition~\eqref{eq:radio-map-solver-ci-all} is comprehensive, it is not practical for computation as-is.
Indeed, directly integrating over the surfaces of the measurement cells would require connecting points on the measurement surface to the source, for example by placing one or more targets within each cell and using the path solver to compute the connecting paths. However, because accurately computing fine-grained radio maps typically demands a large number of target points, this approach is computationally prohibitive.
Instead, we use Monte Carlo integration with a single \gls{SBR} loop to efficiently estimate the value of each cell.
This is achieved by reformulating the radio map definition~\eqref{eq:radio-map-solver-ci-all} so that the integration is performed over the directions of rays originating from the source and the points that spawn ray tubes intersecting the measurement surface, rather than directly over the measurement and diffuse scatterer surfaces. This is enabled by applying the change of variables $dA = \nicefrac{r^2}{\abs{\widehat{\nv}\tp\widehat{\kv}}} d\omega$, where $r$ is the length of the ray tube, $\widehat{\kv}$ is its propagation direction, $\widehat{\nv}$ is the unit normal of either a scatterer or the measurement surface, and $d\omega$ is the infinitesimal solid angle subtended by the ray tube (see Section~\ref{sec:essential-concepts-ray-tube}). Thus, $dA$ represents the intersection area of the ray tube on the measurement surface.
With this change of variables, the radio map definition~\eqref{eq:radio-map-solver-ci-all} can be rewritten as:
\begin{multline}
    \label{eq:radio-map-solver-ci-all-2}
    C_i
    = \\
    \frac{1}{\abs{M_i}}
    \int_{\Omega} \sum_{n_{\Sc}=0}^{\infty}
    \int_{\Omega^{n_{\Sc}}}
    \widetilde e^{(\Sc,n_{\Sc})} \LB \vv_{\Sc}^{(n_{\Sc})},\tv \RB
    \frac{\LB r^{(\Sc, n_{\Sc})} \RB^2}{\abs{\LB \widehat{\nv}^{(\Sc, n_{\Sc} + 1)} \RB \tp \widehat{\kv}_i^{(\Sc, n_{\Sc} + 1)}}}
    \prod_{k=1}^{n_{\Sc}} \frac{\LB r^{(\Sc, k-1)} \RB^2}{\abs{\LB \widehat{\nv}^{(\Sc, k)} \RB \tp \widehat{\kv}_i^{(\Sc, k)}}}
    d\omega^{n_{\Sc}} \LB \widehat{\kv}_{\Sc}^{(n_{\Sc})}\RB
    d\omega \LB \widehat{\kv}_0\RB
\end{multline}
where $\Omega$ is the unit sphere,
$\widehat{\kv}_{\Sc}^{(n_{\Sc})} = \LB \widehat{\kv}^{(\Sc, 1)}, \dots, \widehat{\kv}^{(\Sc, n_{\Sc})} \RB$ is the sequence of scattered propagation directions at the diffuse reflection points,
$\LB \widehat{\kv}_i^{(\Sc, 1)}, \dots, \widehat{\kv}_i^{(\Sc, n_{\Sc})} \RB$ is the sequence of incident propagation directions at those points,
and $\widehat{\kv}_0$ is the propagation direction of the ray tube originating from the source.
The outer integral is therefore over ray tubes originating from the source, whereas the inner integral is over ray tubes originating from the diffuse reflection points.
Here, $r^{(\Sc, i)}$ is the length of the $i$-th ray tube originating from the $i$-th diffuse reflection point, $r^{(\Sc, 0)}$ is the length of the ray tube originating from the source, and $\widehat{\nv}^{(\Sc, i)}$ is the unit normal of the $i$-th diffusely reflecting surface.
Finally, $\widehat{\nv}^{(\Sc, n_{\Sc} + 1)}$ denotes the unit normal to the measurement surface, and $\widehat{\kv}_i^{(\Sc, n_{\Sc} + 1)}$ is the incident propagation direction on that surface.

Expression~\eqref{eq:radio-map-solver-ci-all-2} can be estimated as follows. Paths are generated by launching $N_S$ rays from the source $\sv$ and tracing their interactions with the scene geometry, thus performing a Monte Carlo integration over the surfaces of the scatterers. Intuitively, when a ray intersects a diffuse reflecting surface, it samples a surface element $\Delta A$ that corresponds to the ray tube intersection area, given by $\Delta A = \LB \nicefrac{r^2}{\abs{\widehat{\nv}\tp\widehat{\kv}}} \RB \Delta \omega$.
The source emits $N_S$ ray tubes, each subtending a solid angle of $\nicefrac{4\pi}{N_S}$, while each diffuse reflection point spawns a single ray tube, uniformly sampled over the hemisphere, with a solid angle of $2\pi$.
The maximum path depth is set to $L < \infty$, and since a single path may intersect the measurement surface at multiple depths, all path depths $\ell = 0, 1, \ldots, L$ are considered. Interaction types (specular reflection, refraction, or diffuse reflection) are sampled according to~\eqref{eq:path-solver-used-dist-event}.
This process leads to the following Monte Carlo estimator for the radio map~\eqref{eq:radio-map-solver-ci-all-2}:
\begin{multline}
    \label{eq:radio-map-solver-ci-all-2-estimator}
    \widehat{C}_i
    =
    \frac{4\pi}{N_S \abs{M_i}}
    {}\cdot \\
    \sum_{n=1}^{N_S}
    \sum_{\ell = 0}^L
    \widetilde{e}^{(\Sc, n_{\Sc})} \LB p_n^{(\ell)} \RB
    \mathbb{I}\LB p_n^{(\ell)} \RB
    \LB \Pr \LB \boldsymbol{\chi}_n^{(\ell)} \RB \RB^{-1}
    \LB 2\pi \RB^{n_{\Sc}}
    \frac{\LB r^{(\Sc, n_\Sc)}_n \RB^2}{\abs{\LB \widehat{\nv}^{(\Sc, n_\Sc + 1)}_n \RB \tp \widehat{\kv}^{(\Sc, n_\Sc + 1)}_{i,n}}}
    \prod_{k=1}^{n_{\Sc}} \frac{\LB r^{(\Sc, k-1)}_n \RB^2}{\abs{\LB \widehat{\nv}^{(\Sc, k)}_n \RB \tp \widehat{\kv}_{i,n}^{(\Sc, k)}}}
\end{multline}
where $p_n^{(\ell)}$ denotes the $n$-th path at depth $\ell$, $\mathbb{I}\LB p_n^{(\ell)} \RB$ is an indicator function equal to $1$ if $p_n^{(\ell)}$ intersects the measurement surface at depth $\ell$ and $0$ otherwise, and $\Pr \LB \boldsymbol{\chi}_n^{(\ell)} \RB$ is the probability of sampling the specific sequence of interaction types $\boldsymbol{\chi}_n^{(\ell)}$ corresponding to $p_n^{(\ell)}$. The number of diffuse reflections $n_{\Sc}$ depends on the path, but its depedency on $p_n^{(\ell)}$ is not explicitly stated for simplicity.

\subsection{Computation of Radio Map due to Diffraction}
\label{sec:radio-map-solver-diffraction}

\begin{figure}[ht!]
    \centering
    \includegraphics[width=0.85\textwidth]{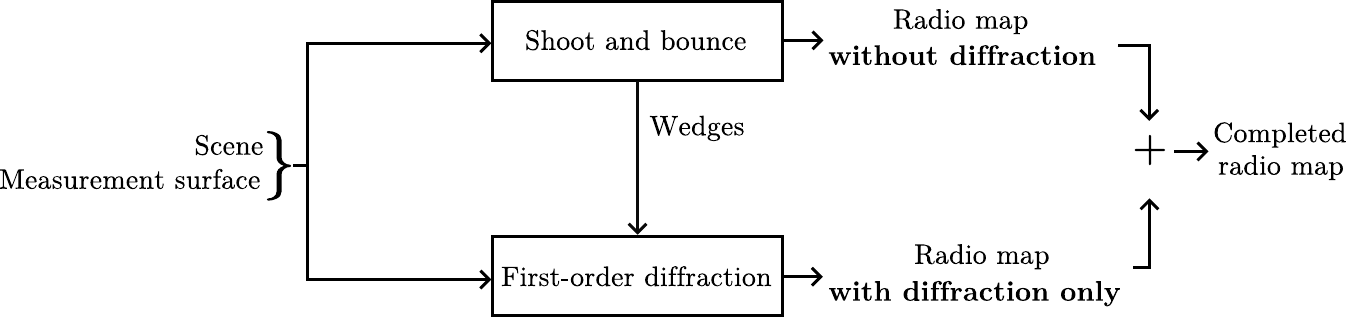}
    \caption{Computation of the radio map in two steps: First, the radio map is computed for all interaction types except diffraction using a single \gls{SBR} loop. Second, only diffracted paths are computed. The two radio maps are then summed to obtain the final radio map.\label{fig:radio-map-solver-steps}}
\end{figure}

\gls{SBR} is not applicable to paths that include diffraction: as explained in Section~\ref{sec:path-solver-sbr}, the probability of a ray intersecting an edge is zero.
To address this, \SRT{} computes the radio map in two steps, as illustrated in Figure~\ref{fig:radio-map-solver-steps}.
First, the radio map is computed for all interaction types except diffraction as described previously. Wedges near the source are detected and stored during this step.
Second, only diffracted paths are computed.
Currently, \SRT{} supports only first-order diffraction for radio map computation, i.e., paths connecting the source to the measurement surface through a single diffracting wedge and no other interactions with the scene.
The two radio maps are then summed to obtain the final result.
The following describes the computation of radio map due to diffraction.

For clarity, we consider diffraction on a single edge $\Ec$, defined by an origin $\ov$, an edge vector $\widehat{\ev}$, and length $L_{\Ec}$. \SRT{} processes all such wedges found near the source in parallel.
Let $\vv \in \Ec$ be a diffraction point, which is parametrized by the distance along the edge $x$ and the Keller cone azimuth $\phi$, i.e., $\vv = \ov + x \widehat{\ev}$.
Then for a measurement point $\tv \in M_i$, define:
\begin{equation}
    w(\tv, \vv) = \underbrace{\frac{\LB \vv - \sv \RB \tp \widehat{\ev}}{\norm{\vv - \sv}}}_{\cos \beta_0'} - \underbrace{\frac{\LB \tv - \vv \RB \tp \widehat{\ev}}{\norm{\tv - \vv}}}_{\cos \beta_0}
\end{equation}
where $\beta_0'$ and $\beta_0$ are the angles between the edge and the incident and diffracted wave, respectively (see Figure~\ref{fig:path-solver-diffraction-model}).
The Keller cone condition is $w(\tv, \vv) = 0$, which can be shown to be satisfied only for a single diffraction point on the edge, denoted by $\vv_{\Dc}(\tv)$.
For a single diffraction wedge, as the diffraction point is unique, the radio map definition is similar to~\eqref{eq:radio-map-solver-ci-ell}:
\begin{equation}
    \label{eq:radio-map-solver-ci-D}
    C_i^{(\Dc)} = \frac{1}{\abs{M_i}} \int_{M_i} e^{(\Dc)} \LB p \RB \Vc\LB p \RB dA(\tv)
\end{equation}
where $p = \LB \sv, \vv_{\Dc}(\tv), \tv \RB$ is the path from the source to the measurement point $\tv$ via the diffraction point $\vv_{\Dc}(\tv)$.

\begin{figure}[ht!]
    \centering
    \begin{subfigure}[b]{0.40\textwidth}
        \centering
        \includegraphics[width=\textwidth]{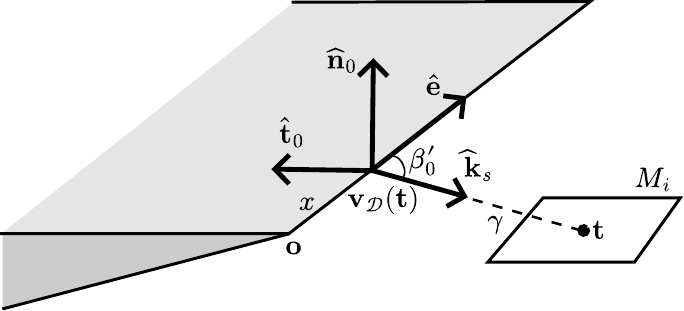}
        \caption{3D view}
        \label{fig:radio-map-solver-diffraction-frame-3d}
    \end{subfigure}
    \hfill
    \begin{subfigure}[b]{0.40\textwidth}
        \centering
        \includegraphics[width=\textwidth]{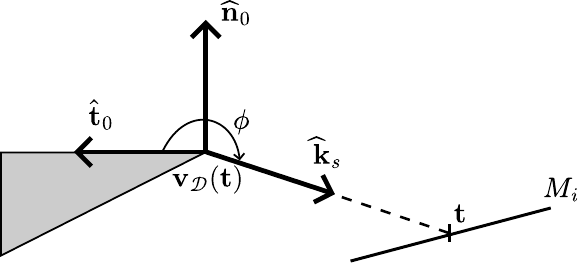}
        \caption{2D cross-section}
        \label{fig:radio-map-solver-diffraction-frame-2d}
    \end{subfigure}
    \caption{
    A measurement point $\tv$ reachable by a diffracted ray from a wedge is determined by a distance along the edge $x$ and the Keller cone azimuth $\phi$, while the cone opening angle is $\beta_0'$.
    \label{fig:radio-map-solver-diffraction-frame}}
\end{figure}

Direct integration over the measurement surface is inefficient, as it would require connecting surface points to the source (for example, by placing targets within each cell and running the path solver).
To overcome this, consider an orthogonal basis $\LB \widehat{\tv}_0, \widehat{\nv}_0, \widehat{\ev} \RB$, where $\widehat{\nv}_0$ is the normal to a wedge face and $\widehat{\tv}_0 \times \widehat{\nv}_0 = \widehat{\ev}$, as illustrated in Figure~\ref{fig:radio-map-solver-diffraction-frame}. For a given angle of incidence $\beta_0'$, any point on $M_i$ that can be reached by a diffracted ray can be written as:
\begin{equation}
    \label{eq:radio-map-solver-ci-D-reparametrization}
    \tv = \underbrace{\ov + x \widehat{\ev}}_{\vv_{\Dc}(\tv)} + \gamma \underbrace{\LB \sin\beta_0' \cos\phi \widehat{\tv}_0 + \sin\beta_0' \sin\phi \widehat{\nv}_0 + \cos\beta_0' \widehat{\ev} \RB}_{\widehat{\kv}_s}.
\end{equation}
Here, $\widehat{\kv}_s$ is the direction of the diffracted ray (with $\phi$ the Keller cone azimuth), and $\gamma > 0$ is the distance for which the ray from $\vv_{\Dc}(\tv)$ in direction $\widehat{\kv}_s$ intersects the measurement cell $M_i$ (see Figure~\ref{fig:radio-map-solver-diffraction-frame}). Since a ray and a plane intersect at most once, $\tv$ is uniquely determined by $x$ and $\phi$, and thus $\tv(x, \phi)$ forms a reparametrization of the subset of $M_i$ that can be reached by a diffracted ray when the angle of incidence is $\beta_0'$.

With this reparametrization,~\eqref{eq:radio-map-solver-ci-D} becomes:
\begin{equation}
    \label{eq:radio-map-solver-ci-D-3}
    C_i^{(\Dc)} = \frac{1}{\abs{M_i}} \int_{0}^{L_{\Ec}} \int_0^{n\pi} e \LB p \RB \Vc\LB p \RB \norm{ \frac{\partial \tv}{\partial x} \times \frac{\partial \tv}{\partial \phi} } dx d\phi.
\end{equation}

To estimate~\eqref{eq:radio-map-solver-ci-D-3}, $N_S'$ samples are drawn on the wedge: $x$ uniformly in $[0, L_{\Ec}]$ and $\phi$ uniformly in $[0, n\pi]$. This yields the Monte Carlo estimator:
\begin{equation}
    \label{eq:radio-map-solver-ci-D-est}
    \widehat{C}_i^{(\Dc)} = \frac{1}{\abs{M_i}} \frac{L_{\Ec}n\pi}{N_S'}\sum_{n=1}^{N_S'} e \LB p_n \RB \mathbb{I}\LB p_n \RB \norm{ \frac{\partial \tv}{\partial x} \times \frac{\partial \tv}{\partial \phi} }
\end{equation}
where $p_n$ is the $n$-th path, $\mathbb{I}\LB p_n \RB$ is 1 if $p_n$ intersects the measurement surface and is not occluded by other scatterers and 0 otherwise, and the weighting factor $\norm{ \frac{\partial \tv}{\partial x} \times \frac{\partial \tv}{\partial \phi} }$ is computed as described in Appendix~\ref{sec:diffraction-radio-map-weighting-factor}.

\subsection{Non-Coherent vs. Coherent Radio Maps}

\begin{figure}[ht!]
    \centering
    \includegraphics[width=0.75\textwidth]{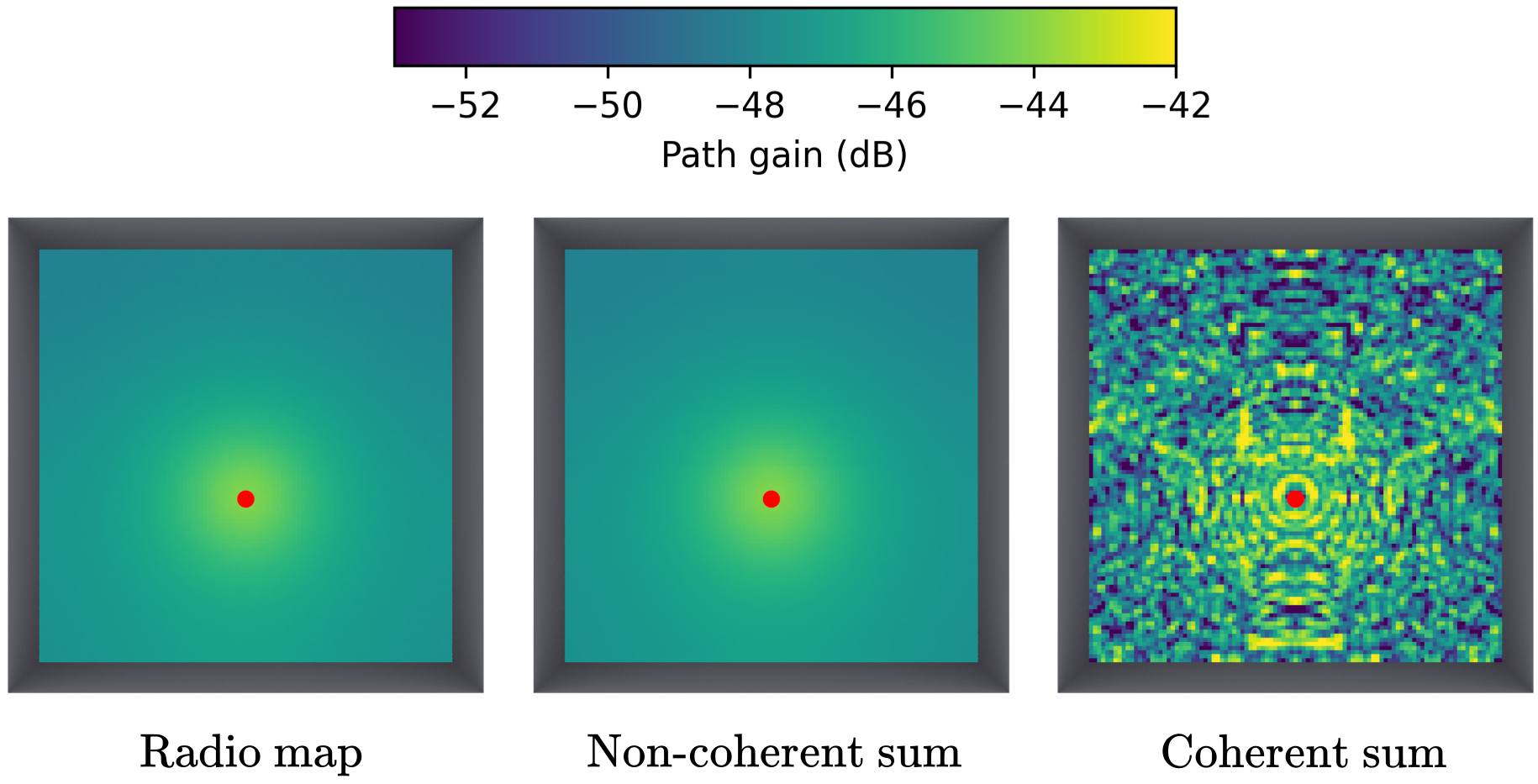}
    \caption{Comparison of radio maps: one computed using the radio map solver, one by non-coherently summing path coefficients from the path solver, and one by coherently summing path coefficients from the path solver. The transmitter is depicted as a red ball. All radio maps are computed with a maximum depth of $L = 3$, cell sizes of $10 \times 10$ cm, using isotropic antenna patterns on both the transmitter and receiver, and for the ``box'' scene built in \SRT{}. The ``non-coherent sum'' and ``coherent sum'' radio maps are generated by placing a single target point at the center of each measurement cell.\label{fig:radio-map-solver-fast-fading}}
\end{figure}

The proposed definition of radio maps involves a non-coherent summation of path gains~\eqref{eq:radio-map-solver-path-gain}. As a result, it does not capture fast fading effects, which result from the constructive and destructive interference of multiple paths.
Figure~\ref{fig:radio-map-solver-fast-fading} demonstrates this by comparing three scenarios: a radio map generated using the radio map solver, a radio map created by non-coherently summing the path coefficients~\eqref{eq:path-coefficient} calculated with the path solver (Section~\ref{sec:path-solver}) by placing a single target at the center of each measurement cell, and a radio map produced by coherently summing the path coefficients~\eqref{eq:path-coefficient} calculated with the path solver, also by placing a single target at the center of each measurement cell. The radio map obtained using the radio map solver aligns with the non-coherent summation of path coefficients calculated using the path solver, as anticipated. However, it does not account for fast fading effects, which require the coherent summation of path coefficients.
Also, because the diffracted contributions are accumulated non-coherently with the specular and \gls{LoS} contributions, field-level continuity at incident and reflection shadow boundaries is not preserved.

\subsection{Handling Multiple Transmit Antennas}
\label{sec:radio-map-solver-antenna-arrays}

The \SRT{} radio map solver supports antenna arrays of any size at the transmitter and allows for user-defined precoding vectors. However, it only supports synthetic arrays. In this setup, paths are generated from a single source located at the center of the array, and the channel response for each individual antenna is calculated by applying the appropriate phase shifts synthetically. This approximation requires plan wave approximation, which is valid only when $D^2 \ll R \lambda_0$, where $D$ is the array size and $R$ is the distance from the array center to the first interaction point or measurement plane.

Using synthetic arrays enables efficient scaling with respect to the number of antennas. To see this, let $N_A^{\text{tx}}$ denote the number of antennas at the transmitter, and let $\uv_P \in \CC^{N_A^{\text{tx}}}$ be the precoding vector. For a given path connecting the source to the measurement plane, let $\Em \in \CC^2$ represent the electric field observed at the measurement plane. By employing a synthetic array, the corresponding electric fields for each antenna element are obtained by applying the array response vector~\eqref{eq:path-solver-array-response-vector}, denoted as $\uv_A^{\text{tx}}$, leading to
\begin{equation}
    \Em_A = \Em \LB \uv_A^{\text{tx}} \RB\tp.
\end{equation}
Here, $\Em_A$ is a matrix of size $2 \times N_A^{\text{tx}}$, where each column represents the electric fields radiated by each transmit antenna. The contribution of this path to the radio map is given by
\begin{equation}
    \norm{\Em_A \uv_P}^2.
\end{equation}
This expression allows the path's contribution to the radio map to be represented as
\begin{equation}
    \norm{\Em_A \uv_P}^2 = \norm{\LB \Em \LB \uv_A^{\text{tx}} \RB \tp \uv_P \RB}^2 = \norm{\Em \alpha}^2
\end{equation}
where $\alpha = \LB \uv_A^{\text{tx}} \RB\tp \uv_P$ is a scalar, which depends on the departure direction of the path from the transmitter and thus varies for paths originating from the source. By precomputing $\alpha$ for each path, the effect of the transmitter array can be efficiently simulated by multiplying the electric field of each path by $\alpha$ whenever the path intersects the measurement surface. This operation does not depend on the size of the array.

\begin{figure}[htbp]
    \centering
    \begin{subfigure}[b]{0.3\textwidth}
        \centering
        \includegraphics[width=\textwidth]{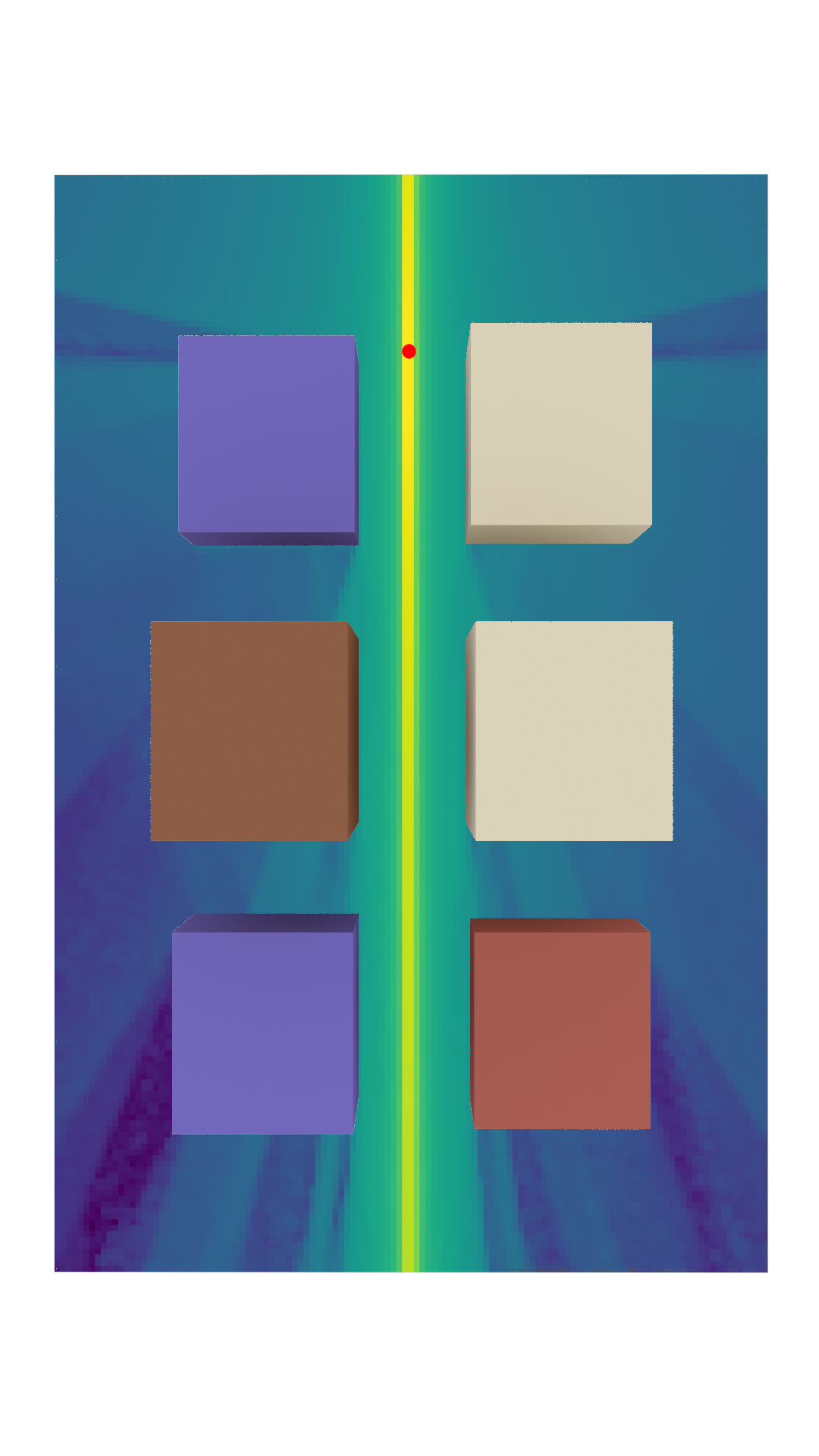}
        \caption{Scene with the radio map.\label{fig:radio-map-solver-sbr-scene}}
    \end{subfigure}
    \hfill
    \begin{subfigure}[b]{0.65\textwidth}
        \centering
        \includegraphics[width=\textwidth]{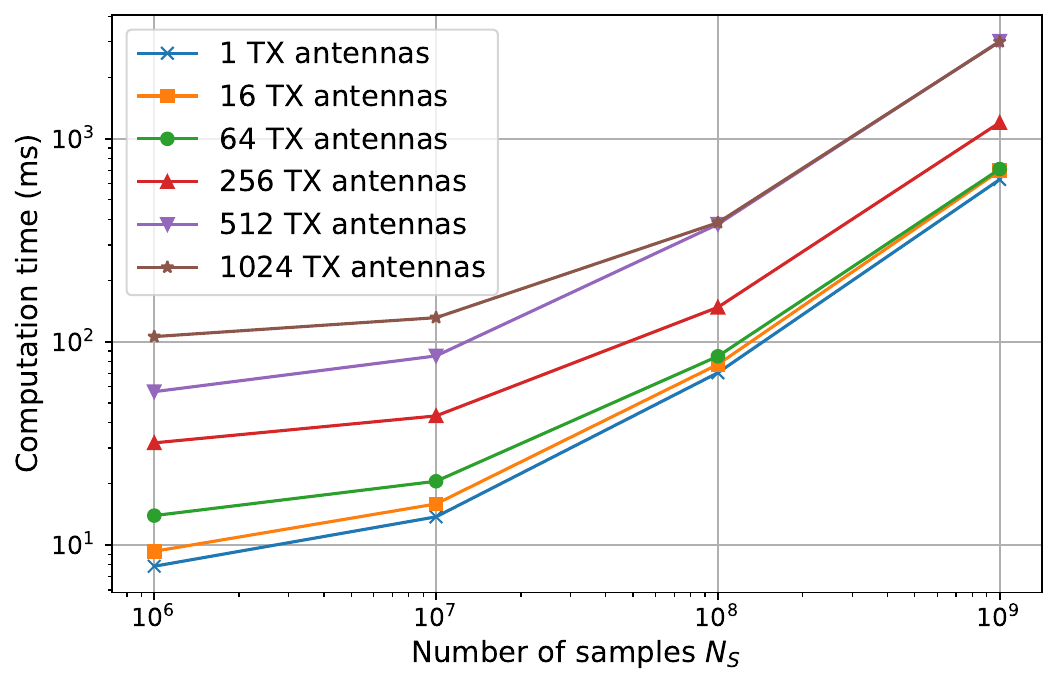}
        \caption{Computation time vs. number of samples $N_S$.\label{fig:radio-map-solver-sbr-scaling}}
    \end{subfigure}
    \caption{Computation time for calculating radio maps using an NVIDIA RTX 4090 GPU for a single transmitter with a linear array of 1 to 1024 antennas.}
\end{figure}

Figure~\ref{fig:radio-map-solver-sbr-scaling} illustrates the computation time needed to generate a radio map for a single source as the number of samples $N_S$ and the number of transmit antennas $N_A^{\text{tx}}$ are varied. The setup, depicted in Figure~\ref{fig:radio-map-solver-sbr-scene}, employs the ``simple street canyon'' scene built into \SRT{}. The red ball indicates the source, positioned 20 meters above the ground. The radio map in Figure~\ref{fig:radio-map-solver-sbr-scene} is calculated using $10^8$ samples and a linear array of 512 transmit antennas. Notably, the radio map solver can compute radio maps with $10^9$ samples on a single NVIDIA RTX 4090 GPU.
The computation time increases with both the number of samples and antennas, exceeding one second only for the most demanding computations. It is important to note that while the computation time increases with the number of antennas due to the precomputation of the array weighting factor $\alpha$, it does not increase the memory usage of the radio map solver.

\paragraph{Handling of Dual-Polarized Transmit Antennas}

The radio map solver supports dual-polarized transmit antennas by considering two independent electric field vectors for each path. These are processed as if they were separate antenna elements (see Section~\ref{sec:radio-map-solver-antenna-arrays}).

\subsection{Practical Notes}

\paragraph{Mesh-based Measurement Surface}
\SRT{} supports mesh-based measurement surfaces, which are defined as collections of triangles (see Section~\ref{sec:essential-concepts-scene-objects-and-meshes}). In this case, each triangle serves as a measurement cell. Meshes can possess arbitrary topology, making this feature well suited for computing radio maps over complex terrains.

\paragraph{Early Termination with Russian Roulette and Gain Thresholding}
The radio map solver allows for early deactivation of paths using two techniques: Russian roulette and gain thresholding.
Both features can be enabled simultaneously and are applied only after a user-defined depth is reached.
With gain thresholding, a path with ray tube length $r$ is deactivated when the propagation gain falls below a user-defined threshold.
With Russian roulette, at each iteration, a path continues with a probability given by
\begin{equation}
    \label{eq:radio-map-solver-rr-prob}
    p_{\text{rr}} = \min\LB r^2 \norm{\Em}^2, p_{\text{rr,max}} \RB
\end{equation}
where $0 < p_{\text{rr,max}} \leq 1$ is the maximum probability, set by the user, for keeping the path active, and $r^2\norm{\Em}^2$ is the distance-normalized propagation gain.
The surviving paths are reweighted by the inverse square root of the probability of being kept active, i.e., $1/\sqrt{p_{\text{rr}}}$, to ensure the estimator is unbiased.
Unlike compensated Russian roulette, gain thresholding removes the weakest contributions and therefore introduces bias.

\paragraph{Extension to other Types of Radio Maps}
The \SRT{} solver currently supports the calculation of channel gain maps, which are radio maps that provide an estimate of the channel gain for each cell. It is possible to extend the solver to calculate other types of radio maps, such as root-mean-square delay spread maps and angular spread maps~\cite{instant_rm}.

\clearpage

\begin{highlightbox}{Radio Map Solver: Supported Path Families, Array Model, and Accumulation}

    Let $L$ denote the user-configured maximum path depth. The radio map
    solver supports the following path families:

    \renewcommand{\arraystretch}{1.25}
    \begin{center}
    \begin{tabular}{@{}p{0.28\linewidth}p{0.66\linewidth}@{}}
    \hline
    \textbf{Path family} & \textbf{Supported interaction sequences} \\
    \hline

    \Gls{LoS}
    &
    A path of depth zero, containing no interaction.
    \\

    Non-diffracted paths
    &
    Any sequence of specular reflections, refractions, and diffuse
    reflections, with total depth at most $L$.
    \\

    Diffracted paths
    &
    Direct first-order diffraction paths of the form
    \[
        \sv \longrightarrow \text{diffraction} \longrightarrow \tv,
    \]
    where $\sv$ is the source (transmitter), $\tv$ is a point on the measurement surface.
    \\

    \hline
    \end{tabular}
    \end{center}
    \renewcommand{\arraystretch}{1.0}
    Higher-order diffraction and paths combining diffraction with other
    interaction types are currently not supported.

    \medskip
    \noindent
    \textbf{Array model.}
    The radio map solver supports only synthetic transmit arrays. Paths are
    traced from the center of the transmit array, and the contributions from
    the individual antenna elements are obtained by applying synthetic phase
    shifts and the user-provided precoding vector.
    The approximation requires the wavefront to be approximately planar across the array. A commonly used sufficient far-field criterion is $D^2 \ll R \lambda_0$, where $D$ is the array total aperture and $R$ is the distance from the array center to the first interaction point or measurement plane.

    \medskip
    \noindent
    \textbf{Non-coherent accumulation.}
    The fields radiated by the transmit-array elements are combined coherently
    through the precoding vector for each individual propagation path.
    Different propagation paths, however, are accumulated non-coherently:
    Radio maps represent sums of individual path gains and do
    not include cross terms between paths. They therefore do not capture
    fast fading caused by constructive and destructive multipath interference.

    \end{highlightbox}

\newpage
\appendix

\section{Primer on Electromagnetism}
\label{sec:primer-em}
This section provides useful background for the general understanding of ray tracing for wireless propagation modeling. In particular, our goal is to provide a concise definition of a ``channel impulse response'' between a transmitting and receiving antenna, as in~\cite[Ch. 2 \& 3]{geng2013planungsmethoden}.

\subsection{Coordinate System, Rotations, and Vector Fields}

We consider a \gls{GCS} with Cartesian canonical basis $\hat{\mathbf{x}}$, $\hat{\mathbf{y}}$, $\hat{\mathbf{z}}$.
The spherical unit vectors are defined as
\begin{equation}
    \begin{aligned}
        \label{eq:spherical_vecs}
        \hat{\mathbf{r}}          (\theta, \varphi) &= \sin(\theta)\cos(\varphi) \hat{\mathbf{x}} + \sin(\theta)\sin(\varphi) \hat{\mathbf{y}} + \cos(\theta)\hat{\mathbf{z}}\\
        \hat{\boldsymbol{\theta}} (\theta, \varphi) &= \cos(\theta)\cos(\varphi) \hat{\mathbf{x}} + \cos(\theta)\sin(\varphi) \hat{\mathbf{y}} - \sin(\theta)\hat{\mathbf{z}}\\
        \hat{\boldsymbol{\varphi}}(\theta, \varphi) &=            -\sin(\varphi) \hat{\mathbf{x}} +             \cos(\varphi) \hat{\mathbf{y}}.
    \end{aligned}
\end{equation}
For an arbitrary unit norm vector $\hat{\mathbf{v}} = (x, y, z)$, the zenith and azimuth angles $\theta$ and $\varphi$ can be computed as
\begin{equation}
    \begin{aligned}
        \theta  &= \cos^{-1}(z) \\
        \varphi &= \mathop{\text{atan2}}(y, x)
    \end{aligned}
\end{equation}
where $\mathop{\text{atan2}}(y, x)$ is the two-argument inverse tangent function~\cite{atan2}. As any vector uniquely determines $\theta$ and $\varphi$, we sometimes also write $\hat{\boldsymbol{\theta}}(\hat{\mathbf{v}})$ and $\hat{\boldsymbol{\varphi}}(\hat{\mathbf{v}})$ instead of $\hat{\boldsymbol{\theta}} (\theta, \varphi)$ and $\hat{\boldsymbol{\varphi}}(\theta, \varphi)$.

A 3D rotation with yaw, pitch, and roll angles $\alpha$, $\beta$, and $\gamma$, respectively, is expressed by the matrix
\begin{equation}
    \label{eq:rotation}
    \mathbf{R}(\alpha, \beta, \gamma) = \mathbf{R}_z(\alpha)\mathbf{R}_y(\beta)\mathbf{R}_x(\gamma)
\end{equation}
where $\mathbf{R}_z(\alpha)$, $\mathbf{R}_y(\beta)$, and $\mathbf{R}_x(\gamma)$ are rotation matrices around the $z$, $y$, and $x$ axes, respectively, which are defined as
\begin{equation}
    \begin{aligned}
        \mathbf{R}_z(\alpha) &= \begin{pmatrix}
                        \cos(\alpha) & -\sin(\alpha) & 0\\
                        \sin(\alpha) & \cos(\alpha) & 0\\
                        0 & 0 & 1
                      \end{pmatrix}\\
        \mathbf{R}_y(\beta) &= \begin{pmatrix}
                        \cos(\beta) & 0 & \sin(\beta)\\
                        0 & 1 & 0\\
                        -\sin(\beta) & 0 & \cos(\beta)
                      \end{pmatrix}\\
        \mathbf{R}_x(\gamma) &= \begin{pmatrix}
                            1 & 0 & 0\\
                            0 & \cos(\gamma) & -\sin(\gamma)\\
                            0 & \sin(\gamma) & \cos(\gamma)
                      \end{pmatrix}.
    \end{aligned}
\end{equation}
A closed-form expression for $\mathbf{R}(\alpha, \beta, \gamma)$ can be found in~\cite[Sec.~7.1-4]{TR38901}.
The inverse rotation is simply defined by $\mathbf{R}^{-1}(\alpha, \beta, \gamma)=\mathbf{R}^\mathsf{T}(\alpha, \beta, \gamma)$.
A vector $\mathbf{x}$ defined in a first coordinate system is represented in a second coordinate system rotated by $\mathbf{R}(\alpha, \beta, \gamma)$ with respect to the first one as $\mathbf{x}'=\mathbf{R}^\mathsf{T}(\alpha, \beta, \gamma)\mathbf{x}$.
If a point in the first coordinate system has spherical angles $(\theta, \varphi)$, the corresponding angles $(\theta', \varphi')$ in the second coordinate system can be found to be
\begin{equation}
    \label{eq:theta_phi_prime}
    \begin{aligned}
        \theta' &= \cos^{-1}\left( \hat{\mathbf{z}}^\mathsf{T} \mathbf{R}^\mathsf{T}(\alpha, \beta, \gamma)\hat{\mathbf{r}}(\theta, \varphi)          \right)\\
        \varphi' &= \arg\left( \left( \hat{\mathbf{x}} + j\hat{\mathbf{y}}\right)^\mathsf{T} \mathbf{R}^\mathsf{T}(\alpha, \beta, \gamma)\hat{\mathbf{r}}(\theta, \varphi) \right).
    \end{aligned}
\end{equation}

For a vector field $\mathbf{F}'(\theta',\varphi')$ expressed in local spherical coordinates
\begin{equation}
    \mathbf{F}'(\theta',\varphi') = F_{\theta'}(\theta',\varphi')\hat{\boldsymbol{\theta}}'(\theta',\varphi') + F_{\varphi'}(\theta',\varphi')\hat{\boldsymbol{\varphi}}'(\theta',\varphi')
\end{equation}
that are rotated by $\mathbf{R}=\mathbf{R}(\alpha, \beta, \gamma)$ with respect to the \gls{GCS}, the spherical field components in the \gls{GCS} can be expressed as
\begin{equation}
    \label{eq:F_prime_2_F}
    \begin{bmatrix}
        F_\theta(\theta, \varphi) \\
        F_\varphi(\theta, \varphi)
    \end{bmatrix} =
    \begin{bmatrix}
        \hat{\boldsymbol{\theta}}(\theta,\varphi)^\mathsf{T}\mathbf{R}\hat{\boldsymbol{\theta}}'(\theta',\varphi') & \hat{\boldsymbol{\theta}}(\theta,\varphi)^\mathsf{T}\mathbf{R}\hat{\boldsymbol{\varphi}}'(\theta',\varphi') \\
        \hat{\boldsymbol{\varphi}}(\theta,\varphi)^\mathsf{T}\mathbf{R}\hat{\boldsymbol{\theta}}'(\theta',\varphi') & \hat{\boldsymbol{\varphi}}(\theta,\varphi)^\mathsf{T}\mathbf{R}\hat{\boldsymbol{\varphi}}'(\theta',\varphi')
    \end{bmatrix}
    \begin{bmatrix}
        F_{\theta'}(\theta', \varphi') \\
        F_{\varphi'}(\theta', \varphi')
    \end{bmatrix}
\end{equation}
so that
\begin{equation}
    \mathbf{F}(\theta,\varphi) = F_{\theta}(\theta,\varphi)\hat{\boldsymbol{\theta}}(\theta,\varphi) + F_{\varphi}(\theta,\varphi)\hat{\boldsymbol{\varphi}}(\theta,\varphi).
\end{equation}

Sometimes, it is also useful to find the rotation matrix that maps a unit vector $\hat{\mathbf{a}}$ to $\hat{\mathbf{b}}$.
This can be achieved with the help of Rodrigues' rotation formula~\cite{Wikipedia_Rodrigues} which defines the matrix
\begin{equation}
    \label{eq:rodrigues_matrix}
    \mathbf{R}(\hat{\mathbf{a}}, \hat{\mathbf{b}}) = \mathbf{I} + \sin(\theta)\mathbf{K} + (1-\cos(\theta)) \mathbf{K}^2
\end{equation}
where
\begin{align}
    \mathbf{K} &= \begin{bmatrix}
                            0 & -\hat{k}_z &  \hat{k}_y \\
                    \hat{k}_z &          0 & -\hat{k}_x \\
                   -\hat{k}_y &  \hat{k}_x &          0
                 \end{bmatrix}\\
    \hat{\mathbf{k}} &= \frac{\hat{\mathbf{a}} \times \hat{\mathbf{b}}}{\norm{\hat{\mathbf{a}} \times \hat{\mathbf{b}}}}\\
    \theta &= \arccos\left(\hat{\mathbf{a}}^\mathsf{T}\hat{\mathbf{b}}\right)
\end{align}
such that $\mathbf{R}(\hat{\mathbf{a}}$, $\hat{\mathbf{b}})\hat{\mathbf{a}}=\hat{\mathbf{b}}$.
In the degenerate case $\hat{\mathbf{a}}=\hat{\mathbf{b}}$, the rotation matrix is the identity.
If $\hat{\mathbf{a}}=-\hat{\mathbf{b}}$, the rotation is by $\pi$ about any unit vector orthogonal to $\hat{\mathbf{a}}$.

\subsection{Planar Time-Harmonic Waves}

A time-harmonic planar-wave electric field $\mathbf{E}(\mathbf{x}, t)\in\mathbb{C}^3$ traveling in a homogeneous medium with wave vector $\mathbf{k}\in\mathbb{C}^3$ can be described at position $\mathbf{x}\in\mathbb{R}^3$ and time $t$ as
\begin{align}
    \mathbf{E}(\mathbf{x}, t) &= \mathbf{E}_0 e^{j(\omega t-\mathbf{k}\tp\mathbf{x})}\\
                                &= \mathbf{E}(\mathbf{x}) e^{j\omega t}
\end{align}
where $\mathbf{E}_0\in\mathbb{C}^3$ is the field phasor. The wave vector can be decomposed as $\mathbf{k}=k \hat{\mathbf{k}}$, where $\hat{\mathbf{k}}$ is a unit norm vector, $k=\omega\sqrt{\varepsilon\mu}$ is the wave number, and $\omega=2\pi f$ is the angular frequency. The permittivity $\varepsilon$ and permeability $\mu$ are defined as
\begin{align}
    \varepsilon &= \eta \varepsilon_0 \label{eq:epsilon}\\
    \mu &= \mu_r \mu_0 \label{eq:mu}
\end{align}
where $\eta$ and $\varepsilon_0$ are the complex relative and vacuum permittivities, and $\mu_r$ and $\mu_0$ are the relative and vacuum permeabilities.
The complex relative permittivity $\eta$ is given as
\begin{equation}
    \label{eq:eta}
    \eta = \varepsilon_r - j\frac{\sigma}{\varepsilon_0\omega}
\end{equation}
where $\varepsilon_r$ is the real relative permittivity of a non-conducting dielectric and $\sigma$ is the conductivity.

With these definitions, the phase velocity in the medium is given as~\cite[Eq. 4-28d]{Balanis}
\begin{equation}
    v_\mathrm{p}=\frac{1}{\sqrt{\varepsilon_0\varepsilon_r\mu}}\left\{\frac12\left(\sqrt{1+\left(\frac{\sigma}{\omega\varepsilon_0\varepsilon_r}\right)^2}+1\right)\right\}^{-\frac{1}{2}}
\end{equation}
where the factor in curly brackets becomes one for non-conducting materials. The speed of light in vacuum is denoted $c_0=\frac{1}{\sqrt{\varepsilon_0 \mu_0}}$, the free-space wavelength is $\lambda_0=\frac{c_0}{f_c}$, and the vacuum wave number is $k_0=\frac{\omega}{c_0}=\frac{2\pi}{\lambda_0}$. In conducting materials, the wave number is complex which translates to propagation losses.

The associated magnetic field $\mathbf{H}(\mathbf{x}, t)\in\mathbb{C}^3$ is
\begin{equation}
    \mathbf{H}(\mathbf{x}, t) = \frac{\hat{\mathbf{k}}\times  \mathbf{E}(\mathbf{x}, t)}{Z} = \mathbf{H}(\mathbf{x})e^{j\omega t}
\end{equation}
where $Z=\sqrt{\mu/\varepsilon}$ is the wave impedance. The vacuum impedance is denoted by $Z_0=\sqrt{\mu_0/\varepsilon_0}\approx 376.73\,\Omega$.

The time-averaged Poynting vector is defined as
\begin{equation}
    \mathbf{S}(\mathbf{x}) = \frac{1}{2} \Re\left\{\mathbf{E}(\mathbf{x})\times  \mathbf{H}^*(\mathbf{x})\right\}
                           = \frac{1}{2} \Re\left\{\frac{1}{Z} \right\} \norm{\mathbf{E}(\mathbf{x})}^2 \hat{\mathbf{k}}
\end{equation}
which describes the directional energy flux [\si{\watt\per\meter\squared}], i.e. energy transfer per unit area per unit time.
Note that the actual electromagnetic waves are the real parts of $\mathbf{E}(\mathbf{x}, t)$ and $\mathbf{H}(\mathbf{x}, t)$.

\subsection{Far Field of a Transmitting Antenna}
\label{sec:primer-em-tx-far-field}

We assume that the electric far field of an antenna in free space can be described by a spherical wave originating from the center of the antenna:
\begin{equation}
    \mathbf{E}(r, \theta, \varphi, t) = \mathbf{E}(r,\theta, \varphi) e^{j\omega t} = \mathbf{E}_0(\theta, \varphi) \frac{e^{-jk_0r}}{r} e^{j\omega t}
\end{equation}
where $\mathbf{E}_0(\theta, \varphi)$ is the electric field phasor, $r$ is the distance (or radius), $\theta$ the zenith angle, and $\varphi$ the azimuth angle.
In contrast to a planar wave, the field strength decays as $1/r$.

The complex antenna field pattern $\mathbf{F}(\theta, \varphi)$ is defined as
\begin{align}
        \label{eq:F}
    \mathbf{F}(\theta, \varphi) = \frac{ \mathbf{E}_0(\theta, \varphi)}{\max_{\theta,\varphi}\norm{\mathbf{E}_0(\theta, \varphi)}}.
\end{align}
The time-averaged Poynting vector for such a spherical wave is

\begin{equation}
    \label{eq:S_spherical}
    \mathbf{S}(r, \theta, \varphi) = \frac{1}{2Z_0}\norm{\mathbf{E}(r, \theta, \varphi)}^2 \hat{\mathbf{r}}
\end{equation}
where $\hat{\mathbf{r}}$ is the radial unit vector. It simplifies for an ideal isotropic antenna with input power $P_\text{T}$ to
\begin{equation}
    \mathbf{S}_\text{iso}(r, \theta, \varphi) = \frac{P_\text{T}}{4\pi r^2} \hat{\mathbf{r}}.
\end{equation}

The antenna gain $G$ is the ratio of the maximum radiation power density of the antenna in radial direction and that of an ideal isotropic radiating antenna:
\begin{equation}
    \label{eq:G}
    G = \frac{\max_{\theta,\varphi}\norm{\mathbf{S}(r, \theta, \varphi)}}{ \norm{\mathbf{S}_\text{iso}(r, \theta, \varphi)}}
          = \frac{2\pi}{Z_0 P_\text{T}} \max_{\theta,\varphi}\norm{\mathbf{E}_0(\theta, \varphi) }^2.
\end{equation}
One can similarly define a gain with directional dependency by ignoring the computation of the maximum in the last equation:
\begin{equation}
    \label{eq:Gdir}
    G(\theta, \varphi) = \frac{2\pi}{Z_0 P_\text{T}} \norm{\mathbf{E}_0(\theta, \varphi)}^2 = G \norm{\mathbf{F}(\theta, \varphi)}^2.
\end{equation}

If one uses in the last equation the radiated power $P=\eta_\text{rad} P_\text{T}$, where $\eta_\text{rad}$ is the radiation efficiency, instead of the input power $P_\text{T}$, one obtains the directivity $D(\theta,\varphi)$.
Both are related through $G(\theta, \varphi)=\eta_\text{rad} D(\theta, \varphi)$.

\begin{highlightbox}{Antenna Pattern}
    Since $\mathbf{F}(\theta, \varphi)$ contains no information about the maximum gain $G$ and $G(\theta, \varphi)$ does not carry any phase information, we define the ``antenna pattern'' $\mathbf{C}(\theta, \varphi)$ as
    \begin{equation}
        \label{eq:C}
        \mathbf{C}(\theta, \varphi) = \sqrt{G}\mathbf{F}(\theta, \varphi)
    \end{equation}
    such that $G(\theta, \varphi)= \norm{\mathbf{C}(\theta, \varphi)}^2$.
    Using the spherical unit vectors $\hat{\boldsymbol{\theta}}\in\mathbb{R}^3$ and $\hat{\boldsymbol{\varphi}}\in\mathbb{R}^3$, we can rewrite $\mathbf{C}(\theta, \varphi)$ as
    \begin{equation}
        \mathbf{C}(\theta, \varphi) = C_\theta(\theta,\varphi) \hat{\boldsymbol{\theta}} + C_\varphi(\theta,\varphi) \hat{\boldsymbol{\varphi}}
    \end{equation}
    where $C_\theta(\theta,\varphi)\in\mathbb{C}$ and $C_\varphi(\theta,\varphi)\in\mathbb{C}$ are the ``zenith pattern'' and ``azimuth pattern'', respectively.
\end{highlightbox}

Combining $F$ and $G$, we can obtain the following expression of the electric far field
\begin{equation}
    \label{eq:E_T}
    \mathbf{E}_\text{T}(r,\theta_\text{T},\varphi_\text{T}) = \sqrt{ \frac{P_\text{T} G_\text{T} Z_0}{2\pi}} \frac{e^{-jk_0 r}}{r} \mathbf{F}_\text{T}(\theta_\text{T}, \varphi_\text{T})
\end{equation}
where we have added the subscript $\text{T}$ to all quantities that are specific to the transmitting antenna.

The input power $P_\text{T}$ of an antenna with (conjugate matched) impedance $Z_\text{T}$, fed by a voltage source with complex amplitude $V_\text{T}$, is given by (see e.g.,~\cite{Wikipedia_power}),
\begin{equation}
    \label{eq:P_T}
    P_\text{T} = \frac{|V_\text{T}|^2}{8\Re\{Z_\text{T}\}}.
\end{equation}

\begin{highlightbox}{Normalization of Antenna Patterns}
    The radiated power $\eta_\text{rad} P_\text{T}$ of an antenna can be obtained by integrating the Poynting vector over the surface of a closed sphere of radius $r$ around the antenna:
    \begin{align}
        \eta_\text{rad} P_\text{T} &=  \int_0^{2\pi}\int_0^{\pi} \mathbf{S}(r, \theta, \varphi)^\mathsf{T} \hat{\mathbf{r}} r^2 \sin(\theta)d\theta d\varphi \\
                        &= \int_0^{2\pi}\int_0^{\pi} \frac{1}{2Z_0} \norm{\mathbf{E}(r, \theta, \varphi)}^2 r^2\sin(\theta)d\theta d\varphi \\
                        &= \frac{P_\text{T}}{4 \pi} \int_0^{2\pi}\int_0^{\pi} G(\theta, \varphi) \sin(\theta)d\theta d\varphi.
    \end{align}

    We can see from the last equation that the directional gain of any antenna must satisfy
    \begin{equation}
        \int_0^{2\pi}\int_0^{\pi} G(\theta, \varphi) \sin(\theta)d\theta d\varphi = 4 \pi \eta_\text{rad}.
    \end{equation}
\end{highlightbox}

\subsection{Modeling of a Receiving Antenna}

Although the transmitting antenna radiates a spherical wave $\mathbf{E}_\text{T}(r,\theta_\text{T},\varphi_\text{T})$, we assume that the receiving antenna observes a planar incoming wave $\mathbf{E}_\text{R}$ that arrives from the angles $\theta_\text{R}$ and $\varphi_\text{R}$ which are defined in the local spherical coordinates of the receiving antenna. The Poynting vector of the incoming wave $\mathbf{S}_\text{R}$ is therefore given by~\eqref{eq:S_spherical}
\begin{equation}
    \label{eq:S_R}
    \mathbf{S}_\text{R} = -\frac{1}{2Z_0} \norm{\mathbf{E}_\text{R}}^2 \hat{\mathbf{r}}(\theta_\text{R}, \varphi_\text{R})
\end{equation}
where $\hat{\mathbf{r}}(\theta_\text{R}, \varphi_\text{R})$ is the radial unit vector in the spherical coordinate system of the receiver.

The aperture or effective area $A_\text{R}$ of an antenna with gain $G_\text{R}$ is defined as the ratio of the available received power $P_\text{R}$ at the output of the antenna and the absolute value of the Poynting vector, i.e. the power density:
\begin{equation}
    \label{eq:A_R}
    A_\text{R} = \frac{P_\text{R}}{\norm{\mathbf{S}_\text{R}}} = G_\text{R}\frac{\lambda_0^2}{4\pi}
\end{equation}
where $\frac{\lambda_0^2}{4\pi}$ is the aperture of an isotropic antenna. In the definition above, it is assumed that the antenna is ideally directed towards and polarization matched to the incoming wave.
For an arbitrary orientation of the antenna (but still assuming polarization matching), we can define a direction dependent effective area
\begin{equation}
    \label{eq:A_dir}
    A_\text{R}(\theta_\text{R}, \varphi_\text{R}) = G_\text{R}(\theta_\text{R}, \varphi_\text{R})\frac{\lambda_0^2}{4\pi}.
\end{equation}
The available received power at the output of the antenna can be expressed as
\begin{equation}
    \label{eq:P_R}
    P_\text{R} = \frac{|V_\text{R}|^2}{8\Re\{Z_\text{R}\}}
\end{equation}
where $Z_\text{R}$ is the impedance of the receiving antenna and $V_\text{R}$ the open circuit voltage.
We can now combine~\eqref{eq:P_R},~\eqref{eq:A_dir}, and~\eqref{eq:A_R} to obtain the following expression for the absolute value of the voltage $|V_\text{R}|$ assuming matched polarization:
\begin{align}
    |V_\text{R}| &= \sqrt{P_\text{R} 8\Re\{Z_\text{R}\}}\\
                    &= \sqrt{\frac{\lambda_0^2}{4\pi} G_\text{R}(\theta_\text{R}, \varphi_\text{R}) \frac{8\Re\{Z_\text{R}\}}{2 Z_0} \norm{\mathbf{E}_\text{R}}^2}\\
                    &= \sqrt{\frac{\lambda_0^2}{4\pi} G_\text{R} \frac{4\Re\{Z_\text{R}\}}{Z_0}} \norm{\mathbf{F}_\text{R}(\theta_\text{R}, \varphi_\text{R})}\norm{\mathbf{E}_\text{R}}.
\end{align}

By extension of the previous equation, we can obtain an expression for $V_\text{R}$ which is valid for arbitrary polarizations of the incoming wave and the receiving antenna:
\begin{equation}
    \label{eq:V_R}
    V_\text{R} = \sqrt{\frac{\lambda_0^2}{4\pi} G_\text{R} \frac{4\Re\{Z_\text{R}\}}{Z_0}} \mathbf{F}_\text{R}(\theta_\text{R}, \varphi_\text{R})^{\mathsf{H}}\mathbf{E}_\text{R}.
\end{equation}

\begin{highlightbox}{Recovering Friis Equation}
   In the case of free space propagation, we have $\mathbf{E}_\text{R}=\mathbf{E}_\text{T}(r,\theta_\text{T},\varphi_\text{T})$.
    Combining~\eqref{eq:V_R},~\eqref{eq:P_R}, and~\eqref{eq:E_T}, we obtain the following expression for the received power:
    \begin{equation}
        P_\text{R} = \left(\frac{\lambda_0}{4\pi r}\right)^2 G_\text{R} G_\text{T} P_\text{T} \left|\mathbf{F}_\text{R}(\theta_\text{R}, \varphi_\text{R})^{\mathsf{H}} \mathbf{F}_\text{T}(\theta_\text{T}, \varphi_\text{T})\right|^2.
    \end{equation}

    It is important that $\mathbf{F}_\text{R}$ and $\mathbf{F}_\text{T}$ are expressed in the same coordinate system for the last equation to make sense.
    For perfect orientation and polarization matching, we can recover the well-known Friis transmission equation
    \begin{equation}
        \frac{P_\text{R}}{P_\text{T}} = \left(\frac{\lambda_0}{4\pi r}\right)^2 G_\text{R} G_\text{T}.
    \end{equation}
\end{highlightbox}

\subsection{General Propagation Path}
\label{sec:primer-em-path}

A single propagation path consists of a sequence of scattering processes, where a scattering process can be anything that prevents the wave from propagating as in free space. This includes specular reflection, refraction, diffraction, and diffuse reflection. For each scattering process, one needs to compute a relationship between the incoming field at the scatter center and the created far field at the next scatter center or the receiving antenna.
We can represent this cascade of scattering processes by a single matrix $\widetilde{\mathbf{T}}$ that describes the transformation that the radiated field $\mathbf{E}_\text{T}(r, \theta_\text{T}, \varphi_\text{T})$ undergoes until it reaches the receiving antenna:
\begin{equation}
    \label{eq:E_R}
    \mathbf{E}_\text{R} = \sqrt{ \frac{P_\text{T} G_\text{T} Z_0}{2\pi}} \widetilde{\mathbf{T}} \mathbf{F}_\text{T}(\theta_\text{T}, \varphi_\text{T}).
\end{equation}
Note that we have obtained this expression by replacing the free space propagation term $\frac{e^{-jk_0r}}{r}$ in~\eqref{eq:E_T} by the matrix $\widetilde{\mathbf{T}}$. This requires that all quantities are expressed in the same coordinate system which is also assumed in the following expressions. Further, it is assumed that the matrix $\widetilde{\mathbf{T}}$ includes the necessary coordinate transformations. In some cases, e.g., for diffuse reflection (see~\eqref{eq:scattered_field} in Section~\ref{sec:primer-em-diffuse}), the matrix $\widetilde{\mathbf{T}}$ depends on the incoming field and is not a linear transformation.

Plugging~\eqref{eq:E_R} into~\eqref{eq:V_R}, we can obtain a general expression for the received voltage of a propagation path:
\begin{equation}
    V_\text{R} = \sqrt{\left(\frac{\lambda_0}{4\pi}\right)^2 G_\text{R}G_\text{T}P_\text{T} 8\Re\{Z_\text{R}\}} \,\mathbf{F}_\text{R}(\theta_\text{R}, \varphi_\text{R})^{\mathsf{H}}\widetilde{\mathbf{T}} \mathbf{F}_\text{T}(\theta_\text{T}, \varphi_\text{T}).
\end{equation}
If the electromagnetic wave arrives at the receiving antenna over $N$ propagation paths, we can simply add the received voltages from all paths to obtain
\begin{align}
    \label{eq:V_Rmulti}
    V_\text{R} &= \sqrt{\left(\frac{\lambda_0}{4\pi}\right)^2 G_\text{R}G_\text{T}P_\text{T} 8\Re\{Z_\text{R}\}} \sum_{i=1}^N\mathbf{F}_\text{R}(\theta_{\text{R},i}, \varphi_{\text{R},i})^{\mathsf{H}}\widetilde{\mathbf{T}}_i \mathbf{F}_\text{T}(\theta_{\text{T},i}, \varphi_{\text{T},i})\\
    &= \sqrt{\left(\frac{\lambda_0}{4\pi}\right)^2 P_\text{T} 8\Re\{Z_\text{R}\}} \sum_{i=1}^N\mathbf{C}_\text{R}(\theta_{\text{R},i}, \varphi_{\text{R},i})^{\mathsf{H}}\widetilde{\mathbf{T}}_i \mathbf{C}_\text{T}(\theta_{\text{T},i}, \varphi_{\text{T},i})
\end{align}
where all path-dependent quantities carry the subscript $i$. Note that the matrices $\widetilde{\mathbf{T}}_i$ also ensure appropriate scaling so that the total received power can never be larger than the transmit power.

\subsection{Frequency and Impulse Response}
\label{sec:primer-em-freq-ir}
The channel frequency response $H(f)$ at frequency $f$ is defined as the ratio between the received voltage and the voltage at the input to the transmitting antenna:
\begin{equation}
    \label{eq:H}
    H(f) = \frac{V_\text{R}}{V_\text{T}} = \frac{V_\text{R}}{|V_\text{T}|}
\end{equation}
where it is assumed that the input voltage has zero phase.

It is useful to separate phase shifts due to wave propagation from the transfer matrices $\widetilde{\mathbf{T}}_i$.
If we denote by $d_i$ the total length of path $i$, then assuming propagation at free-space speed $c_0$, its path delay is $\tau_i=d_i/c_0$.
We can now define the new transfer matrix
\begin{equation}
    \label{eq:T_tilde}
    \mathbf{T}_i=\widetilde{\mathbf{T}}_ie^{j2\pi f \tau_i}.
\end{equation}
Using~\eqref{eq:P_T} and~\eqref{eq:T_tilde} in~\eqref{eq:V_Rmulti} while assuming equal real parts of both antenna impedances, i.e. $\Re\{Z_\text{T}\}=\Re\{Z_\text{R}\}$ (which is typically the case), we obtain the final expression for the channel frequency response:
\begin{equation}
    \label{eq:H_final}
    \boxed{H(f) = \sum_{i=1}^N \underbrace{\frac{\lambda_0}{4\pi} \mathbf{C}_\text{R}(\theta_{\text{R},i}, \varphi_{\text{R},i})^{\mathsf{H}}\mathbf{T}_i \mathbf{C}_\text{T}(\theta_{\text{T},i}, \varphi_{\text{T},i})}_{\triangleq a_i} e^{-j2\pi f\tau_i}}
\end{equation}
Taking the inverse Fourier transform, we finally obtain the channel impulse response
\begin{equation}
    \label{eq:h_final2}
    \boxed{h(\tau) = \int_{-\infty}^{\infty} H(f) e^{j2\pi f \tau} df = \sum_{i=1}^N a_i \delta(\tau-\tau_i)}
\end{equation}
where the antenna patterns and material-dependant transfer matrices $\Tm_i$ are evaluated at the fixed carrier frequency $f_c$. The resulting path coefficients $a_i$ are therefore assumed constant over the simulated bandwidth.
The baseband equivalent channel impulse response is then defined as~\cite[Eq.~2.28]{Tse}:
\begin{equation}
    \label{eq:h_b}
    h_\text{b}(\tau) = \sum_{i=1}^N \underbrace{a_i e^{-j2\pi f_c \tau_i}}_{\triangleq a^\text{b}_i} \delta(\tau-\tau_i).
\end{equation}

\subsection{Specular Reflection and Refraction}
\label{sec:primer-em-spec-refrac}

When a plane wave hits a plane interface which separates two materials, e.g., air and concrete, a part of the wave gets reflected and the other refracted, i.e. it propagates into the other material. We assume in the following description that both materials are uniform non-magnetic dielectrics, i.e. $\mu_r=1$, and follow the definitions as in~\cite{ITU_R_P2040}. The incoming wave phasor $\mathbf{E}_\text{i}$ is expressed by two arbitrary orthogonal polarization components, i.e.
\begin{equation}
    \mathbf{E}_\text{i} = E_{\text{i},s} \hat{\mathbf{e}}_{\text{i},s} + E_{\text{i},p} \hat{\mathbf{e}}_{\text{i},p}
\end{equation}
which are both orthogonal to the incident wave vector, i.e. $\hat{\mathbf{e}}_{\text{i},s}^{\mathsf{T}} \hat{\mathbf{e}}_{\text{i},p}=\hat{\mathbf{e}}_{\text{i},s}^{\mathsf{T}} \hat{\mathbf{k}}_\text{i}=\hat{\mathbf{e}}_{\text{i},p}^{\mathsf{T}} \hat{\mathbf{k}}_\text{i} =0$.

\begin{figure}[ht!]
    \center
    \includegraphics[scale=0.9]{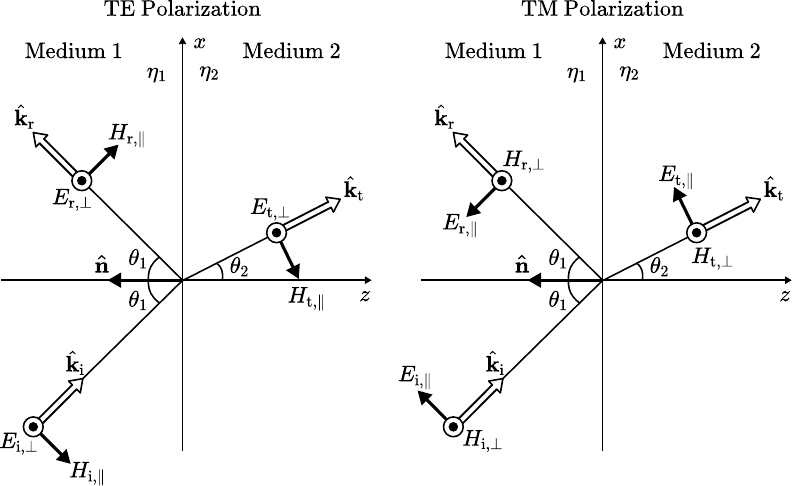}
    \caption{Reflection and refraction of a plane wave at a plane interface between two materials.\label{fig:fig_reflection}}
\end{figure}

Figure~\ref{fig:fig_reflection} shows reflection and refraction of the incoming wave at the plane interface between two materials with relative permittivities $\eta_1$ and $\eta_2$. The coordinate system is chosen such that the wave vectors of the incoming, reflected, and transmitted waves lie within the plane of incidence, which is chosen to be the x-z plane. The normal vector of the interface $\hat{\mathbf{n}}$ is pointing toward the negative z axis.
The incoming wave must be represented in a different basis, i.e. in the form of two different orthogonal polarization components $E_{\text{i}, \perp}$ and $E_{\text{i}, \parallel}$, i.e.
\begin{equation}
    \mathbf{E}_\text{i} = E_{\text{i},\perp} \hat{\mathbf{e}}_{\text{i},\perp} + E_{\text{i},\parallel} \hat{\mathbf{e}}_{\text{i},\parallel}
\end{equation}
where the former is orthogonal to the plane of incidence and called transverse electric (TE) polarization (left), and the latter is parallel to the plane of incidence and called transverse magnetic (TM) polarization (right). In the following, we adopt the convention that all transverse components are coming out of the figure (indicated by the $\odot$ symbol). One can easily verify that the following relationships must hold:
\begin{equation}
    \label{eq:fresnel_in_vectors}
    \begin{aligned}
        \hat{\mathbf{e}}_{\text{i},\perp} &= \frac{\hat{\mathbf{k}}_\text{i} \times \hat{\mathbf{n}}}{\norm{\hat{\mathbf{k}}_\text{i} \times \hat{\mathbf{n}}}} \\
        \hat{\mathbf{e}}_{\text{i},\parallel} &= \hat{\mathbf{e}}_{\text{i},\perp} \times \hat{\mathbf{k}}_\text{i}\\
        \begin{bmatrix}E_{\text{i},\perp} \\ E_{\text{i},\parallel} \end{bmatrix} &=
            \begin{bmatrix}
                \hat{\mathbf{e}}_{\text{i},\perp}^\mathsf{T}\hat{\mathbf{e}}_{\text{i},s} & \hat{\mathbf{e}}_{\text{i},\perp}^\mathsf{T}\hat{\mathbf{e}}_{\text{i},p}\\
                \hat{\mathbf{e}}_{\text{i},\parallel}^\mathsf{T}\hat{\mathbf{e}}_{\text{i},s} & \hat{\mathbf{e}}_{\text{i},\parallel}^\mathsf{T}\hat{\mathbf{e}}_{\text{i},p}
            \end{bmatrix}
         \begin{bmatrix}E_{\text{i},s} \\ E_{\text{i},p}\end{bmatrix} =
         \mathbf{W}\left(\hat{\mathbf{e}}_{\text{i},\perp}, \hat{\mathbf{e}}_{\text{i},\parallel}, \hat{\mathbf{e}}_{\text{i},s}, \hat{\mathbf{e}}_{\text{i},p}\right) \begin{bmatrix}E_{\text{i},s} \\ E_{\text{i},p}\end{bmatrix}
    \end{aligned}
\end{equation}
where we have defined the following matrix-valued function
\begin{align}
    \label{eq:W}
    \mathbf{W}\left(\hat{\mathbf{a}}, \hat{\mathbf{b}}, \hat{\mathbf{q}}, \hat{\mathbf{r}} \right) =
    \begin{bmatrix}
        \hat{\mathbf{a}}^\textsf{T} \hat{\mathbf{q}} & \hat{\mathbf{a}}^\textsf{T} \hat{\mathbf{r}} \\
        \hat{\mathbf{b}}^\textsf{T} \hat{\mathbf{q}} & \hat{\mathbf{b}}^\textsf{T} \hat{\mathbf{r}}
    \end{bmatrix}.
\end{align}
At normal incidence ($\hat{\mathbf{k}}_\text{i}\times\hat{\mathbf{n}}=\mathbf{0}$), the TE direction in~\eqref{eq:fresnel_in_vectors} is undefined, and any orthonormal basis transverse to $\hat{\mathbf{k}}_\text{i}$ may be used.

While the angles of incidence and reflection are both equal to $\theta_1$, the angle of the refracted wave $\theta_2$ is given by Snell's law
\begin{equation}
    \sin(\theta_2) = \sqrt{\frac{\eta_1}{\eta_2}} \sin(\theta_1)
\end{equation}
or, equivalently,
\begin{equation}
    \cos(\theta_2) = \sqrt{1 - \frac{\eta_1}{\eta_2} \sin^2(\theta_1)}.
\end{equation}

The reflected and transmitted wave phasors $\mathbf{E}_\text{r}$ and $\mathbf{E}_\text{t}$ are similarly represented as
\begin{align}
    \mathbf{E}_\text{r} &= E_{\text{r},\perp} \hat{\mathbf{e}}_{\text{r},\perp} + E_{\text{r},\parallel} \hat{\mathbf{e}}_{\text{r},\parallel}\\
    \mathbf{E}_\text{t} &= E_{\text{t},\perp} \hat{\mathbf{e}}_{\text{t},\perp} + E_{\text{t},\parallel} \hat{\mathbf{e}}_{\text{t},\parallel}
\end{align}
where
\begin{equation}
    \label{eq:fresnel_out_vectors}
    \begin{aligned}
        \hat{\mathbf{e}}_{\text{r},\perp} &= \hat{\mathbf{e}}_{\text{i},\perp}\\
        \hat{\mathbf{e}}_{\text{r},\parallel} &= \frac{\hat{\mathbf{e}}_{\text{r},\perp}\times\hat{\mathbf{k}}_\text{r}}{\norm{\hat{\mathbf{e}}_{\text{r},\perp}\times\hat{\mathbf{k}}_\text{r}}}\\
        \hat{\mathbf{e}}_{\text{t},\perp} &= \hat{\mathbf{e}}_{\text{i},\perp}\\
        \hat{\mathbf{e}}_{\text{t},\parallel} &= \frac{\hat{\mathbf{e}}_{\text{t},\perp}\times\hat{\mathbf{k}}_\text{t}}{ \norm{\hat{\mathbf{e}}_{\text{t},\perp}\times\hat{\mathbf{k}}_\text{t} }}
    \end{aligned}
\end{equation}
and
\begin{equation}
    \label{eq:reflected_refracted_vectors}
    \begin{aligned}
        \hat{\mathbf{k}}_\text{r} &= \hat{\mathbf{k}}_\text{i} - 2\left( \hat{\mathbf{k}}_\text{i}^\mathsf{T}\hat{\mathbf{n}} \right)\hat{\mathbf{n}}\\
        \hat{\mathbf{k}}_\text{t} &= \sqrt{\frac{\eta_1}{\eta_2}} \hat{\mathbf{k}}_\text{i} + \left(\sqrt{\frac{\eta_1}{\eta_2}}\cos(\theta_1) - \cos(\theta_2) \right)\hat{\mathbf{n}}.
    \end{aligned}
\end{equation}

The \emph{Fresnel} equations provide relationships between the incident, reflected, and refracted field components for $\sqrt{\left| \eta_1/\eta_2 \right|}\sin(\theta_1)<1$:
\begin{equation}
    \label{eq:fresnel}
    \begin{aligned}
        r_{\perp}     &= \frac{E_{\text{r}, \perp    }}{E_{\text{i}, \perp    }} = \frac{ \sqrt{\eta_1}\cos(\theta_1) - \sqrt{\eta_2}\cos(\theta_2) }{ \sqrt{\eta_1}\cos(\theta_1) + \sqrt{\eta_2}\cos(\theta_2) } \\
        r_{\parallel} &= \frac{E_{\text{r}, \parallel}}{E_{\text{i}, \parallel}} = \frac{ \sqrt{\eta_2}\cos(\theta_1) - \sqrt{\eta_1}\cos(\theta_2) }{ \sqrt{\eta_2}\cos(\theta_1) + \sqrt{\eta_1}\cos(\theta_2) } \\
        t_{\perp}     &= \frac{E_{\text{t}, \perp    }}{E_{\text{i}, \perp    }} = \frac{ 2\sqrt{\eta_1}\cos(\theta_1) }{ \sqrt{\eta_1}\cos(\theta_1) + \sqrt{\eta_2}\cos(\theta_2) } \\
        t_{\parallel} &= \frac{E_{\text{t}, \parallel}}{E_{\text{i}, \parallel}} = \frac{ 2\sqrt{\eta_1}\cos(\theta_1) }{ \sqrt{\eta_2}\cos(\theta_1) + \sqrt{\eta_1}\cos(\theta_2) }.
    \end{aligned}
\end{equation}
Assuming real-valued permittivities, which correspond to lossless media, if $\sqrt{\left| \eta_1/\eta_2 \right|}\sin(\theta_1)\ge 1$, then total reflection occurs, i.e., $\abs{r_{\perp}}=\abs{r_{\parallel}}=1$, and the transmitted field is evanescent with zero normal time-average power flux.
For the case of an incident wave in vacuum, i.e. $\eta_1=1$, the Fresnel equations~\eqref{eq:fresnel} simplify to
\begin{equation}
    \label{eq:fresnel_vac}
    \begin{aligned}
        r_{\perp}     &= \frac{\cos(\theta_1) -\sqrt{\eta_2 -\sin^2(\theta_1)}}{\cos(\theta_1) +\sqrt{\eta_2 -\sin^2(\theta_1)}} \\
        r_{\parallel} &= \frac{\eta_2\cos(\theta_1) -\sqrt{\eta_2 -\sin^2(\theta_1)}}{\eta_2\cos(\theta_1) +\sqrt{\eta_2 -\sin^2(\theta_1)}} \\
        t_{\perp}     &= \frac{2\cos(\theta_1)}{\cos(\theta_1) + \sqrt{\eta_2-\sin^2(\theta_1)}}\\
        t_{\parallel} &= \frac{2\sqrt{\eta_2}\cos(\theta_1)}{\eta_2 \cos(\theta_1) + \sqrt{\eta_2-\sin^2(\theta_1)}}.
    \end{aligned}
\end{equation}

Putting everything together, we obtain the following relationships between incident, reflected, and transmitted waves:
\begin{align}
    \begin{bmatrix}E_{\text{r},\perp} \\ E_{\text{r},\parallel} \end{bmatrix} &=
    \begin{bmatrix}
        r_{\perp} & 0 \\
        0         & r_{\parallel}
    \end{bmatrix}
    \mathbf{W}\left(\hat{\mathbf{e}}_{\text{i},\perp}, \hat{\mathbf{e}}_{\text{i},\parallel}, \hat{\mathbf{e}}_{\text{i},s}, \hat{\mathbf{e}}_{\text{i},p}\right)
    \begin{bmatrix}E_{\text{i},s} \\ E_{\text{i},p}\end{bmatrix} \\
    \begin{bmatrix}E_{\text{t},\perp} \\ E_{\text{t},\parallel} \end{bmatrix} &=
    \begin{bmatrix}
        t_{\perp} & 0 \\
        0         & t_{\parallel}
    \end{bmatrix}
    \mathbf{W}\left(\hat{\mathbf{e}}_{\text{i},\perp}, \hat{\mathbf{e}}_{\text{i},\parallel}, \hat{\mathbf{e}}_{\text{i},s}, \hat{\mathbf{e}}_{\text{i},p}\right)
    \begin{bmatrix}E_{\text{i},s} \\ E_{\text{i},p}\end{bmatrix}.
\end{align}

\subsubsection{Single-Layer Slab}
\label{sec:single-layer-slab}

\begin{figure}
    \center
    \includegraphics[scale=0.2]{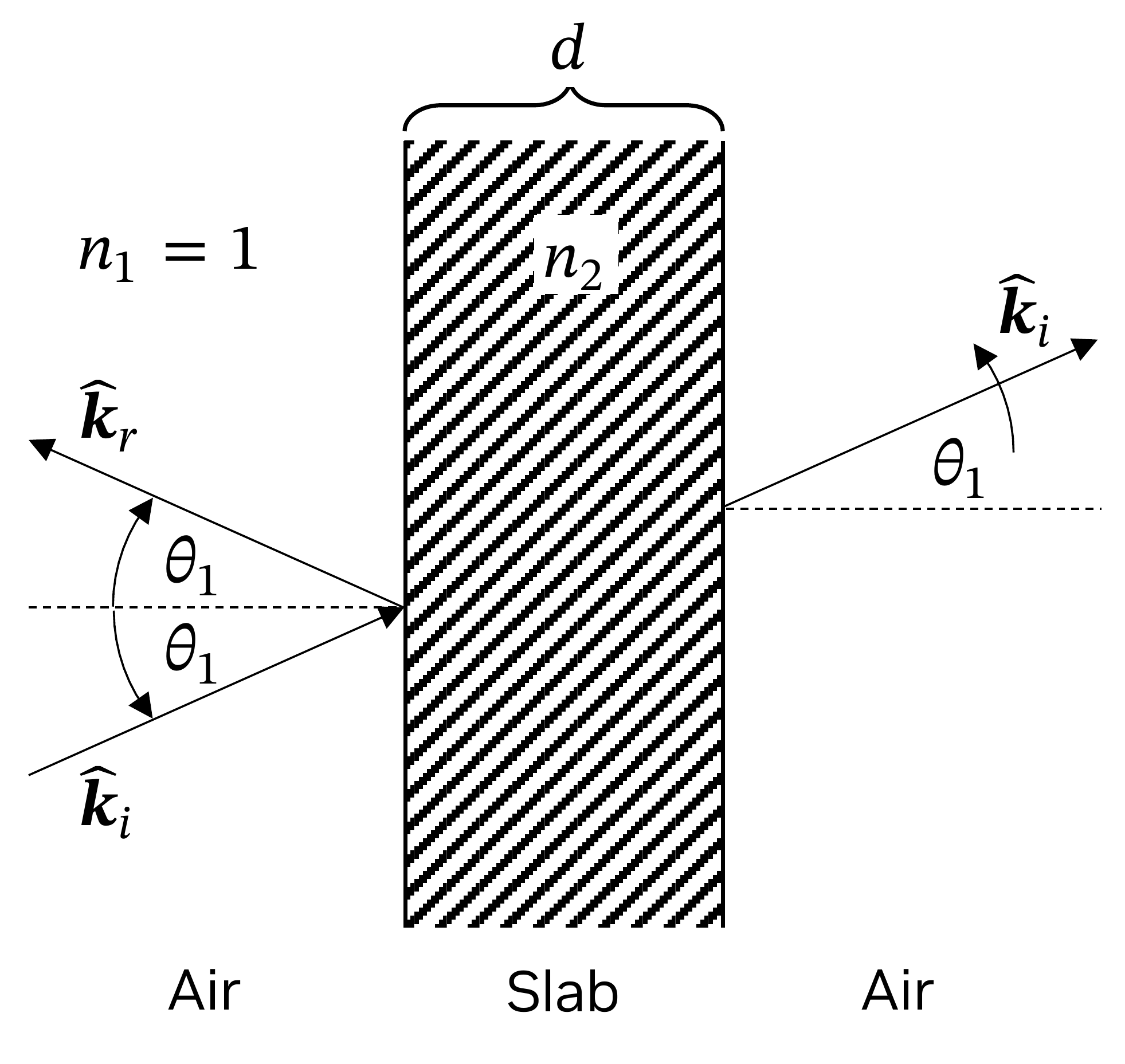}
    \caption{Reflection and refraction of a plane wave at a single-layer slab.\label{fig:primer-em-slab}}
\end{figure}

The reflection and refraction coefficients described above assume that the object reflecting the wave or allowing it to penetrate is of infinite thickness. However, since this is rarely the case, it is often more practical to assume that the object has a finite thickness. In such cases, the object can be modeled as a slab consisting of a single layer made of the same material, as shown in Figure~\ref{fig:primer-em-slab}. The reflection and transmission coefficients, which should be used instead of~\eqref{eq:fresnel_vac}, are then computed as described in~\cite[Section~2.2.2.2]{ITU_R_P2040}:
\begin{align}
    \label{eq:fresnel_slab}
    r &= \frac{r'\left(1-e^{-j2q}\right)}{1-r^{'2}e^{-j2q}} \\
    t &= \frac{\left(1-r^{'2}\right)e^{-jq}}{1-r^{'2}e^{-j2q}}
\end{align}
where
\begin{equation}
    \label{eq:q_fresnel_slab}
    q = \frac{2\pi d}{\lambda_0} \sqrt{\eta - \sin^2 \theta_1}
\end{equation}
$d$ [\si{\meter}] is the thickness of the slab, $\eta$ the complex relative permittivity as defined in~\eqref{eq:eta}, and $r'$ denotes either $r_{\perp}$ or
$r_{\parallel}$ from~\eqref{eq:fresnel_vac}, depending on the polarization of the incident electric field.

\subsection{Diffraction}
\label{sec:primer-em-diffraction}

While modern geometrical optics (GO)~\cite{Kline51,Luneburg64} can accurately describe phase and polarization properties of electromagnetic fields undergoing reflection and refraction (transmission) as described above, it fails to account for the phenomenon of diffraction, e.g., bending of waves around corners. This leads to the undesired and physically incorrect effect that the field abruptly falls to zero at geometrical shadow boundaries (for incident and reflected fields).

Joseph Keller presented in~\cite{Keller:62} a method which allowed the incorporation of diffraction into GO which is known as the geometrical theory of diffraction (GTD). He introduced the notion of diffracted rays that follow the law of edge diffraction, i.e., the diffracted and incident rays make the same angle with the edge at the point of diffraction and lie on opposite sides of the plane normal to the edge. The GTD suffers, however, from several shortcomings, most importantly the fact that the diffracted field is infinite at shadow boundaries.

The uniform theory of diffraction (UTD)~\cite{1451581} alleviates this problem and provides solutions that are uniformly valid, even at shadow boundaries. For a great introduction to the UTD, we refer to~\cite{McNamara90}. While~\cite{1451581} deals with diffraction at edges of perfectly conducting surfaces, it was heuristically extended to finitely conducting wedges in~\cite{1143189}. This solution is also recommended by the ITU~\cite{ITU_R_P526}.
However, both~\cite{1143189} and~\cite{ITU_R_P526} only deal with two-dimensional scenes, where the source and the observation point lie in the same plane, orthogonal to the edge.
This model was later extended to three-dimensional wedges with arbitrary polarization in~\cite{1143019}, and is currently implemented in \SRT{} (see~\cite{9362253} for an extension to interior wedges, which are not supported).

\begin{figure}
    \center
    \includegraphics[scale=0.2]{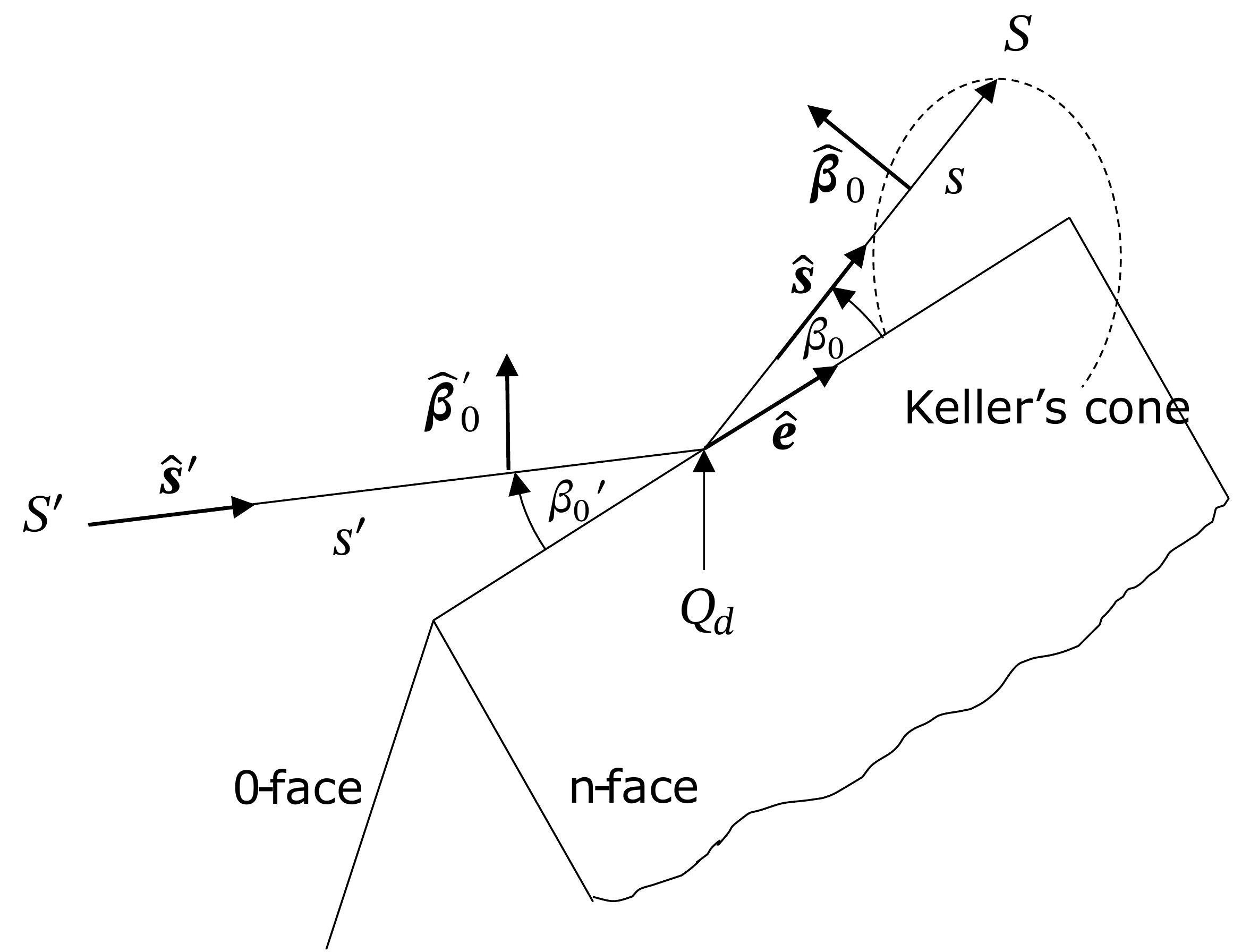}
    \caption{Incident and diffracted rays for an infinitely long wedge in an edge-fixed coordinate system.\label{fig:primer-em-kellers-cone}}
\end{figure}

We consider an infinitely long wedge with unit norm edge vector $\hat{\mathbf{e}}$, as shown in Figure~\ref{fig:primer-em-kellers-cone}. An incident ray of a spherical wave with field phasor $\mathbf{E}_i(S')$ at point $S'$ propagates in the direction $\hat{\mathbf{s}}'$ and is diffracted at point $Q_d$ on the edge. The diffracted ray of interest (there are infinitely many on Keller's cone) propagates
in the direction $\hat{\mathbf{s}}$ towards the point of observation $S$. We denote by $s'=\norm{S'-Q_d}$ and $s=\norm{Q_d - S}$ the lengths of the incident and diffracted path segments, respectively. By the law of edge diffraction, the angles $\beta_0'$ and $\beta_0$ between the edge and the incident and diffracted rays, respectively, satisfy:
\begin{equation}
    \cos{\LB \beta_0' \RB} = \hat{\sv}'^{\textsf{T}}\hat{\ev} = \hat{\sv}\tp\hat{\ev} = \cos(\beta_0).
\end{equation}

To be able to express the diffraction coefficients as a $2 \times 2$ matrix---similar to what is done for reflection and refraction---the incident field must be resolved into two components $E_{i,\phi'}$ and $E_{i,\beta_0'}$, the former orthogonal and the latter parallel to the edge-fixed plane of incidence, i.e., the plane containing $\hat{\ev}$ and $\hat{\sv}'$. The diffracted field is then represented by two components $E_{d,\phi}$ and $E_{d,\beta_0}$ that are respectively orthogonal and parallel to the edge-fixed plane of diffraction, i.e., the plane containing $\hat{\ev}$ and $\hat{\sv}$.
The corresponding component unit vectors are defined as
\begin{align}
    \hat{\boldsymbol{\phi}}' &= \frac{\hat{\sv}' \times \hat{\ev}}{\norm{\hat{\sv}' \times \hat{\ev}} }\\
    \hat{\boldsymbol{\beta}}_0' &=  \hat{\boldsymbol{\phi}}' \times \hat{\sv}' \\
    \hat{\boldsymbol{\phi}} &= -\frac{\hat{\sv} \times \hat{\ev}}{\norm{\hat{\sv} \times \hat{\ev}} } \label{eq:phi_definition}\\
    \hat{\boldsymbol{\beta}}_0 &=  \hat{\boldsymbol{\phi}} \times \hat{\sv} \label{eq:beta_0_definition}.
\end{align}

\begin{figure}
    \center
    \includegraphics[scale=0.2]{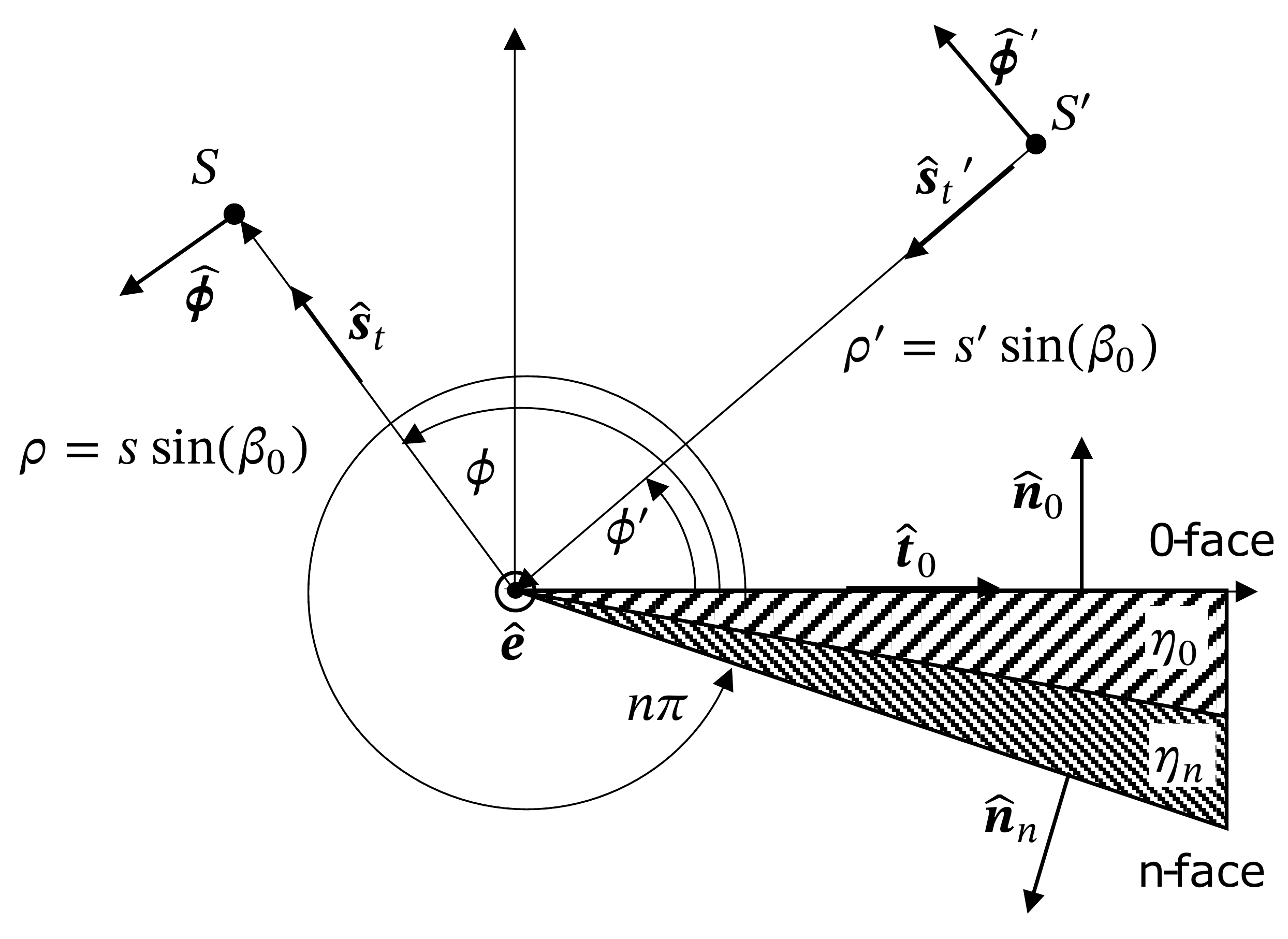}
    \caption{Top view on the wedge with edge vector pointing upwards.\label{fig:primer-em-diffraction}}
\end{figure}

Figure~\ref{fig:primer-em-diffraction} shows the top view on the wedge that we need for some additional definitions.
The wedge has two faces called \emph{0-face} and \emph{n-face}, respectively, with surface normal vectors $\hat{\mathbf{n}}_0$ and $\hat{\mathbf{n}}_n$. The exterior wedge angle is $n\pi$, with $1\le n \le 2$. Note that the surfaces are chosen such that
\begin{equation}
    \hat{\mathbf{e}} = \frac{\hat{\mathbf{n}}_0 \times \hat{\mathbf{n}}_n}{\norm{\hat{\mathbf{n}}_0 \times \hat{\mathbf{n}}_n}}.
\end{equation}
For $n=2$, the wedge reduces to a screen and the choice of the \emph{0-face} and \emph{n-face} is arbitrary as they point in opposite directions.

The incident and diffracted rays have angles $\phi'$ and $\phi$ measured with respect to the \emph{0-face} in the plane perpendicular to the edge.
They can be computed as follows:
\begin{align}
    \phi' & = \pi - \LSB\pi - \arccos\LB-\hat{\sv}_t'^{\textsf{T}} \hat{\tv}_0\RB\RSB \operatorname{sgn}\LB-\hat{\sv}_t'^{\textsf{T}} \hat{\nv}_0\RB \\
    \phi & = \pi - \LSB\pi - \arccos\LB\hat{\sv}_t\tp \hat{\tv}_0\RB\RSB \operatorname{sgn}\LB\hat{\sv}_t\tp \hat{\nv}_0\RB
\end{align}
where
\begin{align}
    \hat{\tv}_0 &= \hat{\nv}_0 \times \hat{\ev}\\
    \hat{\sv}'_t &= \frac{ \hat{\sv}' - \LB\hat{\sv}'^{\textsf{T}}\hat{\ev}\RB\hat{\ev} }{\norm{\hat{\sv}' - \LB\hat{\sv}'^{\textsf{T}}\hat{\ev}\RB\hat{\ev}}  }\\
    \hat{\sv}_t  &= \frac{ \hat{\sv} - \LB\hat{\sv}\tp\hat{\ev}\RB\hat{\ev} }{\norm{\hat{\sv} - \LB\hat{\sv}\tp\hat{\ev}\RB\hat{\ev}}  }
\end{align}
are, respectively, the unit vector tangent to the \emph{0-face}, and the unit vectors pointing in the directions of $\hat{\sv}'$ and $\hat{\sv}$, projected onto the plane perpendicular to the edge. The function $\mathop{\text{sgn}}(x)$ is defined in this context as
\begin{equation}
\mathop{\text{sgn}}(x) = \begin{cases}
                            1  &, x \geq 0\\
                            -1 &, x< 0.
                            \end{cases}
\end{equation}

With these definitions, the diffracted field at point $S$ can be computed from the incoming field at point $S'$ as follows~\cite{9362253}:
\begin{align}
    \label{eq:diffracted-field}
    \begin{bmatrix}
        E_{d,\phi} \\
        E_{d,\beta_0}
    \end{bmatrix} (S)
    = -\LB \LB -D_1 + D_4 \RB \Rm_0 + \LB -D_2 + D_3 \RB \Rm_n \RB
    \begin{bmatrix}
        E_{i,\phi'} \\
        E_{i,\beta_0'}
    \end{bmatrix}(S') \sqrt{\frac{1}{s's(s'+s)}} e^{-jk(s'+s)}
\end{align}
where $k=2\pi/\lambda_0$ is the free-space wave number and the matrices $\Rm_\nu,\, \nu \in \{0,n\}$, are defined as:
\begin{equation}
    \label{eq:diffraction-matrix-0}
    \Rm_0 =
        \Wm \LB
                \hat{\boldsymbol{\phi}}_{\text{SB},0},
                \hat{\boldsymbol{\beta}}_{0, \text{SB},0},
                \hat{\mathbf{e}}_{r, \perp, 0},
                \hat{\mathbf{e}}_{r, \parallel, 0}
            \RB
        \begin{bmatrix}
            r_{\perp}(\theta_{r,0}, \eta_{0}) & 0\\
            0 & r_{\parallel}(\theta_{r,0}, \eta_{0})
        \end{bmatrix}
        \Wm \LB
                \hat{\mathbf{e}}_{i, \perp, 0},
                \hat{\mathbf{e}}_{i, \parallel, 0},
                \hat{\boldsymbol{\phi}}',
                \hat{\boldsymbol{\beta}}_0'
            \RB
\end{equation}
with $\Wm(\cdot)$ as defined in~\eqref{eq:W}, and
\begin{multline}
    \label{eq:diffraction-matrix-n}
    \Rm_n =\\
    \begin{cases}
        \Wm \LB
                \hat{\boldsymbol{\phi}}_{\text{SB},n},
                \hat{\boldsymbol{\beta}}_{0, \text{SB},n},
                \hat{\mathbf{e}}_{r, \perp, n},
                \hat{\mathbf{e}}_{r, \parallel, n}
            \RB
        \begin{bmatrix}
            r_{\perp}(\theta_{r,n}, \eta_{n}) & 0\\
            0 & r_{\parallel}(\theta_{r,n}, \eta_{n})
        \end{bmatrix}
        \Wm \LB
                \hat{\mathbf{e}}_{i, \perp, n},
                \hat{\mathbf{e}}_{i, \parallel, n},
                \hat{\boldsymbol{\phi}}',
                \hat{\boldsymbol{\beta}}_0'
            \RB
        & \text{if } \pi \leq n\pi \leq \phi' + \pi\\
        -\Id
        & \text{if } \phi' + \pi \leq n\pi \leq 2\pi
    \end{cases}
\end{multline}
where $r_{\perp}(\theta_{r,\nu}, \eta_{\nu})$ and $r_{\parallel}(\theta_{r,\nu}, \eta_{\nu})$ are the Fresnel reflection coefficients from~\eqref{eq:fresnel_vac}, evaluated for the complex relative permittivities $\eta_{\nu}$ and angles $\theta_{r,\nu}$ defined as:
\begin{equation}
    \cos{\theta_{r,\nu}} = \left| \hat{\nv}_\nu\tp \hat{\sv}' \right|.
\end{equation}
Note that it is assumed that $\phi' \leq \pi$.
Using different material properties for the 0- and n-faces in~\eqref{eq:diffraction-matrix-0} and~\eqref{eq:diffraction-matrix-n} is not part of the original model in~\cite{9362253} and is therefore an extension of that work.
The vectors $\hat{\ev}_{i,\perp,\nu}$ and $\hat{\ev}_{i,\parallel,\nu}$ are defined as in~\eqref{eq:fresnel_in_vectors} and~\eqref{eq:fresnel_out_vectors}, but made explicit here for the case of diffraction:
\begin{align}
    \hat{\ev}_{i,\perp,\nu} &= \frac{ \hat{\sv}' \times \hat{\nv}_\nu }{\norm{\hat{\sv}' \times \hat{\nv}_\nu} }\\
    \hat{\ev}_{i,\parallel,\nu} &=  \hat{\ev}_{i,\perp,\nu} \times \hat{\sv}'\\
    \hat{\ev}_{r,\perp,\nu} &=  \hat{\ev}_{i,\perp,\nu}\\
    \hat{\ev}_{r,\parallel,\nu} &=  \hat{\ev}_{i,\perp,\nu} \times \hat{\sv}_{\text{SB},\nu}
\end{align}
where $\hat{\sv}_{\text{SB},\nu} = \hat{\sv}' - 2\LB\hat{\sv}'^{\textsf{T}}\hat{\nv}_\nu\RB\hat{\nv}_\nu$ is the direction associated with the reflection shadow boundary from the $\nu$-face.
The vectors $\hat{\phi}_{\text{SB},\nu}$ and $\hat{\beta}_{0, \text{SB},\nu}$ are defined as in~\eqref{eq:phi_definition} and~\eqref{eq:beta_0_definition}, but for a direction of propagation that matches the direction of specular reflection from the $\nu$-face:
\begin{align}
    \hat{\phi}_{\text{SB},\nu} &= -\frac{\hat{\sv}_{\text{SB},\nu} \times \hat{\ev} }{\norm{\hat{\sv}_{\text{SB},\nu} \times \hat{\ev}} }\\
    \hat{\beta}_{0, \text{SB},\nu} &=  \hat{\phi}_{\text{SB},\nu} \times \hat{\sv}_{\text{SB},\nu}.
\end{align}
In~\eqref{eq:diffraction-matrix-0} and~\eqref{eq:diffraction-matrix-n}, the diffracted-field components are computed in the basis of the corresponding reflection shadow boundary.
Following the heuristic continuation of~\cite{9362253}, the resulting components are interpreted directly in the edge-fixed basis \((\hat{\boldsymbol{\phi}},\hat{\boldsymbol{\beta}}_0)\) of the diffracted ray being evaluated in~\eqref{eq:diffracted-field}.
The diffraction coefficients $D_1,\dots,D_4$ are computed as
\begin{align}
    D_1 &= \frac{-e^{-\frac{j\pi}{4}}}{2n\sqrt{2\pi k} \sin(\beta_0)} \mathop{\text{cot}}\LB \frac{\pi+(\phi-\phi')}{2n} \RB F\LB k L a^+(\phi-\phi')\RB\\
    D_2 &= \frac{-e^{-\frac{j\pi}{4}}}{2n\sqrt{2\pi k} \sin(\beta_0)} \mathop{\text{cot}}\LB \frac{\pi-(\phi-\phi')}{2n} \RB F\LB k L a^-(\phi-\phi')\RB\\
    D_3 &= \frac{-e^{-\frac{j\pi}{4}}}{2n\sqrt{2\pi k} \sin(\beta_0)} \mathop{\text{cot}}\LB \frac{\pi+(\phi+\phi')}{2n} \RB F\LB k L a^+(\phi+\phi')\RB\\
    D_4 &= \frac{-e^{-\frac{j\pi}{4}}}{2n\sqrt{2\pi k} \sin(\beta_0)} \mathop{\text{cot}}\LB \frac{\pi-(\phi+\phi')}{2n} \RB F\LB k L a^-(\phi+\phi')\RB
\end{align}
where
\begin{align}
    L &= \frac{ss'}{s+s'}\sin^2(\beta_0)\\
    a^{\pm}(\beta) &= 2\cos^2\LB \frac{2n\pi N^{\pm}-\beta}{2} \RB\\
    N^{\pm} &= \mathop{\text{round}}\LB \frac{\beta\pm\pi}{2n\pi} \RB\\
    F(x) &= 2j\sqrt{x}e^{jx}\int_{\sqrt{x}}^\infty e^{-jt^2}dt
\end{align}
and $\mathop{\text{round}}(\cdot)$ is the function that rounds to the closest integer. The function $F(x)$ can be expressed with the help of the standard Fresnel integrals~\cite{Wikipedia_Fresnel}.
\begin{align}
    S(x) &= \int_0^x \sin\LB \pi t^2/2 \RB dt \\
    C(x) &= \int_0^x \cos\LB \pi t^2/2 \RB dt
\end{align}
as
\begin{equation}
    F(x) = \sqrt{\frac{\pi x}{2}} e^{jx} \LSB 1+j-2\LB S\LB \sqrt{2x/\pi} \RB +jC\LB \sqrt{2x/\pi} \RB \RB \RSB.
\end{equation}

\subsection{Diffuse Reflection}
\label{sec:primer-em-diffuse}

When an electromagnetic wave impinges on a surface, one part of the energy gets reflected while the other part gets refracted, i.e. it propagates into the surface. We distinguish between two types of reflection, specular and diffuse. The former type is discussed in Section~\ref{sec:primer-em-spec-refrac} and we will focus now on the latter type. When a ray hits a diffuse reflection surface, it is not reflected into a single (specular) direction but rather scattered toward many different directions. Since most surfaces give both specular and diffuse reflections, we denote by $S^2$ the fraction of the reflected energy that is diffusely scattered, where $S\in[0,1]$ is the so-called \emph{scattering coefficient}~\cite{4052607}. Similarly, $R^2$ is the specularly reflected fraction of the reflected energy, where $R\in[0,1]$ is the \emph{reflection reduction factor}. The following relationship between $R$ and $S$ holds:
\begin{equation}
    \label{eq:R}
    R = \sqrt{1-S^2}.
\end{equation}
Whenever a material has a scattering coefficient $S>0$, the Fresnel reflection coefficients in~\eqref{eq:fresnel} must be multiplied by~\eqref{eq:R}.

\begin{figure}[ht!]
    \center
    \includegraphics[scale=0.2]{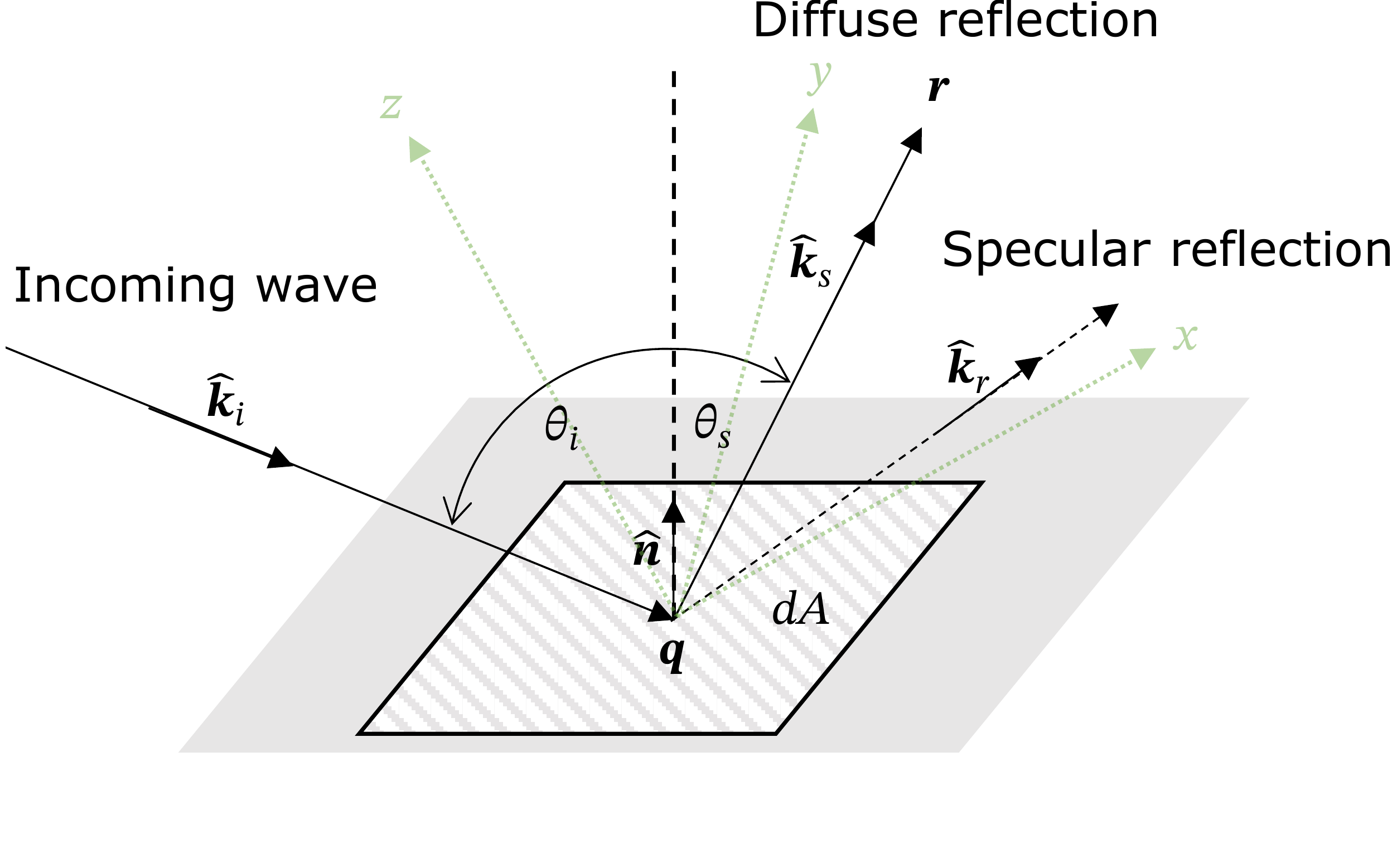}
    \caption{Diffuse and specular reflection of an incoming wave.\label{fig:primer-em-scattering}}
\end{figure}

Let us consider an incoming locally planar linearly polarized wave with field phasor $\mathbf{E}_\text{i}(\mathbf{q})$ at the scattering point $\mathbf{q}$ on the surface, as shown in Figure~\ref{fig:primer-em-scattering}. We focus on the field scattered by an infinitesimally small surface element $dA$ in the direction $\hat{\mathbf{k}}_\text{s}$. Note that the surface normal $\hat{\mathbf{n}}$ has an arbitrary orientation with respect to the global coordinate system, whose $(x,y,z)$ axes are shown in green dotted lines.
The small surface element is related to the incident ray tube solid angle $d\omega$ through
\begin{equation}
    dA = \frac{r^2}{\abs{\hat{\mathbf{n}}\tp\hat{\mathbf{k}}_\text{i}}}d\omega,
\end{equation}
where $r$ is the ray tube length. The denominator is the cosine of the incidence angle.

The incoming field phasor can be represented by two arbitrary orthogonal polarization components (both orthogonal to the incoming wave vector $\hat{\mathbf{k}}_i$):
\begin{align}
    \mathbf{E}_\text{i}(\mathbf{q}) &= E_{\text{i},s} \hat{\mathbf{e}}_{\text{i},s} + E_{\text{i},p} \hat{\mathbf{e}}_{\text{i},p} \\
                        &= E_{\text{i},\perp} \hat{\mathbf{e}}_{\text{i},\perp} + E_{\text{i},\parallel} \hat{\mathbf{e}}_{\text{i},\parallel} \\
                        &= E_{\text{i},\theta} \hat{\boldsymbol{\theta}}(\hat{\mathbf{k}}_\text{i}) + E_{\text{i},\phi} \hat{\boldsymbol{\varphi}}(\hat{\mathbf{k}}_\text{i})
\end{align}
where we have omitted the dependence of the field strength on the position $\mathbf{q}$ for brevity.
The second representation via $(E_{\text{i},\perp}, E_{\text{i},\parallel})$ is used for the computation of the specularly reflected field as explained in Section~\ref{sec:primer-em-spec-refrac}. The third representation decomposes the field into a vertically and horizontally polarized component, where $\hat{\boldsymbol{\theta}}, \hat{\boldsymbol{\varphi}}$ are defined in~\eqref{eq:spherical_vecs}.

According to~\cite[Eq.~9]{5979177}, the diffusely scattered field $\mathbf{E}_\text{s}(\mathbf{r})$ at the observation point $\mathbf{r}$ can be modeled as $\mathbf{E}_\text{s}(\mathbf{r})=E_{\text{s}, \theta}\hat{\boldsymbol{\theta}}(\hat{\mathbf{k}}_\text{s}) + E_{\text{s}, \varphi}\hat{\boldsymbol{\varphi}}(\hat{\mathbf{k}}_\text{s})$, where the orthogonal field components are computed as
\begin{align}
    \label{eq:scattered_field}
    \begin{bmatrix}E_{\text{s}, \theta} \\ E_{\text{s}, \varphi}
    \end{bmatrix} (\rv) &= \frac{S \Gamma}{\norm{\mathbf{r} - \mathbf{q}
   }}  \sqrt{f_\text{s}\left(\hat{\mathbf{k}}_i, \hat{\mathbf{k}}_s,
    \hat{\mathbf{n}}\right) \cos(\theta_i)dA}
    \begin{bmatrix} \sqrt{1-K_x}e^{j\chi_1} & -\sqrt{K_x}e^{j\chi_1} \\
    \sqrt{K_x}e^{j\chi_2} & \sqrt{1-K_x} e^{j\chi_2} \end{bmatrix}
    \begin{bmatrix}E_{\text{i}, \theta} \\ E_{\text{i}, \varphi}
    \end{bmatrix} (\qv)
    e^{-j k \norm{\mathbf{r} - \mathbf{q}}}.
\end{align}
Here, $\Gamma^2$ is the ratio of the incoming power that is reflected (specularly and diffusely), which can be computed as
\begin{equation}
    \Gamma = \frac{\sqrt{ |r_{\perp} E_{\text{i},\perp} |^2 + |r_{\parallel} E_{\text{i},\parallel} |^2}}
              {\norm{\mathbf{E}_\text{i}(\mathbf{q})}}
\end{equation}
where $r_{\perp}, r_{\parallel}$ are the Fresnel coefficients from~\eqref{eq:fresnel} \emph{before} multiplication by the reflection-reduction factor~\eqref{eq:R}, $\chi_1, \chi_2 \in [0,2\pi]$ are independent random phase shifts, and the quantity $K_x\in[0,1]$ is defined by the scattering cross-polarization discrimination
\begin{equation}
    \label{eq:xpd}
    \text{XPD}_\text{s} = 10\log_{10}\left(\frac{|E_{\text{s}, \text{pol}}|^2}{|E_{\text{s}, \text{xpol}}|^2} \right) = 10\log_{10}\left(\frac{1-K_x}{K_x} \right).
\end{equation}
This quantity determines how much energy gets transferred from one polarization direction into the other through the scattering process. Lastly, $dA$ is the area of the small surface element on the reflecting surface under consideration, and $f_\text{s}\left(\hat{\mathbf{k}}_i, \hat{\mathbf{k}}_s, \hat{\mathbf{n}}\right)$ is the \emph{scattering pattern}, which has similarities with the \gls{BSDF} in computer graphics~\cite[Ch.~4.3.1]{PBRT4e}.
The scattering pattern must be normalized to satisfy the condition
\begin{equation}
    \int_{0}^{\pi/2}\int_0^{2\pi} f_\text{s}\left(\hat{\mathbf{k}}_\text{i}, \hat{\mathbf{k}}_\text{s}, \hat{\mathbf{n}}\right) \sin(\theta_s) d\phi_s d\theta_s = 1
\end{equation}
which ensures the power balance between the incoming, reflected, and refracted fields.

\begin{highlightbox}{Example scattering patterns}
    The authors of~\cite{4052607} derived several simple scattering patterns that were shown to achieve good agreement with measurements when correctly parametrized.

\begin{description}
    \item[Lambertian:] This model describes a perfectly diffuse scattering surface whose \emph{scattering radiation lobe} has its maximum in the direction of the surface normal:
    \begin{equation}
        \label{eq:lambertian_model}
        f^\text{Lambert}_\text{s}\left(\hat{\mathbf{k}}_\text{i}, \hat{\mathbf{k}}_\text{s}, \hat{\mathbf{n}}\right) = \frac{\hat{\mathbf{n}}^\mathsf{T} \hat{\mathbf{k}}_\text{s} }{\pi} = \frac{\cos(\theta_s)}{\pi}
    \end{equation}

    \item[Directive:] This model assumes that the scattered field is concentrated around the direction of the specular reflection $\hat{\mathbf{k}}_\text{r}$ (defined in~\eqref{eq:reflected_refracted_vectors}). The width of the scattering lobe can be controlled via the integer parameter $\alpha_\text{R}=1,2,\dots$:
    \begin{equation}
        \label{eq:directive_model}
        \begin{aligned}
            f^\text{directive}_\text{s}\left(\hat{\mathbf{k}}_\text{i}, \hat{\mathbf{k}}_\text{s}, \hat{\mathbf{n}}\right) &= F_{\alpha_\text{R}}(\theta_i)^{-1} \left(\frac{ 1 + \hat{\mathbf{k}}_\text{r}^\mathsf{T} \hat{\mathbf{k}}_\text{s}}{2}\right)^{\alpha_\text{R}}\\
            F_{\alpha}(\theta_i) &= \frac{1}{2^\alpha} \sum_{k=0}^\alpha \binom{\alpha}{k} I_k,\qquad \theta_i =\cos^{-1}(-\hat{\mathbf{k}}_\text{i}^\mathsf{T}\hat{\mathbf{n}})\\
            I_k &= \frac{2\pi}{k+1} \begin{cases}
                                1 & k \text{ even} \\
                                \cos(\theta_i) \sum_{w=0}^{(k-1)/2} \binom{2w}{w} \frac{\sin^{2w}(\theta_i)}{2^{2w}}  & k \text{ odd}
                            \end{cases}
        \end{aligned}
    \end{equation}

    \item[Backscattering:] This model adds a scattering lobe to the directive model described above which points toward the direction from which the incident wave arrives (i.e. $-\hat{\mathbf{k}}_\text{i}$). The width of this lobe is controlled by the parameter $\alpha_\text{I}=1,2,\dots$. The parameter $\Lambda\in[0,1]$ determines the distribution of energy between both lobes. For $\Lambda=1$, this model reduces to the directive model.
    \begin{equation}
        \label{eq:backscattering_model}
        \begin{aligned}
            f^\text{bs}_\text{s}\left(\hat{\mathbf{k}}_\text{i}, \hat{\mathbf{k}}_\text{s}, \hat{\mathbf{n}}\right) &= F_{\alpha_\text{R}, \alpha_\text{I}}(\theta_i)^{-1} \left[ \Lambda \left(\frac{ 1 + \hat{\mathbf{k}}_\text{r}^\mathsf{T} \hat{\mathbf{k}}_\text{s}}{2}\right)^{\alpha_\text{R}} + (1-\Lambda) \left(\frac{ 1 - \hat{\mathbf{k}}_\text{i}^\mathsf{T} \hat{\mathbf{k}}_\text{s}}{2}\right)^{\alpha_\text{I}}\right]\\
            F_{\alpha, \beta}(\theta_i) &= \Lambda F_\alpha(\theta_i) + (1-\Lambda)F_\beta(\theta_i)
        \end{aligned}
    \end{equation}
\end{description}
\end{highlightbox}

\section{A Brief Overview of Importance Sampling}
\label{sec:importance-sampling}
This section provides a brief overview of importance sampling~\cite{Kloek1978, Wikipedia_Importance_Sampling} in the context of ray tracing for the simulation of radio wave propagation.
We consider the channel gain $g$~\eqref{eq:channel-gain}.
To account for a continuum of paths, such as those arising from diffuse reflections, we generalize the gain from a discrete sum to a path integral:
\begin{equation}
    g = \int_{\Pc} \abs{a(p)}^2 d\mu(p).
\end{equation}
Here, $\Pc$ represents the high-dimensional space of all possible propagation paths from the transmitter to the receiver.
A single element $p \in \Pc$ denotes a specific path.
The term $a(p)$ is the path coefficient associated with that path, and $\mu(p)$ is the measure on the path space $\Pc$.
For a more in-depth discussion of path integrals within ray tracing, see~\cite[Chapter 8]{Veach:1997:RMC}.

We aim to estimate the channel gain $g$ using $N_S$ samples, by sampling paths following a distribution $q(p)$, such that $q(p) > 0$ if $\abs{a(p)}^2 > 0$ and $\int_{\Pc} q(p) d\mu(p) = 1$.
For a ray-tracing algorithm, this full path density is induced jointly by the source-direction distribution, the conditional interaction-type probabilities, the conditional outgoing-direction distributions (e.g., for diffuse reflection and diffraction), the mapping from sampled directions to surface intersections, and the target-connection and candidate-selection rules.
The estimate of the channel gain is then given by:
\begin{equation}
    \widehat{g} = \frac{1}{N_S} \sum_{n=1}^{N_S} \frac{\abs{a(p_n)}^2}{q(p_n)}. \label{eq:channel-gain-estimate}
\end{equation}
An important consideration is that the estimate $\widehat{g}$ is unbiased, i.e., $\mathbb{E}[\widehat{g}] = g$ for any sampling distribution $q(\cdot)$.
Therefore, we aim to choose the sampling distribution $q(\cdot)$ such that the variance of the estimate $\widehat{g}$ is minimized.
Practically, this would result in estimating the channel gain with a smaller number of samples $N_S$, i.e., achieving a higher sample efficiency.
The variance of the estimator $\widehat{g}$ is given by:
\begin{align}
    \text{Var}(\widehat{g}) &= \mathbb{E}[\widehat{g}^2] - g^2 \\
                            &= \frac{1}{N_S^2}\mathbb{E}\left[ \sum_{n=1}^{N_S} \sum_{m=1}^{N_S} \frac{\abs{a(p_n)}^2 \abs{a(p_m)}^2}{q(p_n) q(p_m)} \right] - g^2 \\
                            &= \frac{1}{N_S}\left(\mathbb{E} \left[ \left( \frac{\abs{a(p)}^2}{q(p)} \right)^2 \right] - g^2\right).
\end{align}
where the last equality holds because the samples $p_n$ are independent.
Observe that
\begin{equation}
    \mathbb{E} \left[ \left( \frac{\abs{a(p)}^2}{q(p)} \right)^2 \right]
    = \int_{\Pc} q(p) \left( \frac{\abs{a(p)}^2}{q(p)} \right)^2 d\mu(p)
    = \int_{\Pc} \frac{\abs{a(p)}^4}{q(p)} d\mu(p).
\end{equation}
If we choose the sampling distribution
\begin{equation}
    q^*(p) \coloneqq \frac{\abs{a(p)}^2}{g}
\end{equation}
then
\begin{equation}
    \mathbb{E} \left[ \left( \frac{\abs{a(p)}^2}{q^*(p)} \right)^2 \right]
    = \int_{\Pc} g \abs{a(p)}^2 d\mu(p)
    = g^2.
\end{equation}
This implies that the variance of the estimator $\widehat{g}$ is zero, i.e., $\text{Var}(\widehat{g}) = 0$, which is optimal. A zero-variance estimator would allow us to determine the channel gain exactly from a single sample. In practice, however, this is unattainable because evaluating the optimal distribution $q^*(\cdot)$ requires prior knowledge of the path coefficients $a(p)$ and the channel gain $g$, i.e., the quantity we aim to estimate. Nevertheless, this result provides useful guidance: an effective sampling distribution $q(\cdot)$ should give more importance to paths that contribute more to the channel gain $g$.
The distribution $\Qc$, introduced in~\eqref{eq:path-solver-used-dist-event}, is designed as a practical choice by selecting interaction types based on the squared magnitudes of the corresponding reflection and refraction coefficients. However, it does not account for the antenna pattern, the free-space propagation loss, or the scattering pattern from diffuse reflection, and is thus suboptimal. Developing effective, practical importance sampling strategies remains an open and challenging problem.

\section{Closed-Form Solution for First-Order Diffraction}
\label{sec:diffraction-first-order}
For first-order diffraction, the position of the diffraction point can be determined in closed form.
Consider a source $\sv$, a target $\tv$, and an edge defined by its direction $\widehat{\ev}$.
For convenience, we translate the coordinate system so that the origin $\ov=\zerov$ lies on the edge.
It is assumed that the source and target don't lie on the edge.

According to Fermat's principle, the diffraction point $\vv$ on the edge minimizes the total path length $\sv \rightarrow \vv \rightarrow \tv$. This corresponds to minimizing the function
\begin{equation}
    \Lc \LB x \RB = \norm{\sv - \vv} + \norm{\vv - \tv}
\end{equation}
where
\begin{equation}
    \vv = x \widehat{\ev}.
\end{equation}
As shown in~\cite{4200889}, $\Lc\LB \cdot \RB$ is strictly convex and thus has a unique minimizer.

\begin{figure}[ht!]
    \center
    \begin{subfigure}[b]{0.45\textwidth}
        \centering
        \includegraphics[width=0.35\textwidth]{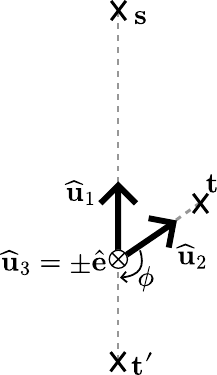}
        \caption{View in the plane $\LB \ov, \widehat{\uv}_1, \widehat{\uv}_2 \RB$, which is perpendicular to the edge $\widehat{\ev}$.}
        \label{fig:diffraction-first-order-a}
    \end{subfigure}
    \hfill
    \begin{subfigure}[b]{0.45\textwidth}
        \centering
        \includegraphics[width=0.5\textwidth]{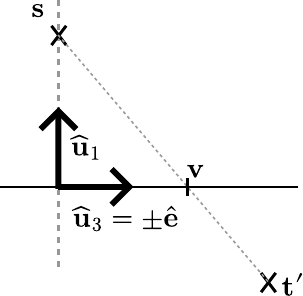}
        \caption{View in the plane $\LB \ov, \widehat{\uv}_1, \widehat{\ev} \RB$, containing $\sv$, the edge, and the rotated target $\tv'$.}
        \label{fig:diffraction-first-order-b}
    \end{subfigure}
    \caption{Illustration of rotating the target $\tv$ about the edge $\widehat{\ev}$ by the angle $\phi$ to obtain the rotated target $\tv'$, which is coplanar with the source $\sv$ and the edge. The diffraction point $\vv$ is such that $\sv$, $\vv$, and $\tv'$ are collinear.}
    \label{fig:diffraction-first-order}
\end{figure}

We introduce the set of vectors $\LB \widehat{\uv}_1, \widehat{\uv}_2, \widehat{\uv}_3 \RB$, where
\begin{align}
    \widehat{\uv}_1 &= \frac{\sv - \LB\widehat{\ev}\tp \sv\RB \widehat{\ev}}{\norm{\sv - \LB\widehat{\ev}\tp \sv\RB \widehat{\ev}}}\\
    \widehat{\uv}_2 &=\frac{\tv - \LB\widehat{\ev}\tp \tv\RB \widehat{\ev}}{\norm{\tv - \LB\widehat{\ev}\tp \tv\RB \widehat{\ev}}}\\
    \widehat{\uv}_3 &=
    \begin{cases}
        \frac{\widehat{\uv}_1 \times \widehat{\uv}_2}{\norm{\widehat{\uv}_1 \times \widehat{\uv}_2}} & \text{ if } \widehat{\uv}_1 \times \widehat{\uv}_2 \neq \zerov\\
        \widehat{\ev} & \text{ if } \widehat{\uv}_1 \times \widehat{\uv}_2 = \zerov
    \end{cases}
    .
\end{align}
Note that $\widehat{\uv}_3 = \pm \widehat{\ev}$, and that the source and the edge lie within the plane $\LB \ov, \widehat{\uv}_1, \widehat{\ev} \RB$.
A key observation is that rotating the target $\tv$ about the edge does not alter the path length.
Specifically, let $\Rm_{\widehat{\uv}_3} \LB \phi \RB$ denote the rotation matrix for an angle $\phi$ around the edge $\LB \ov, \widehat{\ev} \RB$. Then,
\begin{align}
    \Lc \LB x \RB &= \norm{\sv - \vv} + \norm{\vv - \tv}\\
    &= \norm{\sv - \vv} + \lVert\vv - \underbrace{\Rm_{\widehat{\uv}_3} \LB \phi \RB \tv}_{\tv'}\rVert_2\\
\end{align}
holds for any $\phi \in \RR$.
Especially, by rotating the target around the axis ($\ov, \widehat{\uv}_3$) by the angle
\begin{equation}
    \phi = \pi - \arccos \left( \widehat{\uv}_1\tp \widehat{\uv}_2 \right),
\end{equation}
the rotated target $\tv'$ is placed in the same plane $\LB \ov, \widehat{\uv}_1, \widehat{\ev} \RB$ as the source and the edge, but on the side opposite to the source with respect to the edge, as depicted in Figure~\ref{fig:diffraction-first-order}.
The rotation matrix $\Rm_{\widehat{\uv}_3} \LB \phi \RB$ is given by Rodrigues' rotation formula~\cite{Wikipedia_Rodrigues}.

Minimizing $\Lc\LB x \RB$ with the rotated target $\tv'$ is equivalent to $\sv$, $\vv$, and $\tv'$ being collinear, i.e.
\begin{align}
    (\vv - \sv) \times \overrightarrow{\sv\tv'} &= \zerov \\
    \Leftrightarrow\quad x \left( \widehat{\ev} \times \overrightarrow{\sv\tv'} \right) &= \sv \times \overrightarrow{\sv\tv'}
\end{align}
where $\overrightarrow{\sv\tv'} = \tv' - \sv$.
Since $\widehat{\ev} \times \overrightarrow{\sv\tv'}$ and $\sv \times \overrightarrow{\sv\tv'}$ are parallel, the solution for $x$ is
\begin{equation}
    x = \operatorname{sgn} \left( \left[ \widehat{\ev} \times \overrightarrow{\sv\tv'} \right]\tp \left[\sv \times \overrightarrow{\sv\tv'}\right] \right)
    \frac{ \norm{ \sv \times \overrightarrow{\sv\tv'} } }
         { \norm{ \widehat{\ev} \times \overrightarrow{\sv\tv'} } }.
\end{equation}

\section{Weighting Factor for Diffraction Radio Maps}
\label{sec:diffraction-radio-map-weighting-factor}
This section details the computation of the weighting factor $\norm{ \frac{\partial \tv}{\partial x} \times \frac{\partial \tv}{\partial \phi} }$, required for the calculation of the radio map due to diffraction as described in Section~\ref{sec:radio-map-solver-diffraction}.
Recall that $\tv(x, \phi)$ denotes the reparametrization of a point on the measurement cell $M_i$ reached by a diffracted ray originating from the diffraction point $\vv$ on the edge $\Ec$, with $\phi$ representing the Keller cone azimuth.

We begin by rewriting~\eqref{eq:radio-map-solver-ci-D-reparametrization} as
\begin{equation}
    \label{eq:diffraction-radio-map-weighting-factor-reparametrization}
    \tv(x,\phi) = \vv(x) + \gamma \widehat{\kv}_s(x, \phi),
\end{equation}
where $\vv(x) = \ov + x \widehat{\ev}$ is the diffraction point along the edge $\Ec$, and $\widehat{\kv}_s$ denotes the direction of the diffracted ray, defined by the Keller cone azimuth $\phi$ and the opening angle $\beta_0'$:
\begin{equation}
    \label{eq:diffraction-radio-map-weighting-factor-kv-s}
    \widehat{\kv}_s(x, \phi) = \sin\beta_0' \cos\phi\, \widehat{\tv}_0 + \sin\beta_0' \sin\phi\, \widehat{\nv}_0 + \cos\beta_0'\, \widehat{\ev}.
\end{equation}
The dependencies of $\vv$ and $\widehat{\kv}_s$ on $x$ and $\phi$ are shown explicitly for clarity.
Let $\widehat{\nv}$ denote the normal to the measurement cell $M_i$, and $\qv$ a point on the plane containing the measurement cell.
Since $\tv$ lies on $M_i$, $\gamma$ can be found by requiring
\begin{equation}
    \widehat{\nv}\tp \LB \tv(x, \phi) - \qv \RB = 0
\end{equation}
which yields
\begin{equation}
    \label{eq:diffraction-radio-map-weighting-factor-gamma}
    \gamma = \frac{\widehat{\nv}\tp \LB \qv - \vv(x) \RB}{\widehat{\nv}\tp \widehat{\kv}_s(x, \phi)}.
\end{equation}
Additionally, for a fixed source position $\sv$, the angle of incidence $\beta_0'$ depends on the location of the diffraction point $\vv(x)$ along edge $\Ec$
\begin{equation}
    \begin{aligned}
        \cos\beta_0' &= \frac{\LB\vv(x) - \sv\RB\tp\widehat{\ev}}{\norm{\vv(x) - \sv}}, \\
        \sin\beta_0' &= \frac{\norm{\sv - \sv_{\Ec}}}{\norm{\vv(x) - \sv}}
    \end{aligned}
    \label{eq:diffraction-radio-map-weighting-factor-beta-0}
\end{equation}
where $\sv_{\Ec}$ denotes the projection of the source point $\sv$ onto the edge $\Ec$.

Equations~\eqref{eq:diffraction-radio-map-weighting-factor-reparametrization}, \eqref{eq:diffraction-radio-map-weighting-factor-kv-s}, \eqref{eq:diffraction-radio-map-weighting-factor-gamma}, and \eqref{eq:diffraction-radio-map-weighting-factor-beta-0} together provide the reparametrization of $\tv$ in terms of $x$ and $\phi$, for a given source position $\sv$.
In \SRT{}, the derivatives of $\tv$ with respect to $x$ and $\phi$ are computed via automatic differentiation.
Specifically, Dr.Jit~\cite{Jakob2022DrJit} is employed to obtain $\nicefrac{\partial \tv}{\partial x} \in \RR^3$ and $\nicefrac{\partial \tv}{\partial \phi} \in \RR^3$ by differentiating through these equations. The weighting factor is then evaluated as the norm of the cross product of these two derivatives.
One could also have implemented the derivatives by hand, but using automatic differentiation simplifies the implementation without incurring significant overhead.

\newpage


\section*{Version History}

\renewcommand{\arraystretch}{1.4}
\begin{tabular}{cll}
\hline
\textbf{Version} & \textbf{Date} & \textbf{Description} \\
\hline
2.1 & September 2026 & Apply corrections and overall improvements \\
1.2 & September 2025 & Support for first-order diffraction and path solver improvements \\
1.1 & June 2025 & Support for mesh-based measurement surfaces \\
1.0 & April 2025 & \textbf{Major release:} Complete re-architecture of Sionna~RT \\
\hline
\end{tabular}
\renewcommand{\arraystretch}{1.0}

\newpage

\bibliographystyle{IEEEtran}
\bibliography{IEEEabrv,bibliography}

\end{document}